\documentclass[12pt]{article}

\usepackage{amsmath}
\usepackage{amssymb}
\usepackage{parskip}

\newcommand{\be}{\begin{equation}}
\newcommand{\ee}{\end{equation}}
\newcommand{\bea}{\begin{eqnarray}}
\newcommand{\eea}{\end{eqnarray}}
\newcommand{\bi}{\begin{itemize}}
\newcommand{\ei}{\end{itemize}}
\newcommand{\cH}{{\cal H}}

\newcommand{\R}{\mathbb R}
\newcommand{\Z}{\mathbb Z}
\newcommand{\NN}{\mathbb N}
\newcommand{\sfrac}[2]{\mbox{$\frac{#1}{#2}$}}
\newcommand\ointint{\begingroup
\displaystyle \unitlength 1pt
\int\mkern-12.2mu
\begin{picture}(0,3)
\put(0,1){\oval(5,5)}
\end{picture}\mkern 6mu
\endgroup}

\usepackage{etoolbox}
\apptocmd\normalsize{%
 \abovedisplayskip=20pt plus 3pt minus 9pt
 \abovedisplayshortskip=8pt plus 3pt
 \belowdisplayskip=20pt plus 3pt minus 9pt
 \belowdisplayshortskip=15pt plus 3pt minus 4pt
}{}{}

\begin{document}
{\center
{\normalsize Yerevan State University}
\\[150pt]
{\large Armen Saghatelian}
\\[25pt]
{\huge Action-Angle Variables In Conformal Mechanics}
\\[25pt]
{\large PhD Thesis}
\\
{\large 01.04.02 - Theoretical Physics}
\\[25pt]
{{\normalsize \it Adviser:  Professor  Armen Nersessian}}
\\[230pt]
{\large Yerevan - 2014}

}

\thispagestyle{empty}
\mbox{}
\newpage

\tableofcontents
\newpage
\section*{INTRODUCTION}
\addcontentsline{toc}{section}{INTRODUCTION}

It is  well-known that, for the systems with finite motion, one can introduce the distinguished set of phase space variables (the ``action-angle" variables),   such that the ``angle" variables parameterize a torus, while their conjugated ``action" variables are functions of constants of motion only \cite{arnold}. As a consequence, the  Hamiltonian depends only on action variables. The formulation of an integrable system in these variables gives us a comprehensive  geometric description of its dynamics. Such a formulation defines a useful tool for the developing of perturbation theory \cite{arnold,goldstein}, since the ``action" variables define adiabatic invariants of the system.  The action-angle formulation is important from the quantum-mechanical point of view as well,  since in action-angle variables the Bohr-Sommerfeld quantization  is equivalent to the canonical quantization, with trivial expressions for the wavefunctions.  Hence,  evaluation of quantum-mechanical aspects of such system becomes quite simple in this approach.

Besides the practical importance, the action-angle formulation has an academic interest as well. From the academic viewpoint, it gives a precise indication of the (non)equivalence of different Hamiltonian systems. Indeed, gauging the integrable system by action-angle variables, we preserve the freedom only in the functional dependence of the Hamiltonian from the action variables, $H=H({\bf I})$, and in the range of validity of the action variables, $I_i\in [\beta^{-}_i, \beta ^{+}_i]$. Hence formulating the systems in terms of action-angle variables, we can indicate the (non)equivalence of different integrable systems. Let us refer, in this respect, to the recent paper \cite{lny}, in which, particularly, the global equivalence of $A_2$ and $G_2$ rational Calogero models, and their global equivalence with a free particle on the circle, has been established in this way.

Due to the recent progress in nanotechnology, now the fabrication of various
low-dimensional systems of complicated geometric form (nanotubes,
nanofibers, spherical and cylindrical layers) is becoming possible \cite{nano}. In these context the  methods of quantum mechanics on
curved space should be relevant for the description of the physics of
nanostructures. The common method for the localizing of the particle in the disc or in the cylinder is that of the two-dimensional oscillator for  the role of the confinement potential.
Similarly, for the localization of
 the particle in quantum lens (e.g. $GaAs/In_{1-x}Ga_xAs$, see \cite{lens})
  one can use the Higgs model of the spherical oscillator
  defined by the potential $V_{Higgs}=\frac 12 \omega^2r^2_0\tan^2\theta $  \cite{higgs}.  Another confinement potential which could be used for the localization of the (quasi)particles
  in quantum lens, is the potential of the so-called $CP^1$ oscillator
  $V_{Higgs}=2 \omega^2r^2_0\tan^2\theta/2 $ \cite{Bellucci}.
The advantage of the latter potential is that a system with such confinement potential preserves the exact solvability after inclusion of magnetic field, which has a constant magnitude on the sphere surface.

The fabrication of semiconductor ring-shaped systems \cite{qr}, presently referred to as  quantum rings (e.g. $In(Ga)As$ - two-dimensional quantum rings), led to the use of the singular oscillator potential with the role of the confinement one. A pioneering  work on the  theoretical study of the impact of the magnetic field on the  electron properties in a quantum ring was  written by Chakraborty and Pietelainen \cite{qdch}.  In that paper the shifted oscillator potential $V_{\rm ChP}=\beta(r-r_0)^2$ was choosen as the confinement potential restricting the motion of electrons in the quantum ring. The results obtained within this approximation are in  a good correspondence with experimental data. The quantum ring model of Chakraborty  and Pietelainen  is not exactly solvable in the general case, it assumes  the use of numerical  simulations.  The quantum ring model based on the singular oscillator system  \cite{qdsw} has been suggested   as an analytically  solvable alternative to the Chakraborty-Pietelainen model. Although calculations performed within the Chakraborty-Pietelainen model are in  better correspondence with experimental data than those within the singular oscillator potential \cite{comp}, the latter has its own place in the study of quantum rings (see,e.g. \cite{hsarkisyan}).

In analogy with the above models, one can suppose, that  singular versions of two-dimensional Higgs and $CP^1$ oscillators may be appropriate candidates for the confinement  potential localizing the motion of the electron in the ring of a spherical quantum layer.

In Chapter \ref{aavar} we start with presenting the general procedure of constructing the action-angle variables for an arbitrary system with finite motion. Then, in Section \ref{quantumrings}, action-angle variables are used for the study of a quantum ring model with two-dimensional singular oscillator potential, and of its two spherical generalizations, based on the Higgs and $CP^1$  spherical oscillator potentials. It is easy to observe that the (singular) Higgs oscillator does not preserve its exact solvability in the presence of a constant magnetic field, in contrast to the Euclidean one. While the study of quantum dot systems in a magnetic field is of a special physical importance. In contrast with the Higgs oscillator, the singurar $CP^1$ oscillator preserves the exact solvability property upon inclusion of the constant magnetic field. These tell us the area of application of the Higgs oscillator potential and of the $CP^1$ oscillator one. The Higgs model is useful for the behavior of the quantum dots systems in the external potential field, e.g, in the electric field. The $CP^1$ model should be applied for the study of the behavior of a spherical quantum dots model in the external magnetic field.

The aim of Chapter \ref{oscilAndCoulomb} is to construct integrable generalizations of the well-known oscillator and Coulomb systems on $N$-dimensional Euclidian space $R^N$, sphere $S^N$ and hyperboloid $H^N$. We show, that if we have an “angular” (spherical) Hamiltonian, we can add a “radial” part to it, thus increasing the dimension by one (the phase space dimension is increased by two). We compute the explicit expressions for action-angle variables for systems with oscillator and Coulomb potentials. Using this formulae, we prove the superintegrability of  Tremblay-Turbiner-Winternitz (TTW, \cite{TTW, TTWothers})  and Post-Winternitz (PW, \cite{PW}) models. Then we construct the spherical and pseudospherical generalizations of the TTW and PW systems, write down their hidden constants of motion, thus demonstrating the superintegrability of these new systems. Additionally, we provide the action-angle variables for a free particle on the $(N{-}1)$-dimensional sphere, which yields the complete set of action-angle variables for the $N$-dimensional oscillator and Coulomb systems as well as their spherical and pseudospherical analogs.

Conformal invariance plays an important role in many areas of the quantum field theory and condensed matter physics, especially in the string theory, the theory of critical phenomena, low-dimensional integrable models, spin and fermion lattice systems. The term ``conformal mechanics'' denotes a system whose Hamiltonian ${H}$, together with the dilatation generator ${D}$ and the generator ${K}$ of conformal boosts forms, with respect to Poisson brackets, the conformal $so(2,1)$ algebra. This property allows one to introduce a coordinate transformation which transforms the Hamiltonian into the "conventional" Hamiltonian of (one-dimensional) conformal mechanics \cite{Alfaro}. This means, that with the corresponding selection of new radial coordinate and momentum, we split the generic conformal mechanical system into a “radial” and an “angular” part. The latter defines a new Hamiltonian system on the orbit of the conformal group, with Casimir function $\mathcal{I}$  of the conformal algebra $so(2,1)$ in the role of the Hamiltonian. Casimir function $\mathcal{I}$ does not depend on the new radial coordinate and momentum, and is called the spherical or angular part of the Hamiltonian. It commutes with all generators of $so(2,1)$ and defines a constant of motion of the initial Hamiltonian ${H}$. So, although conformal symmetry is not a symmetry of the Hamiltonian, it equippes the system with the additional (to the Hamiltonian) constant of motion.

Chapter \ref{confmech} is devoted to the study of conformal mechanics. We develop a general approach to the constants of motion for conformal mechanics, based on $so(3)$ representation theory. We present the procedure of separating the “radial” and “angular” parts of conformal mechanics. Then we study the angular part as a new Hamiltonian system (master mechanics) with finite motion. We find the constants of motion of master mechanics from the constants of motion of the initial conformal system. We illustrate the effectiveness of our method on the example of the rational $A_3$ Calogero model and its spherical mechanics (which defines the cuboctahedric Higgs oscillator). For the latter we construct a complete set of functionally independent constants of motion, proving its superintegrability.

The suggested method, i.e., separating the radial and angular parts of conformal mechanics and studying the angular part using the action-angle variables, was effectively implemented in the study of extremal black holes.

The black hole solutions allowed in supersymmetric field theories have an extremality property, that is, the inner and outer horizons of the black hole coalesce. In this case one can pass to the near-horizon limit, which brings us to new solutions of Einstein equations. In this limit (near-horizon extremal black hole) the solutions become conformal invariant. The conformal invariance was one of the main reasons why the extremal black holes have been payed so much attention to for the last fifteen years. Indeed, due to conformal invariance black hole solutions are a good research area for studying conformal field theories and AdS/CFT correspondence (for the recent review see \cite{bkls}). The simplest way to research this type of configurations is to study the motion of a (super)particle in this background. The first paper that considered such a problem is \cite{cdkktp}, where the motion of particle near horizon of extremal Reissner-Nordstr\"om black hole has been considered. Later similar problems in various extremal black hole backgrounds were studied by several authors (see \cite{ikn,bk,galaj} and refs therein).

In Chapter \ref{BlackHoles1} we study the conformal mechanics associated with near-horizon motion of massive relativistic particle in the field of extremal black holes in arbitrary dimensions. In \ref{BlackHolesConformal} we prove, that by applying a proper canonical transformation one can bring the above mentioned model to the conventional conformal mechanics form. Important information about the $d$-dimensional system, is thus imprinted in the $(d-2)$--dimensional spherical mechanics. In \ref{nearHorizonGeneral} we demonstrate the near-horizon limit of extremal black holes on the cases of 4-dimensional Kerr black hole and higher-dimensional Myers-Perry black holes.

Section \ref{4dimens} is devoted to study of the following two 4-dimensional exactly
solvable systems:
\begin{itemize}
\item
Charged particle moving near the horizon of extremal Reissner-Nordstr\"om  black hole with magnetic momentum,
\item
Particle moving near the horizon of extremal Cl\'ement-Gal'tsov black hole.
\end{itemize}

We construct the action-angle variables for the angular parts of this systems. We show that the angular part of a charged particle moving near the horizon of Reissner-Nordstr\"om  black hole is equivalent to the spherical Landau problem, and has a hidden constant of motion. We find a ``critical point" that divides the different phases of effective periodic  motion. Then we discuss a charged particle moving near the horizon of extremal Cl\'ement-Gal'tsov   black hole. In contrast with Reissner-Nordstr\"om case, this system does not possess hidden constant of motion. We find a critical  point that divide the phases (both effectively two-dimensional ones) of rotations  in opposite directions.

In Chapter \ref{higherdim} we analyse the integrability of spherical mechanics models associated with the near horizon extremal Myers-Perry black hole in arbitrary dimension for the special case that all rotation parameters are equal. We prove the superintegrability of the system and show that the spherical mechanics associated with the black hole in odd dimensions is maximally superintegrable, while its even-dimentsional counterpart lacks for only one constant of the motion to be maximally superintegrable.

\newpage
\setcounter{equation}{0}
\section{ACTION-ANGLE VARIABLES}
\label{aavar}
\subsection{Action-angle variables: General description}
The well-known Liouville theorem gives the exact criterium of integrability of the
$N$-dimensional mechanical system (integrability in the Liouville sense or, so-called, Liouville integrability): that is the existence of   $N$ mutually commuting constants
of motion $F_1=H,\ldots, F_n$: $\{F_i, F_j\}=0$,$i,j=1,\ldots N$.
The theorem also states that if the level surface $M_f=\left( (p_i,q_i):
F_i=const\right)$ is a {\sl compact and connective manifold}, then it
is diffeomorphic to an $N$-dimensional torus $T^N$. The natural angular
coordinates ${\bf \Phi}=(\Phi_1,\ldots, \Phi_N)$ parameterizing that torus satisfy the motion
equations of a free particle moving on a circle. These coordinates form,  with
their conjugate momenta ${\bf I}=(I_1,\ldots, I_N)$, a full set of phase space  variables
called ``action-angle'' variables. One of the
results of the theorem is that the momenta ${\bf I}$ depend on constants of
motion only: ${\bf I}={\bf I}({\bf F})$ (which makes ${\bf I}$ a new set of constants of motion). So, there must be a
canonical transformation to the new variables $({\bf p},{\bf q})\mapsto
({\bf I},{\bf \Phi})$, in which the Hamiltonian depends  on the constants of
motion ${\bf I}$ (which are called action variables) only. Consequently, the
equations of motion read
 \be
\frac{d{\bf I}}{dt}=0,\quad \frac{d{\bf \Phi}}{dt}=\frac{\partial
H(I)}{\partial {\bf I}}%\equiv \omega_i(I),
\qquad \{ I_i, \Phi_j
\}=\delta_{ij},\qquad  \Phi_i\in [0,2\pi ),\quad i,j=1,\ldots, N.
 \ee
 The formulation of the integrable system  in action-angle variables gives us
a comprehensive  geometric description of its dynamics \cite{arnold,goldstein}. Such a
formulation defines a useful tool for the developing of perturbation
theory, since the ``action" variables define adiabatic invariants of
the system.
 The action-angle formulation is important
from the quantum-mechanical point of view as well,
 since in action-angle variables the Bohr-Sommerfeld quantization
 is equivalent to the canonical quantization,
with trivial expressions for the wavefunctions.
 Hence,  evaluation of quantum-mechanical
aspects of such system becomes quite simple in this approach.

Besides the practical importance, the action-angle formulation has an academic interest as well.
From the academic viewpoint, it gives a
precise indication of the (non)equivalence of different
Hamiltonian systems. Indeed, gauging the integrable system by
action-angle variables, we preserve the freedom only in the
functional dependence of the Hamiltonian from the action
variables, $H=H({\bf I})$, and
in the range of validity of the action variables, $I_i\in [\beta^{-}_i, \beta ^{+}_i]$.
Hence formulating the systems in terms of action-angle variables, we can indicate the (non)equivalence
of different integrable systems. Let us refer, in this respect, to the recent paper \cite{lny}, where, particularly,
the global equivalence of $A_2$ and $G_2$ rational Calogero models, and their global equivalence
with a free particle on the circle, has been established in this way.

In action-angle variables the Bohr-Sommerfeld quantization is equivalent to the canonical quantization,
with a quite simple  expression for the wavefunction
\be {\widehat I}_i\Psi(\Phi )= I_i\Psi(\Phi),\qquad
{\widehat I}_i=-\imath\hbar \frac{\partial}{\partial \Phi_i},\quad
\Psi=\frac{1}{({2\pi})^{N/2}}\prod_{i=1}^{N}{\rm e}^{-\imath\; n_{i}\Phi_i},\quad I_i=\hbar  n_{i},
\label{spectra}\ee
where $n_{i}$ are integer numbers taking their values at the range
$[\beta^{-}_i, \beta^{+}_{i}]$.

The general prescription for the construction of action-angle
variables  looks  as follows \cite{arnold}.
 In order to construct the action-angle  variables, we should fix the level
surface of the Hamiltonian
 ${\bf F}={\bf c}$ and then introduce the generating function
 for the canonical transformation $({\bf p},{\bf q})\mapsto({\bf  I},{\bf \Phi})$,
which is defined by the expression
\be S({\bf c}, {\bf q})=\int_{{\bf F}={\bf c}} {\bf p} d{\bf q},\label{sdef} \ee where ${\bf p}$
are expressed via ${\bf c}, {\bf q}$ by the use of the constants of
motion.
The action variables ${\bf I}$ can be obtained from the expression \be
I_i(c)=\frac{1}{2\pi}\oint_{\gamma_i} {\bf p}d{\bf q}, \ee
where $\gamma_i$
is some loop of the level surface ${\bf F}={\bf c}$. Then inverting these relations, we can get the
expressions of ${\bf c}$ via action variables: ${\bf c}={\bf c}({\bf I})$.
The angle variables ${\bf \Phi}$ can be found from the expression \be
{\bf \Phi}= \frac{\partial S({\bf c}({\bf I}), {\bf q})}{\partial {\bf I}}. \ee

\newpage
\subsection{Quantum ring models}
\label{quantumrings}
So, the action-angle variables form a useful tool for the study   of
systems with finite motion. But just such systems presently attract much attention  because of the progress in mesoscopic physics, where
we usually deal with a motion of (quasi)particles localized in quantum dots, quantum layers etc. Due to the recent progress in nanotechnology, now the fabrication of various
low-dimensional systems of complicated geometric form (nanotubes,
nanofibers, spherical and cylindrical layers) become possible \cite{nano}. In these context the  methods of quantum mechanics on
curved space should be relevant for the description of the physics of
nanostructures.

Say, the common method for the localizing of the particle in the disc or in the cylinder is that of the two-dimensional oscillator for  the role of the confinement potential.
Similarly, for the localization of
 the particle in quantum lens (e.g. $GaAs/In_{1-x}Ga_xAs$, see \cite{lens})
  one can use the Higgs model of the spherical oscillator
  defined by the potential $V_{Higgs}=\frac 12 \omega^2r^2_0\tan^2\theta $  \cite{higgs}.

Another confinement potential which could be used for the localization of the (quasi)particles
  in quantum lens, is the potential of the so-called $CP^1$ oscillator
  $V_{Higgs}=2 \omega^2r^2_0\tan^2\theta/2 $ \cite{Bellucci}.
The advantage  of the latter potential is that a system with such confinement potential preservs the exact solvability after inclusion of  magnetic field, which has a constant magnitude on the surface of the sphere.
Such a magnetic field is precisely the magnetic field of a Dirac monopole located at the center of sphere.
So, formally this is not a physical field. However, due to the restriction of the electron in the
segment/ring of the spherical layer, it could be viewed as a physical field generated e.g. by the pole of a magnetic dipole.

The fabrication of semiconductor ring-shaped systems \cite{qr},
presently referred to as  quantum rings (e.g.
$In(Ga)As$ - two-dimensional quantum rings), led to the use of the
singular oscillator potential with the role of the confinement one.
A pioneering  work on the  theoretical study of the
impact of the magnetic field on the  electron properties in a quantum
ring was  written by Chakraborty and Pietelainen \cite{qdch}.
 In that paper the shifted oscillator potential $V_{\rm ChP}=\beta(r-r_0)^2$ was choosen as the confinement potential
restricting the motion of electrons in the quantum ring. The
results obtained within this approximation are in
 a good correspondence with experimental data. The
quantum ring model of Chakraborty  and Pietelainen  is
not exactly solvable in the general case, it assumes
 the use of numerical  simulations.
 The quantum ring model based on the singular oscillator system  \cite{qdsw} has been suggested
  as an analytically  solvable alternative to the Chakraborty-Pietelainen model.
 Although calculations performed within the Chakraborty-Pietelainen model are
 in  better correspondence with experimental data
 than those within the singular oscillator potential \cite{comp}, the latter has its own place
 in the study of quantum rings (see,e.g. \cite{hsarkisyan}).

In analogy with the above models, one can suppose, that
 singular versions of two-dimensional Higgs and $CP^1$ oscillators may
be appropriate candidates for the confinement
 potential localizing the motion of the electron in the ring of a spherical quantum layer.

In this section we present the action-angle formulations of the two-dimensional singular oscillator and its two spherical generalizations based on Higgs and $CP^1$ spherical oscillator models. This section is based on the results of \cite{bnsy}. The goals of this section are to suggest
\begin{itemize}
\item
 To use the
 action-angle variables in  the study of quantum ring models.
\item
 To use singular spherical oscillator models as confinement potentials in spherical quantum rings.
\end{itemize}

For the role of the constant magnetic field, the magnetic field of the Dirac monopole located at the center of the sphere is suggested. Surely, the Dirac monopole is a non-physical object. However, since we assume to use it for the description of the particles localized on a part of the sphere, the non-physical nature of the Dirac monopole can be ignored. The monopole can be considered, e.g. as a pole of the magnetic dipole. The possible impact of the Dirac monopole on the properties of quantum dots model has been considered, e.g., in \cite{monop}. Besides, magnetic monopoles emerge as a class of magnets known as spin ice \cite{si}.

\newpage
\subsubsection*{Singular Euclidean oscillator}
Let us demonstrate our approach with the simplest example of the
singular oscillator on the two-dimensional Euclidean space, which is defined by the
Hamiltonian
\be H=\frac{{\bf
p}^2}{2}+\frac{\alpha^2}{2{\bf r}^2} + \frac{\omega^2 {\bf r}^2}{2}.
\label{singosc} \ee

In polar coordinates this Hamiltonian reads
\be H=\frac{p_r^2}{2}+\frac{p_\varphi^2+\alpha^2}{2 r^2} +
\frac{\omega^2 r^2}{2},\qquad  x=r\cos\varphi,\quad y=r\sin\varphi. \ee
Taking into account that the angular momentum $p_\varphi$ is the constant of motion of this system,
we can represent  its generating function as follows:
$S(p_\varphi, h,\varphi, r )= p_\varphi\varphi +\int_{H=h} p_r dr$.
So,
for the action variables we get the expressions
\be
\begin{split}
I_1 &= \frac{1}{2 \pi} \ointint p_\varphi d \varphi = p_\varphi,
\\
I_2 &=\frac{1}{2 \pi} \ointint p_r d r = \frac{h}{2\omega} - \frac{\widetilde{p}_\varphi}{2}
\\
&{\rm where}\quad\widetilde{p}_\varphi
\equiv\sqrt{p_{\varphi}^2+\alpha^2}
\end{split}
\ee
Respectively, the Hamiltonian takes the form
\be
H_{2d}=\omega\left(2I_2+\sqrt{I^2_1+\alpha^2}\right)%\equiv 2\omega(I_2+\tilde{I_1}),\quad \tilde{I}_1 \in [\alpha,\infty)
\ee
The angle variables read
\be
\begin{split}
&\Phi_1 = \varphi
-\frac{p_\varphi}{2\tilde{p}_\varphi}
\arcsin
\frac{\left(\widetilde{p}_{\varphi} + \omega r^2 \right) \sqrt{2 h r^2-\widetilde{p}_{\varphi}^2-\omega^2 r^4}}{(h+\widetilde{p}_{\varphi} \omega) r^2}
,\\
&\Phi_2 =
-\arcsin\frac{h- r^2 \omega ^2}{\sqrt{h^2-\widetilde{p}_\varphi^2
\omega ^2}}.
\end{split}
\ee
For the reduction of this system to a one-dimensional one, we should put $p_\varphi=0$. In that case the
Hamiltonian takes the form (where we replaced  $r$ by $x$)
$H_{1d}=\omega(2I_2+{\alpha})\equiv 2\omega\tilde{I},\quad \tilde{I} \in [{\alpha}/{2},\infty)$.
So, in the action-angle variable the one-dimensional singular oscillator is locally
equivalent to the nonsingular one. The only difference is in the range of validity of the action variable.

Let us notice that the action variable corresponding to the cyclic coordinate $\varphi$ coincides with the
angular momentum
$I_1=p_\varphi$. However, the respective angle variable $\Phi_1$ is different from the initial angle $\phi$.
In other words, the ``radial" motion, encoded in the dynamic of $I_2$ and $\Phi_2$ variables, has an essential
impact
on the `` angular" motion. While the impact of $\varphi, p_\varphi $ variables in the radial motion is the
shift $\alpha^2\to\alpha^2+p_{\varphi}^2$.

The inclusion of the constant magnetic field in the two-dimensional oscillator system does not
essentially change its properties.
Indeed, it is defined, in the two-dimensional planar system, by the potential
\be {\cal A}=\frac{B_0}{2} (xdy-ydx) = \frac{B_0
r^2}{2} d\varphi \ee
Hence, including the constant magnetic field in the two-dimensional singular oscillator, we shall get
\be H=\frac{p_r^2}{2}+\frac{(p_\varphi-\frac{B_0 r^2}{2})^2}{2
r^2}+\frac{\alpha^2}{2r^2} + \frac{\omega^2 r^2}{2} \qquad \Leftrightarrow \qquad \widetilde{H}=
\frac{p_r^2}{2}+\frac{\widetilde{p}^2_\varphi}{2
r^2}+ \frac{\widetilde{\omega}^2 r^2}{2}, \label{2osc} \ee
where we use the notation
\be {\widetilde{p}_{\varphi}^2} =
{p_{\varphi}^2} + \alpha^2 ,\quad\widetilde{\omega}^2 =
\omega^2+\frac{B_0^2}{4},\quad
 \widetilde{H} =H + \frac{B_0 p_\varphi}{2}  \label{substit}\ee

 Thus, the impact of the magnetic field  in the generating function $S(h,p_\varphi, r, \varphi)$ consists in the replacement (\ref{substit}).
Respectively,
the action variables and Hamiltonian are defined by the expressions
\be
I_1=p_\varphi,\quad
I_2=
\frac{\widetilde{h}}{2\widetilde{\omega}} -
\frac{\widetilde{p}_\varphi}{2} \qquad\Rightarrow\qquad
H=\sqrt{\omega^2+(B_0/2)^2} \left(2I_2+\sqrt{I_1^2+\alpha^2} \right)-\frac{B_0I_1}{2}.
\label{hsaa}\ee
The explicit expressions for angle variables reads
\be
\begin{split}
&\Phi_1 =\varphi - \frac{p_\varphi}{2 \widetilde{p}_{\varphi}}
\arcsin
\frac{\left(\widetilde{p}_{\varphi} + \widetilde{\omega} r^2 \right) \sqrt{2 \tilde{h} r^2-\widetilde{p}_{\varphi}^2-\widetilde{\omega}^2 r^4}}{(\tilde{h} + \widetilde{p}_{\varphi} \widetilde{\omega}) r^2}
,\qquad
\\
&\Phi_2 =
- \arcsin\frac{\widetilde{h}- r^2
\widetilde{\omega}^2}{\sqrt{\widetilde{h}^2-\widetilde{p}_\varphi^2
\widetilde{\omega}^2}}.
\end{split}
\ee
It is seen that the magnetic field yields in the Hamiltonian the term linear on $I_1$,
in addition to the predictable change of the effective frequency $\omega\to\sqrt{\omega^2+B^2_0/4}$.

We have constructed the action-angle variables for the
two-dimensional singular oscillator in the constant magnetic field.
Now we shall consider a similar formulation for the models of singular spherical oscillators.

\newpage
\subsubsection*{Singular Higgs  oscillator}
 Now we shall consider a two-dimensional singular
spherical oscillator defined  by the following Hamiltonian:
\be
H_{\rm Higgs}=\frac{p_{\theta}^2}{2r^2_0}+\frac{
p_{\varphi}^2}{2r^2_0 \sin ^2\theta}+\frac{\alpha^2}{2r^2_0}
\cot^2\theta+\frac{\omega^2r^2_0}{2} \tan^2\theta ,\label{higgs}\ee
where $r_0$ is the radius of the sphere.

This system generalizes the well-known Higgs model of the spherical
oscillator \cite{higgs}, whose
uniqueness is in the closeness
  of all trajectories,
 which reflects   the existence of a number of  hidden symmetries    equal to the those
 of the Euclidean oscillator.
This is the reason why the Higgs oscillator is a convenient background for the
developing of perturbation theory. Particularly, it admits the
anisotropic modification  preserving the integrability of the system \cite{anosc}. Hence, such a model of the spherical ring should be convenient for the study of electrons behavior in external potential fields, e.g., in the electric one. However, it is easy to observe that the (singular) Higgs oscillator does not preserve its exact solvability in the presence of a constant magnetic field, in contrast with the Euclidean one, while the study of quantum dot systems in a magnetic field is of a special physical importance.

In our  consideration we assume the unit radius of the sphere, $r_0=1$. The restoration of the the arbitrary radius can be carried out by the obvious redefinition of the
Hamiltonian and the constants $\alpha,\omega$.

Since the angular momentum $p_\varphi $ is a constant of motion of the system, the
the generating function of the action-angle variables takes the form
\be
S=p_\varphi \varphi+\int p_\theta( h, p_\varphi, \theta ) d\theta ,
\ee
where $H_{\rm Higgs}=h$.
From this generating function we get the action variables
\be
\begin{split}
&I_1= \frac{1}{2 \pi} \ointint p_\varphi d \varphi = p_\varphi\;,
\\
&I_2=\frac{1}{2\pi}\ointint p_\theta d\theta =
\frac{1}{\pi}\int_{\theta_-}^{\theta_+} \sqrt{2\left(h-\frac{p_{\varphi }^2}{2 \sin
^2\theta }-\frac{\alpha^2}{2} \cot^2\theta-\frac{\omega^2}{2} \tan^2\theta\right)}
d\theta ,\label{higgsactang}
\end{split}
\ee
where the integration limits $\theta_\pm $ are defined by the condition
\be
2h=\frac{p_{\varphi }^2}{ \sin
^2\theta_\pm }+{\alpha^2} \cot^2\theta_\pm+{\omega^2} \tan^2\theta_\pm .
\ee
To calculate the integral in the second expression, we introduce the notation
\be
\begin{split}
&a=\sqrt{1-2\frac{p_\varphi^2+\alpha^2+\omega^2}{2h+\alpha^2+\omega^2}+
\left(\frac{p_\varphi^2+\alpha^2-\omega^2}{2h+\alpha^2+\omega^2}\right)^2},
\quad b=-\frac{p_\varphi^2+\alpha^2-\omega^2}{2h+\alpha^2+\omega^2},
\\
&\xi=\frac{{1}}{a}\left[\cos{2\theta}
+{b}\right].
\end{split}
\ee
In this terms the second integral in (\ref{higgsactang}) reads (its value can be found by the use of standard methods,
 see, e.g.\cite{fiht,lny})
\be I_2 = \frac{a^2 \sqrt{2h+\alpha ^2+\omega ^2}}{2\pi}
\int\limits_{-1}^1\frac{\sqrt{1-\xi^2}}{1-\left(a\xi+b\right)^2}d\xi
=  \frac{1}{2}\left(\sqrt{2h+\alpha ^2+\omega ^2}-
\sqrt{p_{\varphi }^2+\alpha ^2}- \omega \right).  \label{h1}\ee
Hence, the functional dependence of the Hamiltonian from the action variables is given by the expression
\be
H=\sfrac12\left(2I_2+\sqrt{I_1^2+\alpha^2}+\omega\right)^2-\frac{\alpha^2+\omega^2}{2}.
\label{h2}\ee
For $\Phi_1$ and $\Phi_2$ we get

%-----------------------------------------------------------
\be
\label{h4}
\begin{split}
\Phi_1 &=
\varphi-\frac{p_\varphi}{\widetilde{p}_\varphi}\arcsin \xi+
\\
&+\frac{p_\varphi}{\widetilde{p}_\varphi}\arctan\frac{1}{2\widetilde{p}_\varphi}\left[
\sqrt{\frac{(2h-p_{\varphi}^{2})^2-4
\omega^2\widetilde{p}_{\varphi}^2}{2 h+\alpha ^2+\omega ^2}} -
\frac{2 h+2\alpha^2+p_{\varphi}^2}{\sqrt{2 h+\alpha ^2+\omega ^2}}
\frac{1+\sqrt{1-\xi^2} }{2 \xi} \right]
\\
\Phi_2 &= -2 \arcsin \xi
\end{split}
\ee
Here, as previously, we use the notation
\be
\widetilde{p}_\varphi=\sqrt{p_\varphi^2+ \alpha^2}
\ee
%-----------------------------------------------------------

We presented the action-angle formulation of the singular Higgs oscillator (\ref{higgs}) on the sphere of unit
 radius $r_0=1$.
 The action-angle formulation of the system on the sphere  with  arbitrary value of $r_0$
 could be easily found from (\ref{h1})-(\ref{h4}) by  the replacement
 \be
 { H}_{r_0}= \frac{H}{r_{0}^2},\quad {\rm with }\quad {\omega}\to\omega r_{0}^2 .
\label{rep} \ee
 In that case the Hamiltonian (\ref{higgs}) is defined, in the action-angle variables,
 by the following expression:
\be
H=\frac{1}{2r^2_0}\left(2I_2+\sqrt{I_1^2+\alpha^2}+
\omega r^2_0\right)^2-\frac{\alpha^2}{2r^{2}_0}-\frac{\omega^2r^2_0}{2}.
\label{h3}
\ee
It is seen that, in the planar limit  $r_0\to\infty$,
it results in the Hamiltonian of the Euclidean singular oscillator (\ref{hsaa}) with $B_0=0$
(i.e. in the absence of constant magnetic field).
However, the singular Higgs oscillator does not respect the inclusion of constant magnetic field,
 in contrast with the Euclidean one.

Indeed, the magnetic field which has a constant magnitude on the sphere,
is the field of a Dirac monopole located at the center of sphere.
It is defined by the following one-form:
\be
A_D=s(1-\cos\theta)d\varphi,\qquad s=B_0r^2_0.
\ee
Hence, the Hamiltonian of the singular  Higgs oscillator
interacting with a constant magnetic field, is defined by the expression
\be H=\frac{p_{\theta}^2}{2r^2_0}+\frac{\left[p_{\varphi }-s \left( 1- \cos
\theta \right) \right]^2}{2r^2_0\sin ^2\theta }+\frac{\alpha^2}{2r^2_0} \cot^2\theta+
\frac{\omega^2r^2_0}{2} \tan^2\theta .\ee
Writing down the corresponding generating function we shall see
 that the impact of the magnetic field cannot be absorbed
by the proper redefinition of constants. Hence, the inclusion of the magnetic field breaks
the exact solvability of the (singular)  Higgs oscillator, so that the presented model is not
suitable for the study of the
 properties of spherical bands and length  in the external magnetic field.
 However, this models is relevant for the consideration of their properties in the external potential,
  e.g. the electric field.
 Moreover, one can further modify the Higgs oscillator potential providing it by the
 anisotropy properties preserving the
 integrability of the system \cite{anosc}. Such a system would be useful to consider
 the quantum dots model restricted
 from the sphere to the spherical segment.

\newpage
\subsubsection*{Singular ${C}P^1 $ oscillator}
As we have already mentioned, the (singular) Higgs oscillator does not preserve its exact solvability in the presence of a constant magnetic field, while the study of quantum dot systems in a magnetic field is of a special physical importance.
For this reason we consider  the  alternative model of the singular spherical oscillator, given by the Hamiltonian \cite{aramyan}
\be
H_{CP^1}= \frac{p_{\theta}^2}{2r^2_0}+\frac{ p_{\varphi}^2}{2r^2_0
\sin
^2\theta}+\frac{\alpha^2}{8r^2_0}\cot^2\frac{\theta}{2}+2\omega^2r^2_0\tan^2\frac{\theta}{2}.
\label{alt}\ee

It is based on the model of the
oscillator on complex
projective spaces \cite{Bellucci} and, in contrast with the (singular)
Higgs oscillator, it respects the inclusion of a constant magnetic
field (of the Dirac monopole). Respectively, its singular version,  defined by the Hamiltonian (\ref{alt}) also remains
exactly solvable in the presence of a constant magnetic field, at least, classically \cite{aramyan}. Since the complex projective plane is  equivalent to the two- dimensional sphere, we can use
this model for the definition of the two-dimensional magnetic oscillator.

Let us notice that a similar model on the
four-dimensional sphere and hyperboloid respects the inclusion of
the BPST instanton field \cite{mn}. Quantum mechanical solutions of
(\ref{alt}) are not constructed yet. But they could be found by
a proper modification of the solutions of the corresponding non-singular
system (third  reference in \cite{Bellucci}). Because of the absence of
hidden symmetries, this model is not convenient for the study of the
system in external potential (e.g. electric ) fields. But it
convenient for the study of the interaction with the external magnetic
field.

Inclusion of the constant magnetic field yields the following
modification of the Hamiltonian (\ref{alt}): \be H=\frac{p_{\theta
}^2}{2r^2_0}+\frac{\left[p_{\varphi }-s \left( 1- \cos \theta
\right) \right]^2}{2r^2_0\sin ^2\theta }+2 \omega^2r^2_0
\tan^2\frac{\theta }{2}+\frac{\alpha^2}{8r^2_0}
\cot^2\frac{\theta}{2},\qquad s=B_0r^2_0 .\label{alt_magn}\ee As
before, we put, without loss of generality, $r_0=1$. The way of the
restoring of $r_0$ is obvious. Then, in a completely similar way as in the
previous cases, we can construct the action-angle variables of this
system. For the action variables $I_1$ and $I_2$ we get
\be
\begin{split}
&I_1=
p_\varphi,
\\
&I_2=\frac{1}{\pi}\int_{\theta_-}^{\theta^+}d\theta \sqrt{2h-\frac{\left[p_{\varphi }-s \left( 1- \cos \theta \right) \right]^2}{\sin ^2\theta }-4 \omega^2 \tan^2\frac{\theta }{2}-\frac{\alpha^2}{4} \cot^2\frac{\theta}{2}}
\end{split}
\ee
where $\theta_\pm$ are defined by the equation
\be
h=\frac{\left[p_{\varphi }+s \left( 1- \cos \theta_\pm \right)
\right]^2}{2\sin ^2\theta_\pm }+2 \omega^2
\tan^2\frac{\theta_\pm}{2}+ \frac{\alpha^2}{8}
\cot^2\frac{\theta_\pm}{2}.
\ee

The explicit expression for the second integral looks as follows:
\be
\begin{split}
 I_2 &=\frac{a^2\sqrt{2h+4\omega ^2+\frac{\alpha^2}{4}+s^2}}{\pi  }
\int\limits_{-1}^1\frac{\sqrt{1-\xi^2}}{1-\left(a\xi+b\right)^2}d\xi=
\\
 &=\sqrt{2 h+ s^2+\frac{\alpha^2}{4}+4 \omega ^2}-\sqrt{\frac{p_\varphi^2+\alpha^2}{4}}
-\sqrt{(\frac{p_\varphi}{2}-s)^2+4 \omega ^2},
\end{split}
\label{hcp}
\ee
where we introduced  the notation
\be
\begin{split}
&\xi =\frac 1a
\left[\cos\theta-{b}\right],\qquad b = \frac{8 \omega
^2+2s^2-2p_\varphi s- \frac{\alpha^2}{2}}{4h+8\omega ^2+
\frac{\alpha^2}{2}+2s^2},
\\
&a=\frac{2}{\sqrt{4h+8\omega^2+\frac{\alpha^2}{2}+2s^2}}
\sqrt{\frac{4h^2-\left(8\omega ^2-p_\varphi s+2s^2\right)
\left(\frac{\alpha^2}{2}+ p_\varphi s\right)}{4h+8\omega
^2+\frac{\alpha^2}{2}+2s^2}-\frac{p_\varphi^2}{2}+p_\varphi s}.
\end{split}
\ee
Hence, from (\ref{hcp}) we get that the explicit expression of
 the Hamiltonian has the following dependence from the
action variables: \be
H=\frac{1}{8}\left(2I_2+\sqrt{I_{1}^2+\alpha^2}+\sqrt{(I_1-2s)^2+16\omega^2}\right)^2-\frac{s^2}{2}-
\frac{\alpha^2}{8}-2\omega^2
\ee

The expressions for the angle variables look as follows:
\be
\begin{split}
\Phi_1  &=
\varphi - \frac{1}{2}\left(\frac{p_\varphi}{\widetilde{p}_\varphi}+
\frac{p_\varphi-2 s}{\sqrt{(p_\varphi-2 s)^2+16 \omega ^2}}\right)
\arcsin \xi +
\\
& +\frac{p_\varphi + s}{\sqrt{(p_\varphi -2 s)^2+16 \omega ^2}}\arctan\eta_+
-\frac{p_\varphi}{\widetilde{p}_\varphi}\arctan\eta_-,
\\
\Phi_2 &=\frac{\partial S}{\partial I_2}  = -\arcsin \xi\;.
\end{split}
\ee
Here we used the notation
 \be
\widetilde{p}_\varphi\equiv\sqrt{p_{\varphi}^2+\alpha^2},\qquad
\eta_\pm\equiv\frac{(1\pm b) \left(\frac{1}{\xi }+
\sqrt{\frac{1}{\xi ^2}-1}\right)\pm a}{\sqrt{(1\pm b)^2-a^2}}
\ee

Finally,  let us restore the radius $r_0$  performing the  replacement (\ref{rep}).
In that case the  Hamiltonian (\ref{alt_magn}) is expressed via action variables as follows:
\be
H=\frac{1}{8 r_0^2}\left(2I_2+\sqrt{I_{1}^2+\alpha^2}+\sqrt{(I_1-2B_0 r^2)^2+
16\omega^2r_0^4}\right)^2-\frac{B_0^2 r_0^2}{2}-\frac{\alpha^2}{8 r_0^2}-2\omega^2r_0^2
\ee
It is seen that, in the planar limit  $r_0\to\infty$,
it results in the Hamiltonian of the Euclidean singular oscillator (\ref{hsaa}).\\[10mm]
So, we presented the action-angle formulation of the model of the spherical singular oscillator
interacting with a constant magnetic field (\ref{alt_magn}).
The  Hamiltonian of the model is  non-degenerate on
both action variables. But it depends on these variables via
elementary functions in the presence of a constant  magnetic field.

These tell us the area of application of the Higgs oscillator potential and of the $CP^1$ oscillator one.
The Higgs model  is useful for the behavior of the quantum dots systems in the external potential
field, e.g, in the electric field.
The $CP^1$ model should be applied for the study of the behavior of a spherical quantum dots model
in the external magnetic field.

\newpage
\setcounter{equation}{0}
\section{INTEGRABLE GENERALIZATIONS OF OSCILLATOR AND COULOMB SYSTEMS}
\label{oscilAndCoulomb}

In this Chapter we construct new integrable conformal mechanical systems by generalizing the known ones \cite{hkln,lny}. Namely, we pick a system with compact phase space and add a {\sl radial\/} part to it, thus increasing the dimension by one (the phase space dimension is increased by two). In the Chapter \ref{aavar} we suggested action-angle variables as a useful tool for the study of systems with compact phase space. So, it is a good idea to use action-angle variables to discribe the motion of the initial "angular" part.

So, we pick an integrable system with a $2(N{-}1)$-dimensional {\sl compact\/} phase space
\be
\cH=\cH(I_i), \qquad \{I_i,\Phi^0_j\}=\delta_{ij},\qquad
\Phi^0_i\in [0, 2\pi), \qquad
i,j=1,\ldots,N{-}1,
\ee
in terms of its action-angle variables, and add a {\sl radial\/} part to it \cite{hlnsy,hlnsy2},
\be
H=\frac{p^{2}_r}{2}+\frac{\cH(I_i)}{r^2}+V(r), \qquad \{p_r, r\}=1,\qquad
r\in[0,\infty) \quad\textrm{or}\quad [0, r_0) .
\label{2}\ee
Here, we introduced a radial coordinate~$r$ and momentum~$p_r$ and obtain
an extended model with $N$ degrees of freedom. The extended configuration space
is a cone over the original compact configuration space. If the latter is just
the sphere~$S^{N-1}$, we can obtain, in particular, the three model spaces of
constant curvature:
\bea
S^N &:& \quad r=r_0\sin\chi, \qquad\ \ p_r=r_0^{-1}p_\chi, \qquad
V(r)\to V(r_0\tan\chi), \label{3} \\[4pt]
\R^N &:& \quad r=r_0\chi, \qquad\qquad\, p_r=r_0^{-1}p_\chi, \qquad
V(r)\to V(r_0\chi), \label{5} \\[4pt]
H^N &:& \quad r=r_0\sinh\chi, \qquad p_r=r_0^{-1}p_\chi, \qquad
V(r)\to V(r_0\tanh\chi), \label{4}
\eea
where $r_0$ is the radial scale and $\{p_\chi,\chi\}=1$ is a dimensionless
canonical pair. Hence, for a particle on the sphere~$S^N$ (the sine-cone
over~$S^{N-1}$) or on the hyperboloid~$H^N$ (the hyperbolic cone over~$S^{N-1}$)
one gets the Hamiltonians
\be
H=\frac{p_{\chi}^2}{2 r_0^2}+\frac{\cH}{r_0^2\sin^2\chi}+V(r_0\tan\chi)
\qquad\textrm{and}\qquad
H=\frac{p_{\chi}^2}{2 r_0^2}+\frac{\cH}{r_0^2\sinh^2\chi}+V(r_0\tanh\chi),
\label{6}\ee
respectively.

As an example, when $\cH$ defines the Landau problem, i.e.~a particle on $S^2$
moving in the magnetic field generated by a Dirac monopole located
at the center of sphere, we arrive at the particle on $\R^3$ interacting with this Dirac monopole.
The extended system remains integrable for two prominent choices of the radial
potential,
\be
V(r)=V_{\textrm{osc}}(r)=\sfrac12\omega^2 r^2 \qquad\textrm{and}\qquad
V(r)=V_{\textrm{cou}}(r)=-\frac{\gamma}{r},
\label{pot}\ee
with frequency $\omega$ and (positive) coupling~$\gamma$, respectively.
For~$\R^N$, these are the familiar oscillator and Coulomb potentials,
while for $S^N$ they have been named Higgs oscillator~\cite{higgs} and
Schr\"odinger-Coulomb~\cite{sch}, respectively.

If the system is spherically symmetric, i.e.~$S^{N-1}$ invariant,
the compact Hamiltonian~$\cH$ is just given by the SO($N$) Casimir function~$J^2$,
which defines the kinetic energy of a free particle on~$S^{N-1}$.
Deviations from spherical symmetry are encoded in~$\cH$. In other words,
replacing $J^2$ by the Hamiltonian of some compact $N{-}1)$-dimensional
integrable system defines a deformation of the $N$-dimensional oscillator
and Coulomb systems.

Particular examples with $N{=}2$ are the so-called
Tremblay-Turbiner-Winternitz (TTW)~\cite{TTW} and Post-Winternitz (PW)~\cite{PW}
models, defined on~$\R^2$, which have attracted some interest recently
(see, e.g.~\cite{TTWothers} and references therein). In this systems the compact subsystem
on the circle~$S^1$ is just the famous P\"oschl-Teller system~\cite{flugge},
\be
\cH=\cH_{\textrm{PT}}=\frac{p^2_\varphi}{2}+\frac{k^2\alpha^2_1}{2\sin^2k\varphi}
+\frac{2k^2\alpha^2_1}{\cos^2k\varphi} \qquad\textrm{with}\qquad k\in\NN.
\label{PT}\ee

Assume now that the compact subsystem is already formulated in terms of
action-angle variables $(I_i,\Phi_i^0)$, with $i=1,\ldots,N{-}1$, while the
radial part is given by~$(p_r,r)$.
We characterize the level sets by $(H{\equiv}E, I_i)$.
The generating function for the extended system~(\ref{2}) then reads
\be
S(E, I_i,r,\Phi_i^0)=\sqrt{2}\int\!dr\ \sqrt{E-\frac{\cH(I)}{r^2}-V(r)}\
+\ \sum_{i=1}^{N-1} I_i\Phi^0_i.
\label{11}\ee
From this function we immediately get the action variables $I_i=I_i$ and
\be
I_r(E,I_i)=\frac{\sqrt{2}}{2\pi}\oint dr\ \sqrt{E-\frac{\cH(I)}{r^2}-V(r)}.
\label{12}\ee
The corresponding angle variables are given by
\be
\begin{split}
\Phi_r=&\frac{1}{\sqrt{2}}\frac{\partial E}{\partial I_r}
\int\!\frac{dr}{\sqrt{E-\frac{\cH(I)}{r^2}-V(r)}} \quad\textrm{and}
\\
\Phi_i=&\Phi^0_i+
\frac{{\partial E}/{\partial I_i}}{\partial E/\partial I_r}\Phi_r-
\frac{1}{\sqrt{2}}\frac{\partial \cH(I)}{\partial I_i}
\int\!\frac{dr}{r^2\sqrt{E-\frac{\cH(I)}{r^2}-V(r)}}.
\label{13}
\end{split}
\ee

Making  in (\ref{11}) and (\ref{13}) the replacements described
in (\ref{3}) or~(\ref{4}),
we shall get the system on the $N$-sphere or -pseudosphere.
Of course, for the full construction of the action-angle variables,
we need to provide the action-angle variables of the subsystem~$\cH$.

In this Chapter we start with computing the explicit expressions for action-angle variables for systems with oscillator and Coulomb potentials (\ref{pot}). Then we construct the spherical and pseudospherical generalizations of the TTW and PW systems. We demonstrate the superintegrability of these systems and write down their hidden constants of motion. Additionally, we provide the action-angle variables for a free particle on the $(N{-}1)$-dimensional sphere, which yields the complete set of action-angle variables for the $N$-dimensional oscillator and Coulomb systems as well as their spherical and pseudospherical analogs.

\newpage
\subsection{Deformed oscillator and Coulomb systems}
Here we present the action-angle variables $(I_r, \Phi_r, \Phi_i)$
for the deformed oscillator and Coulomb systems given by the expressions
(\ref{2})--(\ref{pot}).
The action variables $I_i$ of the ``angular Hamiltonian'' $\cH$ remain unchanged,
while the angle variables $\Phi^0_i$ receive corrections, as seen in~(\ref{13}).
For notational simplicity we abbreviate $H({\bf p},{\bf q})=E$, put $r_0=1$
and drop the argument $I_i$ of~$\cH$. In the following, we list the results for
each of the six combinations in the table below:
\\[10mm]
\begin{tabular}{@{}l|cc@{}}
radial potential & oscillator & Coulomb \\ \hline
metric cone:\hfill$\R^N$ & Euclidean osc. & Euclidean Coulomb \\
sine-cone:\hfill$S^N$ & spherical Higgs osc. &
spherical Schr\"odinger-Coulomb \\
hyperbolic cone:\hfill$H^N$ & pseudospherical Higgs osc. &
pseudospherical Schr\"odinger-Coulomb
\end{tabular}
\\[10mm]
\subsubsection*{Euclidean oscillator}
\be
\begin{split}
&H_{\textrm{osc}}=\frac{p_r^2}{2}+\frac{{\cal H}}{r^2} + \frac{\omega^2 r^2}{2}
=\omega \big(2 I_r + \sqrt{2 {\cal H}}\big),
\\
&I_r=\frac{E}{2\omega} -
\sqrt{\frac{{\cal H}}{2}}
\\
&\Phi_r=-\arcsin\Big(\frac{E-r^2\omega^2}{\sqrt{E^2-2{\cal H}\omega ^2}}\Big),
\\
&\Phi_i =\Phi^0_i+
\frac{1}{2\sqrt{2\cH}}~\frac{\partial\cH}{\partial I_i}\biggl[\Phi_r-
\arcsin\Big(\frac{E\,r^2-2\cH}{r^2\sqrt{E^2-2\omega^2\cH}}\Big)\biggr].
\end{split}
\ee
\\[5mm]
\subsubsection*{Euclidean Coulomb}
\be
\begin{split}
&H_{\textrm{cou}}=\frac{p_r^2}{2}+\frac{{\cal H}}{r^2} - \frac{\gamma}{r}
= -\frac{\gamma^2}{2\big(I_r+\sqrt{2 {\cal H}}\big)^2},
\\
&I_r = \frac{\gamma}{\sqrt{-2 E}}-\sqrt{2{\cal H}}
\\
&\Phi_r= -\frac2\gamma\sqrt{E {\cal H} - E\,r(E\,r+\gamma)}-
\arcsin\Big(\frac{2 E\,r+\gamma }{\sqrt{4 E {\cal H}+\gamma^2}}\Big),
\\
&\Phi_i=\Phi^0_i+\sqrt{\frac{2}{{\cal H}}}~\frac{\partial {\cal H}}{\partial I_i}
\biggl[\Phi_r-\frac{1}{2}\arcsin\Big(\frac{\gamma\,r-2\cH}{r\sqrt{4E\cH+\gamma^2}}
\Big)\biggr].
\end{split}
\ee
\\[5mm]
\subsubsection*{Spherical Higgs oscillator }
\be
\begin{split}
&H_{\textrm{s-higgs}}=\frac{p_{\chi}^2}{2}+\frac{{\cal H}}{\sin^2\chi} +
\frac{\omega^2\tan^2\chi}{2}
=\frac{1}{2}\left(2 I_\chi+\sqrt{2 {\cal H}}+\omega\right)^2-\frac{\omega^2}{2},
\\
&I_\chi=\sfrac12 \Big(\sqrt{2 E + \omega^2}-\sqrt{2\cH}-\omega\Big)
\\
&\Phi_\chi = - 2 \arcsin \Big(
\frac{(2E+\omega^2)\cos 2\chi+2\cH-\omega^2}
{\sqrt{(2E+\omega^2)^2-2(2\cH+\omega^2)(2E+\omega^2)+(2\cH-\omega^2)^2}}\Big),
\\
&\Phi_i=\Phi^0_i+
\frac{1}{2\sqrt{2 {\cal H}}}~\frac{\partial\cH}{\partial I_i}
\biggl[\Phi_\chi+
\\
&+\arctan\Big(
\frac{(E+{\cal H}) \cos2\chi-E+3 {\cal H}}
{\sqrt{2\cH}~\sqrt{2E-4\cH-\omega^2-(4\cH-2\omega^2)\cos2\chi -
(2E+\omega^2)\cos^2 2\chi}}\Big) \biggr].
\end{split}
\ee
\\[5mm]
\subsubsection*{Spherical Schr\"{o}dinger-Coulomb}
\be
\begin{split}
&H_{\textrm{s-sch-cou}}=\frac{p_{\chi}^2}{2}+\frac{{\cal H}}{\sin^2\chi} -
\gamma\cot\chi
=\sfrac12\Big(I_\chi+\sqrt{2 {\cal H}}\Big)^2-
\frac{\gamma ^2}{2 (I_\chi+\sqrt{2 {\cal H}})^2},
\\
&I_\chi=\sqrt{E+\sqrt{E^2+\gamma^2}} -\sqrt{2 {\cal H}}
\\
&\Phi_\chi= \textrm{Im}\biggl[
\frac{2\sqrt{{\cal H}(E+\textrm{i}\gamma )}}
{\sqrt{E+\sqrt{E^2+\gamma^2}}}\log\zeta \biggr]
\\
&\Phi_i=\Phi^0_i+
\frac{1}{\sqrt{2 {\cal H}}}~\frac{\partial {\cal H}}{\partial I_i}
\biggl[\Phi_\chi+ \arcsin\Big(\frac{2 {\cal H} \cot\chi-\gamma}
{\sqrt{4 (E-{\cal H}) {\cal H}+\gamma ^2}}\Big)\biggr],
\\
&\textrm{where}\qquad
\zeta = \frac{4}{\sqrt{{\cal H}}} \textrm{e}^{\textrm{i}(\chi+\frac{\pi}{2})}
\sin\chi\Big(1+\sqrt{
\vphantom{\Big|}E-\smash{\frac{{\cal H}}{\sin^2\chi}}+\gamma\cot\chi}\Big)+
\frac{(4{\cal H}-2 \textrm{i}\gamma)\sqrt{E+\textrm{i}\gamma}}
{\sqrt{{\cal H}(E^2+\gamma^2)}}.
\end{split}
\ee
\\[5mm]
\subsubsection*{Pseudospherical Higgs oscillator}
\be
\begin{split}
&H_{\textrm{ps-higgs}}=\frac{p_{\chi}^2}{2}+\frac{{\cal H}}{\sinh^2\chi} +
\frac{\omega^2\tanh^2\chi}{2}
=\frac{\omega^2}{2}-\frac{1}{2}\big(2I_\chi+\sqrt{2 {\cal H}}-\omega\big)^2,
\\
& I_\chi=\frac{1}{2}\Big(\omega-\sqrt{2 {\cal H}} -\sqrt{\omega^2 -2 E}\Big),
\\
& \Phi_\chi = - 2 \arctan\Big(\frac{(1-\sqrt{1-\eta^2})
\left(E+{\cal H}-\omega^2\right)}{\eta ~ \omega  \sqrt{\omega^2 - 2
E}}+\frac{\sqrt{(E+{\cal H})^2-2 {\cal H} \omega^2}}{\omega
\sqrt{\omega^2 - 2 E}}\Big),
\\
&\Phi_i=\Phi^0_i+\frac{1}{2\sqrt{2{\cal H}}}~\frac{\partial{\cal H}}{\partial I_i}
\biggl[\Phi_\chi-
\\
&-2 \arctan\Big( \frac{(1-\sqrt{1-\eta^2})
\left(E+{\cal H}\right)}{\eta~\omega\sqrt{2{\cal H}}}+
\frac{\sqrt{(E+{\cal H})^2-2 {\cal H} \omega^2}}{\omega \sqrt{2 {\cal H}}}
\Big)\biggr],
\\
&\textrm{where}\qquad\qquad
\eta=\frac{\omega^2\tanh^2\chi-(E+\cH)}{\sqrt{(E+\cH)^2-2\cH\omega^2}}.
\end{split}
\ee
\\[5mm]
\subsubsection*{Pseudospherical Schr\"{o}dinger-Coulomb}

\be
\begin{split}
&H_{\textrm{ps-sch-cou}}=\frac{p_{\chi}^2}{2}+\frac{{\cal H}}{\sinh^2\chi} -
\gamma\coth\chi
=-\sfrac12\Big(I_\chi+\sqrt{2 {\cal H}}\Big)^2-
\frac{\gamma ^2}{2(I_\chi+\sqrt{2 {\cal H}})^2},
\\
&I_\chi=\frac1{\sqrt{2}}\Big(\sqrt{-E+\gamma}-\sqrt{-E-\gamma}-2\sqrt{\cH}\Big)
\\
&\Phi_\chi=\frac{\sqrt{-E+\gamma}}{\sqrt{2}(\sqrt{-E-\gamma}-\sqrt{-E+\gamma })}
\arctan\Big(\frac{\sqrt{4{\cal H}(E+{\cal H})+\gamma^2}+(\gamma-2{\cal H})\eta}
{2 \sqrt{{\cal H} (-E-\gamma ) }\sqrt{1-\eta ^2}}\Big) -
\\
&- \frac{\sqrt{-E-\gamma}}{\sqrt{2}(\sqrt{-E-\gamma}-\sqrt{-E+\gamma })}
\arctan\Big(\frac{\sqrt{4{\cal H}(E+{\cal H})+\gamma^2}+(\gamma+2{\cal H})\eta}
{2 \sqrt{{\cal H} (-E+\gamma ) }\sqrt{1-\eta ^2}}\Big),
\\
&\Phi_i=\Phi^0_i+
\frac{1}{\sqrt{2 {\cal H}}}~\frac{\partial {\cal H}}{\partial I_i}
\biggl[\Phi_\chi+ \frac{1}{\sqrt{2}}\arcsin\Big(
\frac{2\cH\cot\chi-\gamma}{\sqrt{4(E+\cH)\cH+\gamma^2}}\Big)\biggr].
\end{split}
\ee

\newpage
\subsection{Generalized Tremblay-Turbiner-Winternitz and Post-Winternitz systems}
As mentioned in the Introduction, action-angle variables elegantly
explain the superintegrability of the recently suggested deformation of the
two-dimensional oscillator system introduced by
Tremblay-Turbiner-Winternitz (TTW)~\cite{TTW} and also of the
Coulomb versions treated by Post-Winternitz (PW)~\cite{PW}.
They also allow us to construct
analogous deformations of other superintegrable systems.

Our generalizations of the TTW and PW systems are defined
by (\ref{2})--(\ref{pot}) with $N{=}2$,
where the one-dimensional ``angular" Hamiltonian $\cH$ is
given by the generalized P\"oschl-Teller system on the circle
(\ref{PT})~\cite{flugge}. The action-angle variables of this subsystem
are given by \cite{lny}
\be
I_{\textrm{PT}}=\sfrac{1}{k}\sqrt{2\cH_{\textrm{PT}}}-(\alpha _1+\alpha _2)
\qquad\textrm{and}\qquad
\Phi_{\textrm{PT}}=\sfrac12\arcsin\Bigl\{\sfrac{1}{a}\bigl[\cos{2k\varphi}
+b\bigr]\Bigr\},
\label{che}\ee
where
\be
a=\sqrt{1-\frac{{k}^2(\alpha_1^2+\alpha_2^2)}{\cH_{\textrm{PT}}}
+\left(\frac{{k}^2(\alpha _1^2-\alpha _2^2)}{2\cH_{\textrm{PT}}}\right)^2}
\qquad\textrm{and}\qquad
b=\frac{{k}^2(\alpha_2^2-\alpha_1^2)}{2\cH_{\textrm{PT}}},
\ee
so that the Hamiltonian reads
\be
\cH_{\textrm{PT}}=\sfrac12 (k{\tilde I_{\textrm{PT}}})^2
\qquad\textrm{with}\qquad
{\tilde I}_{\textrm{PT}}\equiv I_{PT}+\alpha_1+\alpha_2 \
\in[\alpha_1{+}\alpha_2,\infty).
\ee
Clearly, in action-angle variables, the P\"oschl-Teller Hamiltonian coincides
with the Hamiltonian of a free particle on a circle of radius~$k$,
but with a different domain for the action variable. Hence, choosing the
potential in~(\ref{2}) to be of oscillator or Coulomb type, the extended system
will be superintegrable. More precisely, in the variables $\big(p_r,r,
I_{\textrm{PT}}(p_\varphi,\varphi),\Phi_{\textrm{PT}}(p_\varphi,\varphi)\big)$,
this system takes the form of a conventional two-dimensional
oscillator or Coulomb system on the cone. Hence, for rational
values of~$k$ these systems possess hidden symmetries.
For the oscillator case, the hidden constants of motion
have been constructed in~\cite{gonera}. Here, we extend their results
to the Coulomb case~\cite{PW} as well as to the TTW- and PW-like systems
on spheres and pseudospheres. %along the lines of~\cite{gonera}.

For the three spaces of constant curvature and for the oscillator
potential, the action-angle Hamiltonians are
\be
H_\omega=
\left\{\begin{array}{ccc}
\omega\,(2 I_\chi + k{\tilde I}_{\textrm{PT}}) \qquad\!\qquad
& \qquad {\rm for}& \R^2\\[2pt] \phantom{-}
\sfrac12(2I_\chi+k{\tilde I}_{\textrm{PT}}+\omega)^2-\sfrac{\omega^2}{2}
& \qquad {\rm for}& S^2\\[2pt]
-\sfrac12(2I_\chi+k{\tilde I}_{\textrm{PT}}-\omega)^2+\sfrac{\omega^2}{2}
& \qquad {\rm for}& H^2
\end{array} \right.
\ee
and depend only on the combination $2 I_\chi{+}k{\tilde I}_{\textrm{PT}}$.
Thus, the evolution of the angle variables is given by
\be
\Phi_\chi(t) = 2\,\Omega\,t \qquad\textrm{and}\qquad
\Phi_\varphi(t) = k\,\Omega\,t \qquad\textrm{with}\qquad
\Omega=\frac{d H_\omega}{d(2 I_\chi{+}k{\tilde I}_{\textrm{PT}})}.
\ee
For rational values of~$k$ the trajectories are closed.
It then follows that the hidden constant of motion is
\be
I_{\textrm{hidden}}=\cos \big(m\Phi_\chi{-}2n\Phi_\varphi\big)
\qquad\textrm{for}\quad k=m/n.
\ee
Explicitly, this hidden constant of motion reads:
\subsubsection*{Euclidean TTW system}

\be
\begin{split}
I_{add}=CM_m\left(\frac{E r^2-2{\cal H}_{PT}}{r^2\sqrt{E^2-2  \omega^2{\cal H}_{PT}}}\right) CM_n \left(\sfrac{1}{a}\left[\cos{2k\varphi}
+b\right]\right)
\\
\phantom{.}\qquad
+SM_m\left(\frac{E r^2-2{\cal H}_{PT}}{r^2\sqrt{E^2-2  \omega^2{\cal H}_{PT}}}\right) SM_n \left(\sfrac{1}{a}\left[\cos{2k\varphi}
+b\right]\right)
\end{split}
\ee
where we denoted
\be
\begin{split}
CM_n(x)&=\cos(n \arcsin x)=\sum_{i=0}^{[\frac n2]}(-1)^i C_n^{2 i} x^{2 i}\sqrt{1-x^2}^{n-2 i},
\\
SM_n(x)&=\sin(n \arcsin x)=\sum_{i=0}^{[\frac{n-1}{2}]}(-1)^i C_n^{2 i+1} x^{2 i+1}\sqrt{1-x^2}^{n-2 i-1}.
\end{split}
\ee

\subsubsection*{Spherical  TTW system }
\be
\begin{split}
I_{add}&=CM_m\left(\frac{\xi}{\sqrt{\xi^2+1}}\right) CM_n \left(\sfrac{1}{a}\left[\cos{2k\varphi}+b\right]\right)
\\
&- SM_m\left(\frac{\xi}{\sqrt{\xi^2+1}}\right) SM_n \left(\sfrac{1}{a}\left[\cos{2k\varphi}+b\right]\right)
\end{split}
\ee
where
\be
\xi=\frac{(E+{\cal H}_{PT}) \cos2\chi-E+3 {\cal H}_{PT}}
{\sqrt{2 {\cal H}_{PT}} \sqrt{2 E-4 {\cal H}_{PT}-\omega^2- \left(4 {\cal H}_{PT}-
2 \omega ^2\right)\cos2\chi - \left(2 E+\omega ^2\right) \cos^2 2\chi}}.
\ee

\subsubsection*{Pseudospherical TTW system}

\be
\begin{split}
I_{add}&=CM_{2m}\left(\frac{\xi}{\sqrt{\xi^2+1}}\right) CM_n \left(\sfrac{1}{a}\left[\cos{2k\varphi}+b\right]\right)
\\
&+ SM_{2m}\left(\frac{\xi}{\sqrt{\xi^2+1}}\right) SM_n \left(\sfrac{1}{a}\left[\cos{2k\varphi}+b\right]\right)
\end{split}
\ee

where

\be
\begin{split}
\xi&=\frac{\sqrt{(E+{\cal H}_{PT})^2-2 {\cal H}_{PT} \omega^2}}{\omega
\sqrt{2 {\cal H}_{PT}}} \frac{\omega^2 \tanh^2\chi}{\omega^2 \tanh^2\chi-(E+{\cal H}_{PT})}
\\
&-\frac{
E+{\cal H}_{PT}}{ \omega  \sqrt{2 {\cal H}_{PT}}} \frac{\sqrt{2(E+{\cal H}_{PT})\omega^2
\tanh^2\chi-\omega^4
\tanh^4\chi-2 {\cal H}_{PT} \omega^2}}{\omega^2
\tanh^2\chi-(E+{\cal H}_{PT})}
\end{split}
\ee

Thus, choosing the Higgs oscillator on the (pseudo)sphere, we
get a superintegrable (pseudo)spherical analog of the TTW~oscillator.\\

The construction of superintegrable deformations of the Coulomb system,
i.e.~the PW~model and its generalization to the (pseudo)spherical environment,
proceeds completely similarly. The Hamiltonians
\be
H_\gamma=
\left\{\begin{array}{ccc}
-\sfrac{\gamma^2}{2} (I_\chi + k{\tilde I}_{\textrm{PT}})^{-2}
\qquad\qquad\qquad\qquad
& \qquad {\rm for }& \R^2\\[2pt]
-\sfrac{\gamma^2}{2} (I_\chi + k{\tilde I}_{\textrm{PT}})^{-2}
+\sfrac12(I_\chi+k{\tilde I}_{\textrm{PT}})^2
& \qquad {\rm for}& S^2\\[2pt]
-\sfrac{\gamma^2}{2} (I_\chi + k{\tilde I}_{\textrm{PT}})^{-2}
-\sfrac12(I_\chi+k{\tilde I}_{\textrm{PT}})^2
& \qquad {\rm for}&  H^2
\end{array} \right.
\ee
depend only on the combination $I_\chi{+}k{\tilde I}_{\textrm{PT}}$,
and for rational $k=m/n$ the trajectories are closed, supporting
\be
I_{\textrm{hidden}}=\cos \big(m\Phi_\chi{-}n\Phi_\varphi\big).
\ee
Explicitly this constant of motion reads:

\subsubsection*{Euclidean PW system}

\be
\begin{split}
I_{add}&=CM_{2 m}\left(\frac{\gamma r -2 {\cal H}_{PT} }{r \sqrt{4 E {\cal H}_{PT}+\gamma ^2}}\right) CM_n\left(\sfrac{1}{a}\bigl[\cos{2k\varphi}+b\bigr]\right)
\\
&+SM_{2 m}\left(\frac{\gamma r -2 {\cal H}_{PT} }{r \sqrt{4 E {\cal H}_{PT}+\gamma ^2}}\right) SM_n\left(\sfrac{1}{a}\bigl[\cos{2k\varphi}+b\bigr]\right).
\end{split}
\ee

\subsubsection*{Spherical PW system}
\be
\begin{split}
I_{add}&=CM_{2 m}\left(\frac{2 {\cal H}_{PT} \cot\chi-\gamma}{\sqrt{4 (E-{\cal H}_{PT}) {\cal H}_{PT}+\gamma ^2}}\right) CM_n\left(\sfrac{1}{a}\bigl[\cos{2k\varphi}
+b\bigr]\right)
\\
&-SM_{2 m}\left(\frac{2 {\cal H}_{PT} \cot\chi-\gamma}{\sqrt{4 (E-{\cal H}_{PT}) {\cal H}_{PT}+\gamma ^2}}\right) SM_n\left(\sfrac{1}{a}\bigl[\cos{2k\varphi}
+b\bigr]\right).
\end{split}
\ee

\subsubsection*{Pseudospherical PW system}

\be
\begin{split}
I_{add}&=CM_{2 m}\left(\frac{2 {\cal H}_{PT} \coth\chi-\gamma}{\sqrt{4 (E+{\cal H}_{PT}) {\cal H}_{PT}+\gamma ^2}}\right) CM_n\left(\sfrac{1}{a}\bigl[\cos{2k\varphi}
+b\bigr]\right)
\\
&-SM_{2 m}\left(\frac{2 {\cal H}_{PT} \coth\chi-\gamma}{\sqrt{4 (E+{\cal H}_{PT}) {\cal H}_{PT}+\gamma ^2}}\right) SM_n\left(\sfrac{1}{a}\bigl[\cos{2k\varphi}
+b\bigr]\right).
\end{split}
\ee
Thus, choosing the Schr\"odinger-Coulomb system on the (pseudo)sphere,
we get a superintegrable (pseudo)spherical analog of the PW~model.

\newpage
\subsection{Free particle on $S^{N-1}$}
Here we recollect the  action-angle variables for
the ``angular Hamiltonian" ${\cal H}_{PT}$ appearing in every spherically symmetric
$N$-dimensional system and defining the free motion of a particle on $S^{N-1}$
with radius $r_0=1$.
It is given by the Casimir function  $L^2_N$ of SO($N$),
\be
\cH=\sfrac12 L^2_N.
\label{l2}\ee

The embedding of the unit $(N{-}1)$-sphere into $\R^N$ is given
by a set of polar coordinates,
\be
\left.
\begin{aligned}
x_1&=s_{N-1}\,s_{N-2}\cdots s_3\,s_2\,s_1 \quad \\
x_2&=s_{N-1}\,s_{N-2}\cdots s_3\,s_2\,c_1 \\
x_3&=s_{N-1}\,s_{N-2}\cdots s_3\,c_2 \\[-4pt]
&\vdots \\[-4pt]
x_{N-1}&=s_{N-1}\,c_{N-2} \\
x_N&=c_{N-1}
\end{aligned}
\right\} \qquad\qquad
\begin{aligned}
\textrm{with}\qquad
& s_k:=\sin\theta_k \quad\textrm{and}\quad c_k:=\cos\theta_k \\[4pt]
\textrm{for}\qquad
& \theta_1\in[0,2\pi)\ ,\quad \theta_{k>1}\in[0,\pi) \\[4pt]
\textrm{and}\qquad
& k=1,2,\ldots,N{-}1.
\end{aligned}
\ee
In these coordinates, we have the recursion
\be
L^2_N = p_{N-1}^2+\frac{L^2_{N-1}}{s^2_{N-1}}
\label{reccur}\ee
where
$p_{N-1}$ is the momentum conjugate to $\theta_{N-1}$.
It is easy to see that the $L^2_k$ for $k=1,\ldots,N$ are
in involution with each other and, therefore, can be used for constructing
action-angle variables.
Each variable $\theta_k$ defines an independent homology cycle~$S^1_k$
of the torus~$T^N$.
The level surfaces $L^2_k=\textrm{constant}=:j_k$ are diffeomorphic to~$T^N$.

Following the standard procedure we should compute the $N$ integrals
\be
I_k=\frac{1}{2\pi}\oint\limits_{S^1_k}{\bf p}\cdot d{\bf q}
=\frac{1}{2\pi}\oint\limits_{S^1_k}p_k\,d\theta_k=
\int\limits^{\theta_k^+}\limits_{\theta_k^-}
\sqrt{j_k-\frac{j_{k-1}}{\sin^2{\theta_k}}}\;d\theta_k,
\label{gends}
\ee
where in the second equality we used that the $\theta_k$ are mutually orthogonal
and the cycles $S_k^1$ are independent.
The integration ranges $[\theta_k^-,\theta_k^+]$ are determined from
the condition that the radicants should be non-negative.
Substituting
\be
u_k=\sqrt{\frac{j_k}{j_k-j_{k-1}}}\,\cos{\theta_k},
\ee
we arrive at
\be
I_k=2\,\frac{j_k{-}j_{k-1}}{2\pi\sqrt{j_k}}\int\limits_{-1}^1
\frac{\sqrt{1-u_k^2}\;du_k}{1-\frac{j_k-j_{k-1}}{j_k}u_k^2}=
\sqrt{j_k}-\sqrt{j_{k-1}},
\qquad\textrm{so that}\qquad
\sqrt{j_k}=\sum\limits_{m=1}^k I_m.
\label{acv}\ee
For the generating function we obtain
\be
S=\sum\limits_{l=1}^{N-1}S_l\qquad\textrm{where}\qquad
S_l=\int\!d\theta_l\ {\sqrt{\biggl(\sum\limits_{m=1}^l I_m\biggr)^2-\
\sin^{-2}\theta_l \biggl(\sum\limits_{m=1}^{l-1}I_m\biggr)^2}},
\ee
from which we get the angle variables
\be
\Phi^0_k=\frac{\partial S}{\partial I_k}=
\frac{\partial S_k}{\partial I_k}+
\sum\limits_{l=k+1}^{N-1}\frac{\partial S_l}{\partial I_k}=
\int\!\frac{\sqrt{j_k}\;d\theta_k}{\sqrt{j_k-\frac{j_{k-1}}{\sin^2{\theta_k}}}}+
\sum\limits_{l=k+1}^{N-1}
\int\!\frac{d\theta_l}{\sqrt{j_l-\frac{j_{l-1}}{\sin^2{\theta_l}}}}
\biggl(\sqrt{j_l}-\frac{\sqrt{j_{l-1}}}{\sin^2{\theta_l}}\biggr).
\label{a14} \ee
The first integral can be included in the first part of the sum (as $l{=}k$),
which yields
\be
\sum\limits_{l=k}^{N-1}\int\!\frac{\sqrt{j_l}\;d\theta_l}
{\sqrt{j_l-\frac{j_{l-1}}{\sin^2{\theta_l}}}}=
\sum\limits_{l=k}^{N-1}\arcsin{u_l}.
\ee
After the substitution and abbreviation
\be
u_l=\frac{2t_l}{(1{+}t_l)^2} \qquad\textrm{and}\qquad
a=\sqrt{\frac{j_l-j_{l-1}}{j_l}}\ <1,
\ee
respectively, the second part of the sum in (\ref{a14}) becomes
\be
\begin{split}
&\sum\limits_{l=k+1}^{N-1}\!\!\sqrt{\frac{j_{l-1}}{j_l}}
\int\!\!dt_l\biggl[\frac{1}{(t_l{-}a)^2+1-a^2}+\frac{1}{(t_l{+}a)^2+1-a^2}\biggr]
=
\\
&=\sum\limits_{l=k+1}^{N-1}\!\sqrt{\frac{j_{l-1}}{j_l(1{-}a^2)}}
\left[\arctan{\frac{t_l{-}a}{\sqrt{1{-}a^2}}}
+\arctan{\frac{t_l{+}a}{\sqrt{1{-}a^2}}}\right].
\end{split}
\ee
Pulling all together, we finally find
\be
\Phi^0_k=\sum\limits_{l=k}^{N-1}\arcsin{u_l}+\sum\limits_{l=k+1}^{N-1}
\arctan\biggl({\sqrt{\frac{j_{l-1}}{j_l}}\frac{u_l}{\sqrt{1-u_l^2}}}\biggr).
\label{anv}\ee

To summarize, the action-angle variables for a free particle on $S^{N-1}$
are given by (\ref{acv}) and~(\ref{anv}), with
$j_l=L^2_l(p_1,\ldots,p_l,\theta_1,\ldots,\theta_l)$.
The angular Hamiltonian (\ref{l2}) can be expressed as
\be
\cH=\sfrac12 L^2_N=\sfrac12\biggl(\sum\limits_{m=1}^{N-1} I_m\biggr)^2.
\label{free}\ee

\newpage
\setcounter{equation}{0}
\section{CONFORMAL MECHANICS}
\label{confmech}
Conformal invariance plays an important role in many areas of the quantum field theory and  condensed matter physics, especially in string theory, the theory of critical phenomena, low-dimensional integrable models, spin and fermion lattice systems. Recently, there has been new interest in so-called ``conformal mechanics''. Actually, the conformal group is not an exact symmetry for the conformal mechanical system. It does not commute with the Hamiltonian but, instead, is a symmetry of the action (a symmetry in the field theoretical context\cite{Alfaro}). The Hamiltonian itself forms an $so(1,2)$ algebra together with generators of the dilatation  and conformal boost, with respect to canonical Poisson brackets. It is interesting that, due to the conformal symmetry, the ``angular" part of the Hamiltonian of conformal mechanics is a constant of motion.

The term ``conformal mechanics'' denotes a system whose Hamiltonian ${H}$, together with the dilatation generator ${D}$ and the generator ${K}$ of conformal boosts forms, with respect to Poisson brackets, the conformal algebra:
\be \label{so120}
\{ {H} , {D}\}= {H}, \qquad
\{ {D} , {K} \}={K}, \qquad
\{ {H} , {K}\}=  2 {D}. \ee
Such system can always be presented in the form \cite{hkln}
\be
\label{so2}
{ D}=\frac{p_r r}{2} ,\qquad
{ K}=\frac{r^2}{2},\qquad
{ H}=\frac{{p}^2_r}{2}+ \frac{2 \mathcal{I}(u)}{r^2},
\ee
where the radial coordinates $(r,p_r)$ and the angular coordinates $(u^\alpha)$
obey the basic Poisson brackets
\be
\{p_r, r\}=1,\qquad
\{u^{\alpha}, p_r\}=\{u^{\alpha}, r\}=0.
\ee

The spherical (or angular) part of the Hamiltonian ${H}$,
\be \label{casimir}
\mathcal{I}={KH}-{D}^2,
\ee
commutes with all generators and defines a constant of motion of the Hamiltonian ${H}$. Therefore, although conformal symmetry is not a symmetry of the Hamiltonian, it equipped the system with the additional (to the Hamiltonian) constant of motion $\mathcal{I}$.

The spherical part of the conformal mechanics, determined by $\mathcal{I}$, may be considered as a Hamiltonian system by itself.
We refer to it as "spherical mechanics" or "master mechanics" throughout the paper.
It is obvious that integrability of the initial conformal mechanics
leads to integrability of the ``spherical mechanics'', and vice versa. It is also evident that the constants of
motion of the spherical mechanics are constants of motion for the
conformal mechanics. Yet, the inverse is generally not true, although
there should be a way to construct the ``spherical'' constants of motion
out of the "conformal" ones. This problem was addressed in \cite{hlns}. Some of the results of that paper are presented in this Chapter.

The study of the spherical mechanics is relevant for investigations of the Calogero model \cite{calogero69,calogero71,moser}
and its various extensions and generalizations \cite{algebra}
(for a recent review see \cite{polychronakos}).
Furthermore, the spherical mechanics  of the rational $A_N$ Calogero model
defines the multi-center (Higgs) oscillator system on the $N{-}1$-sphere
\cite{cuboct, sagh-calogero}. The well-known series of hidden constants of motion found
by Wojcechowski \cite{woj83} for the Calogero model has a transparent
explanation in terms of spherical mechanics, and its analog exists in any
integrable conformal mechanical system \cite{hkln}.
Even in the simplest case of $N{=}2$, the one-dimensional spherical mechanics
of the $A_2$ Calogero model shed light on a global aspect of Calogero models,
by elucidating the non-equivalence of different quantizations of the Calogero
model \cite{feher}.
The ${\cal N}{=}4$ superconformal generalizations of the rational $A_2$
Calogero model, constructed via supersymmetrization of spherical mechanics
\cite{bks}, yielded a scheme for lifting any ${\cal N}{=}4$
supersymmetric mechanics to a $D(1,2|\alpha)$ superconformal one \cite{hkln}.
Finally, a formulation in terms of action-angle variables \cite{lny} led
to the equivalence of the rational $A_2$ and $G_2$ Calogero models and provided
restrictions on the ``decoupling'' transformation which maps the Calogero model
to the free-particle system considered in \cite{decoupling,glp1,glp2,glp3, sagh-calogero}.

In \cite{hkln} it was demonstrated that all information
on a conformal mechanics system is encoded in its spherical part.
In particular, the ``conformal'' constants of motion with even conformal
dimension were shown to induce constants of motion for $(\omega_0, \mathcal{I})$.
However, the authors did not find the ``spherical'' constants of motion induced by the odd-dimensional initial constants of motion. This problem was solved in \cite{hlns} with the help of $so(3)$ representation theory, then, was continued in \cite{hln}.

This Chapter contains 3 sections.

In \ref{ConfMech1}, following but extending \cite{hkln},
we relate the symmetries of conformal mechanics to the particular system
of differential equations on the spherical phase space.
The analysis is simplified by the use of $so(3)$ representations,
which clarifies the origin of the spin operators appearing in the final system.

In \ref{ConfMech2} we construct a series of the constants of
motion for the spherical mechanics, which is induced by the constants of
motion (of any conformal dimension) for the conformal system.

In \ref{FourParticleCalogero} we apply our method to the rational $A_3$ Calogero model and derive the complete set of functionally independent constants of motion for the cuboctahedric Higgs oscillator.

\newpage
\subsection{The spherical part of conformal mechanics}
\label{ConfMech1}
Here, following \cite{hkln} and \cite{hlns}, we  relate the constants of motion of the conformal mechanics (\ref{so2}) with certain differential equations on the phase space of the associated spherical mechanics.

For any function $f$ on phase space, define the associated Hamiltonian
vector field by the Poisson bracket action $\hat f =\{f,.\}$. For example, the Hamiltonian vector fields
 corresponding to the generators $H, D, K$ (\ref{so2}), and  Casimir element \eqref{casimir} read
\begin{equation}
\label{so12-vec}
\hat H=
p_r\frac{\partial}{\partial r}
+\frac{4 {\mathcal{I}}}{r^3}\frac{\partial}{\partial p_r}
+\frac{2\hat{\mathcal{I}}}{r^2},
\qquad
\hat K=
-r\frac{\partial}{\partial p_r},
\qquad
\hat D=
\frac r 2 \frac{\partial}{\partial r}
-
\frac{p_r} 2 \frac{\partial}{\partial p_r},
\end{equation}
\begin{equation}
\label{hatI}
\qquad\textrm{and}\qquad
\hat{\mathcal{I}} ={H \hat K  }+{ K \hat H  }-2{ D \hat D}.
\end{equation}
Since the assignment $f\mapsto \hat f$ is a Lie algebra homomorphism, the vector fields $\hat H, \hat K, \hat D $
satisfy the $so(1,2)$ algebra \eqref{so120},  and  the vector field of the Casimir element $\hat{\mathcal{I}}$,
of course, commutes with them.

Any constant of motion is the lowest weight vector of the conformal
algebra (\ref{so120}), since it is annihilated by the Hamiltonian.
Without any restriction, one can choose  it to have a certain conformal dimension (spin):
\begin{equation}
\label{hw}
\hat H I_s= 0,
\qquad
\hat D I_s= -2s I_s.
\end{equation}

A conformal mechanics which describes identical particles and possesses
a permutation-invariant cubic (in momenta, $s{=}3/2$) constant of motion
commuting with the total momentum yields the rational Calogero model,
which is an integrable system \cite{braden}.

In the following, we consider only nonnegative integer and half-integer values of the spin $s$, so that
$I_s$ yields a finite-dimensional (nonunitary) representation
of the $so(1,2)$ algebra \eqref{so12-vec}. This includes the $N$-particle rational Calogero
model and its extensions, whose  Liouville constants of motion  are polynomials in the momenta.

Our goal is to derive the constants of motion for the
``spherical'' Hamiltonian \eqref{casimir} from the constants of motion of the
initial conformal Hamiltonian.
Using \eqref{so2}, \eqref{so12-vec}, and \eqref{hatI}
it is easy to see that the conservation condition \eqref{hw} is equivalent to the equation
\begin{equation}
\label{hatI1}
(\hat{\mathcal{I}}-\hat M)\,I_{s}(p_r,r,u)=0,
\qquad
\text{where}
\quad
 \hat M=2(\hat S_--\mathcal{I}\hat S_+).
\end{equation}
Here, the one-dimensional vector fields $\hat S_\pm$ together with $\hat S_z$
are given by
\begin{equation}
\hat S_+=\frac{1}{r}\frac{\partial}{\partial p_r},
\qquad
\hat S_-=-p_rr^2\frac{\partial}{\partial r},
\qquad
\hat S_z=-\frac{1}{2}\left(r\frac{\partial}{\partial r}
+p_r\frac{\partial}{\partial p_r}\right).
\end{equation}
Interestingly, they form an $so(3)$ algebra,
\begin{equation}
\label{so3}
[\hat S_+,\hat S_-]=2\hat S_z,
\qquad
[\hat S_z,\hat S_\pm]=\pm { \hat  S}_\pm.
\end{equation}
Note that $\hat S_+$ is generated by the Hamiltonian $ S_+=-\log(r)$ while the
other two vector fields are not Hamiltonian.

The integral \eqref{hw} can be presented as a sum of terms with decoupled
radial and angular coordinates and momenta~\footnote{
In comparison to the definition of $f_{s,m}(u)$ in \cite{hkln},
we have multiplied the binomial factor and applied an index shift
$m\to m-s$. This makes more apparent the $so(3)$ properties and
simplifies further relations.},
\be
\label{decomp}
 I_s(p_r, r, u) =\sum_{m=-s}^s f_{s,m}(u)\ R_{s,m}(p_r,r)
\qquad\textrm{with}\quad
R_{s,m}(p_r,r)=  \sqrt{\binom{2s}{s{+}m}}\,\frac{p_r^{s-m}}{r^{s+m}}.
\ee
The radial functions $R_{s,m}$  form a spin $s$-representation
($s=0,\frac12,\dots$) of the $so(3)$ algebra \eqref{so3},
\begin{equation}
\label{spin-s}
\begin{split}
&\hat S_+R_{s,m}=\sqrt{(s{-}m)(s{+}m{+}1)}\,
R_{s,m+1},
\\
&\hat S_-R_{s,m}=\sqrt{(s{-}m{+}1)(s{+}m)}\,
R_{s,m-1},
\\
&\hat S_zR_{s,m}=m\,R_{s,m}.
\end{split}
\end{equation}
Hence, $\mathcal{\hat I}$ acts nontrivially
only on the angular functions,  while the $\hat S_a$
act on the radial ones. Due  to the convolution \eqref{decomp},
one can shift the latter action to the angular functions by transposing the $so(3)$
matrices.
As a result, the action of ${\cal \hat I}$ on the spin-$s$ states
$f_{s,m}$ is given by
\be
\label{recurrent}
\hat{\mathcal{I}} f_{s,m} = \sum_{m'}M_{mm'}f_{s,m'}
=2\bigl(\sqrt{(s{-}m)(s{+}m{+}1)}
f_{s,m+1}-{\mathcal{I}}\sqrt{(s{-}m{+}1)(s{+}m)}f_{s,m-1}\bigr).
\ee
This is a system of $2s{+}1$ first-order linear homogeneous differential equations for the angular functions
$f_{s,m}(u)$. The coefficients depend only on $\mathcal{I}$, which commutes with the differential operator,
and so they can be treated as constants.
Note that all angular coefficients must obey the related $(2s{+}1)$th-order linear homogeneous differential equation
\be
\label{diff-eq}
\text{Det}(\hat{\mathcal{I}} -M)f_{s,m}=0,
\ee
which is, in fact, equivalent to the system \eqref{recurrent}, since any solution $f$ of \eqref{diff-eq}
also generates a solution of the original system.
Indeed, using \eqref{recurrent}, one can recursively express
each function $f_{s,m}$ as a $(s{\pm}m)$th-order polynomial in $\hat{\mathcal{I}}$
acting on the function $f_{s,{\mp}s}$.
Diagonalization of the matrix  $M$ decouples the system \eqref{recurrent}
into independent equations, pertaining to the eigenvalues and eigenvectors
of the vector field $\hat{\mathcal{I}}$.

Consider now some consequences of the relation \eqref{recurrent}.
From a constant of motion of the Hamiltonian, one can construct
other constants with the same conformal spin
by successive application of the vector field generated by the spherical Hamiltonian:
\be
I_s\stackrel{\hat{\mathcal{I}}}{\longrightarrow} I_s^{(1)}\stackrel{\hat{\mathcal{I}}}{\longrightarrow} I_s^{(2)}
\stackrel{\hat{\mathcal{I}}}{\longrightarrow}\ldots\stackrel{\hat{\mathcal{I}}}{\longrightarrow}  I_s^{(k)}
\stackrel{\hat{\mathcal{I}}}{\longrightarrow} \ldots,
\qquad \quad I_s^{(k)}:=\hat{\mathcal{I}}^k I_s.
\ee
In general, the members of this sequence are not in involution.
At most the first $2s{+}1$ integrals can be independent, while the
remaining ones are expressed through them linearly with $\mathcal{I}$-dependent
coefficients, since the vector field $\hat{\mathcal{I}}$ acts on the $(2s{+}1)$-vector
of constants $I_s^{(k)}$ as a square matrix with $\mathcal{I}$-valued entries.
The exact amount of
functionally independent  integrals depends on the $I_s$ as well as on
the concrete realization of the conformal mechanics.

\newpage
\subsection{Constants of motion of the spherical mechanics}
\label{ConfMech2}

The spin-$j$ representation of the rotation group parameterized by
three Euler angles is given by the Wigner $D$-matrix \cite{wigner,angular1,angular2}. Let us remind some formulae about Wigner (small) $d$-matrix. We only need the (small) $d$-matrix, which describes the rotation around the $y$ axis,
\begin{equation}
\label{d-def}
d_{m'm}^s(\beta)=\langle sm'|\exp(-i\beta S_y) |sm\rangle,
\end{equation}
where $m,m'=-s,\dots,s$ are the spin $z$-projection quantum numbers.
Its elements are real and given by \cite{wigner}
\begin{equation}
\begin{split}
d_{m'm}^s(\beta)=\sum_t (-1)^{t+m'-m}
\frac{\sqrt{(s{+}m')!(s{-}m')!(s{+}m)!(s{-}m)!}}
{(j{+}m{-}t)!(m'{-}m{+}t)!(j{-}m'{-}t)!}
\times
\\
\times
\left(\cos\sfrac{\beta}{2}\right)^{2s+m-m'-2t}
\left(\sin\sfrac{\beta}{2}\right)^{m'-m+2t},
\end{split}
\end{equation}

where the sum is over such values of $t$ that the factorials in the denominator
are nonnegative. The elements obey
\begin{equation}
\label{d-rel}
d_{m'm}^s(\beta)=d_{mm'}^s(-\beta)=(-1)^{m-m'}d_{mm'}^s(\beta)=d_{-m-m'}^s(\beta).
\end{equation}
For $\beta=\pi/2$, the above expression simplifies to
\begin{equation}
\label{d-pi2}
d_{m'm}^s(\pi/ 2)= 2^{-s} \sum_t (-1)^{t+m'-m}
\frac{\sqrt{(s{+}m')!(s{-}m')!(s{+}m)!(s{-}m)!}}
{(s{+}m{-}t)!(m'{-}m{+}t)!(s{-}m'{-}t)!}\;.
\end{equation}
Further simplifications occur when one of the spin-projection quantum
numbers vanishes, which is possible for integer spins only:
\begin{equation}
\label{d-m0}
\begin{split}
d_{m0}^s(\pi/2)&=(-1)^\frac{s+m}{2} \delta_{s-m,2\Z}
\frac{\sqrt{(s{-}m)!(s{+}m)!}}{2^s\left(\frac{s+m}{2}\right)!\left(\frac{s-m}{2}\right)!}
\\
&=(-1)^\frac{s+m}{2} \delta_{s-m,2\Z}
\sqrt{\frac{(s{+}m{-}1)!!(s{-}m{-}1)!!}{(s{+}m)!!(s{-}m)!!}}
\\
d_{0m}^s(\pi/2)&=(-1)^\frac{s-m}{2} \delta_{s-m,2\Z}
\frac{\sqrt{(s{-}m)!({s}+m)!}}{2^s\left(\frac{s+m}{2}\right)!\left(\frac{s-m}{2}\right)!}
\\
&=(-1)^\frac{s-m}{2} \delta_{s-m,2\Z}
\sqrt{\frac{(s{+}m{-}1)!!(s{-}m{-}1)!!}{(s{+}m)!!(s{-}m)!!}}
\\
d_{00}^s(\pi/2)&=(-1)^\frac{s}{2} \delta_{s,2\Z} \frac{(s{-}1)!!}{s!!}
\end{split}
\end{equation}
The factor $\delta_{s-m,2\Z}$ excludes odd values of $s{-}m$, for which the
matrix elements vanish.
For $\beta=\pi/2$, the relations are supplemented by
\be
\label{d-rel-pi2}
d_{m'm}^s(\pi/2)=(-1)^{s+m'}d_{m'-m}^s(\pi/2)=(-1)^{s-m}d_{-m'm}^s(\pi/2),
\ee
which can be obtained from $d^s_{m'm}(\pi)=(-1)^{s-m}\delta_{m',-m}$.

Now we shall present the construction of the constants of motion for
the spherical mechanics $(\omega_0, \mathcal{I})$ from those for the initial
conformal mechanics, based on $so(3)$ group representations.
This method yields constants of motion of any conformal dimension
and recovers the expressions found in \cite{hkln}.

Any constant of motion $I_s$ of the original Hamiltonian is given by its coefficients in the
decomposition \eqref{decomp}.
The related conservation condition \eqref{hatI1},
\eqref{recurrent}, or \eqref{diff-eq} is decoupled
into independent equations upon diagonalization of the matrix $M$,
\be
\label{hatI3}
\hat M=4\sqrt{-\mathcal{I}}\; \hat U\,\hat S_z\,\hat U^{-1},
\qquad\textrm{where}\quad
\hat U=(-\mathcal{I})^{\frac12 \hat S_z}e^{-\frac{i\pi}{2} \hat S_y}
\quad\textrm{with}\quad \hat S_y=\sfrac1{2i}(\hat S_+-\hat S_-).
\ee
Thus, up to an $\mathcal{I}$-valued factor, the vector field $\hat M$ is equivalent to the usual spin-$z$ projection operator.
The operator $\exp(-\frac{i\pi}2\hat S_y)$ maps $\hat S_z$ to $\hat S_x$. The latter is then transformed
to $\hat M$ by the operator $(-I)^{\hat S_z/2}$, which, for the present, means a formal power series.
Together with the factor $i\sqrt{\mathcal{I}}$ it contains square roots of $\mathcal{I}$. Thus $\hat M$  is, in general, complex and multi-valued.
When the potential is positive, as is the case in Calogero models, the spherical
part is strictly positive, and the operator \eqref{hatI3}
is complex but single-valued.  In any case, all square roots will cancel in
the final expressions for the constants of motion.

Define now the rotated basis for the algebra \eqref{so3}, which is formed by the eigenstates of
the operator $\hat M$. Using \eqref{hatI3}, we obtain
\be
\begin{split}
\label{Rx}
& {\widetilde R}_{s,m}=(\hat U R)_{s,m}=\sum_{m'} U_{m'm} R_{s,m'},
 \\
 & U_{m'm} = d^s_{m'm}(\pi/2)(-\mathcal{I})^\frac{m'}2,
\\
& \hat M {\widetilde R}_{s,m}= m {\widetilde R}_{s,m},
\end{split}
\ee
where  $d^s_{m'm}(\beta)$ is the  Wigner's small $d$-matrix, which describes
the rotation around the $y$ axis in the usual spin-$s$ representation \eqref{spin-s}.

Note that the functions ${\widetilde R}_{s,m}$ now depend on the
angular variables also through $\mathcal{I}$.
The integral \eqref{decomp} of the original Hamiltonian can be presented
in terms of the rotated functions as
\be
\label{decomp2}
 I_s(p_r, r, u) =\sum_{m=-s}^s {\widetilde f}_{s,m}(u){\widetilde R}_{s,m}(p_r,r,\mathcal{I}(u)).
\ee
The new coefficients are expressed in terms of old ones by means of the
inverse transformation [compare \eqref{decomp} with \eqref{decomp2} and \eqref{Rx}]:
\be
\label{fx}
 {\widetilde f}_{s,m}=\sum_{m'}U^{-1}_{mm'}f_{s,m'}
=\sum_{m'} (-\mathcal{I})^{-\frac {m'}2} d_{m'm}^s(\pi/2)f_{s,m'}.
\ee
In the second equation, we have applied the orthogonality condition of the $d$-matrix.
Substituting the decomposition \eqref{decomp2} into \eqref{hatI1}
and using the eigenvalue-eigenfunction equation form \eqref{decomp2},
we arrive at a similar eigensystem equation for the vector field $\hat{\mathcal{I}}$
and the rotated angular coefficients:
\be
\label{eigen}
\hat{\mathcal{I}} {\widetilde f}_{s,m}(u)=4m\sqrt{-\mathcal{I}(u)} {\widetilde f}_{s,m}(u).
\ee
This provides a formal solution to the system \eqref{recurrent}. For systems
with positive spherical part, the eigenvalue is a well-defined purely imaginary
function, and the evolution of the coefficients driven by the spherical Hamiltonian
oscillate with a frequency proportional to $m$,
\be
{\widetilde f}_{s,m}(t)=e^{iw_m(t-t_0)  }{\widetilde f}_{s,m}(t_0)
\qquad\textrm{with}\quad
\omega_m=4m\sqrt{\mathcal{I}}.
\label{lny}\ee

Various combinations of these quantities give rise to constants of motion
for the spherical Hamiltonian.  In particular, for integer spin $s$,
the zero-frequency coefficient ${\widetilde f}_{s,0}(u)$ is an integral itself.
Using the explicit expression of the Wigner $d$-matrix
for this case \eqref{d-m0}, one can express it in terms of the original
coefficients:
\be
\label{int-lin}
\begin{split}
\mathcal{J}_s(u)= \mathcal{I}(u)^{\frac{s}2}{\widetilde f}_{s,0}(u)=
 \sum_{m=-s}^s\frac{(s{+}m{-}1)!!(s{-}m{-}1)!!}{\sqrt{(2s)!}}\delta_{s{-}m,2\Z}\mathcal{I}(u)^{\frac{s{-}m}{2}}f_{s,m}(u)
\\
=\sum_{\ell=0}^{s}\frac{(2\ell{-}1)!!(2s{-}2\ell{-}1)!!}{\sqrt{(2s)!}}\mathcal{I}(u)^{\ell} f_{s,2\ell{-}s}(u).
\end{split}
\ee
Here, $\Z$ denotes the set of integer numbers, so that
$\delta_{k,2\Z}=1$ for even values of $k$ and vanishes for the odd values.
The supplementary $\mathcal{I}$-dependent factor in front of the angular coefficient
eliminates the fractional powers of $\mathcal{I}$, leaving only integral powers of
$\mathcal{I}$ in the final expression.
Up to a normalization factor, \eqref{int-lin} coincides with the expression (5.2) of \cite{hkln}.

For integer values of~$s$,
the same integral can also be obtained from the equivalent higher-order differential
equation \eqref{diff-eq}. Indeed, due to \eqref{hatI3} or \eqref{eigen}, the related
differential operator takes the following form:
\be
\text{Det}(\hat{\mathcal{I}} -M)=
\prod_{m=-s}^s({\hat{\mathcal{I}}} -4 m \sqrt{-\mathcal{I}})
=:
{\displaystyle
\begin{cases}
 \hat{\mathcal{I}}\hat\Delta_s &
\textrm{for $s\in\Z$},
\\[5pt]
\hat\Delta_s &
\textrm{for $s\in\Z{+}\sfrac12$},
\end{cases}
}
\label{det}
\ee
\be
\nonumber
\textrm{with}\quad
\hat\Delta_s=\prod_{0<m\le s}(\hat{\mathcal{I}}^2 +16m^2\mathcal{I}).
\ee
Therefore, for integer spin value, \eqref{diff-eq}  is reduced to
\be
\hat{\mathcal{I}} \hat\Delta_s f_{s,m}\equiv\prod_{m=1}^s(M^2 +16m^2\mathcal{I})f_{s,m}=0,
\ee
which implies that $\hat\Delta_s f_{s,m}$ is an integral of motion of $\mathcal{I}$. The operator
$\hat\Delta_s$ projects out all but one of the eigenfunctions ${\widetilde f}_{s,m}$,
\be
\hat\Delta_s {\widetilde f}_{s,m}=\delta_{m0}(s!)^2(16\mathcal{I})^s{\widetilde f}_{s,m}.
\ee
 Therefore, the above integral has to be proportional to \eqref{int-lin}.
This can be verified independently if we apply $\hat\Delta_s$ to both
sides of the inversion of \eqref{fx} and use \eqref{Rx}, \eqref{int-lin}, \eqref{d-m0}:
\be
\label{f-fx}
\begin{split}
\hat\Delta_s f_{s,m}
=U_{m0}\hat\Delta_s {\widetilde f}_{s,0}
=\delta_{s-m,2\Z}\,c_{s,m}\,\mathcal{I}^{\frac{s+m}2} {\cal J}_s
\\
\textrm{with}\quad
c_{s,m}=(-8i)^ss!\binom{s}{\frac{s+m}2}\sqrt{(s{-}m)!(s{+}m)!}\,.
\end{split}
\ee

How to construct an integral of $\mathcal{I} $ from an integral of $H$ with  half-integral conformal spin?
The corresponding representation has no $m{=}0$ state, but one can consider such a state
in the integral $I_{2s}=I_s^2$, which has integral spin value equal to $2s$.
In general, integrals of
 $\mathcal{I}$ can be built also by bilinear combinations of $f_{s,m}(u)$ with opposite values
 of the spin projection. In fact, any state
\be
\label{Jsm}
\begin{split}
{\cal J}_{s}^{m}
&= (-\mathcal{I})^s {\widetilde f}_{s,m} {\widetilde f}_{s,-m}
=\sum_{m',m''}
 i^{4s+m''-m'}d_{m''m}^s(\pi/2)d_{m'm}^s(\pi/2)\;
 \mathcal{I}^{s-\frac{m'+m''}{2}}f_{s,m'}f_{s,m''}
 \\
&=  \sum_{m',m''}
 \delta_{m''-m',2\Z}\,(-1)^{2s+\frac{m''-m'}{2}}d_{m''m}^s(\pi/2)d_{m'm}^s(\pi/2)\;
 \mathcal{I}^{s-\frac{m'+m''}{2}}f_{s,m'}f_{s,m''}
\end{split}
\ee
is  an integral of $\mathcal{I}$. In the first equation, we have used the symmetry property
\eqref{d-rel-pi2} of the $d$-matrix.
The Kronecker delta appears after symmetrization over the two summation indices in the first double sum, with the help of
\be
\sfrac12(i^{m''-m'}+i^{m'-m"})
= i^{m''-m'}\,\sfrac12(1+(-1)^{m'-m''})=i^{m''-m'}\delta_{m''-m',2\Z}.
\ee
Therefore, the constant of motion ${\cal J}_{s}^{m}$ of the spherical Hamiltonian is a real polynomial of order $2s$ in $\mathcal{I}$.

There is a clear interpretation of the constructed integrals in
terms of representation theory.
Take some set of angular functions satisfying
\eqref{hatI1} or \eqref{recurrent}, which means that
the related quantity $I_s$ \eqref{decomp}  is
an integral of $H$. Then, according to the tensor
product of $so(3)$ representations, one can
construct other sets of angular functions,
\be
\label{fs'm}
f_{S,m}(u)= \sum_{m_1+m_2=m} C_{s,m_1,s,m_2}^{S,m} f_{s,m_1}(u)f_{s,m_2}(u)
\qquad\textrm{with}\quad
S=2s,2s-2,\ldots,
\ee
which satisfy a similar equation.

Clebsch-Gordan coefficients are the expansion coefficients of total-spin
eigenstates $|SM\rangle$ in terms of the product basis $|s_1m_1s_2m_2\rangle$
of eigenstates of the two coupled spins,
\be
\label{CG-def}
C_{s_1,m_1,s_2,m_2}^{S,M}=\langle s_1m_1s_2m_2|SM\rangle.
\ee
The general expression is complicated, but special cases are often quite simple
like for the highest total-spin value:
\be
\label{CG-1}
  C_{s_1,m_1,s_2,m_2}^{s_1+s_2,m_1+m_2}=
  \sqrt{\frac{
  \binom{2s_1}{s_1-m_1}\binom{2s_2}{s_2-m_2}}{\binom{2s_1+2s_2}{s_1+s_2-m_1-m_2}
  }}.
\ee
The Clebsch-Gordan coefficients have an even-odd exchange symmetry
depending on the total-spin value,
\be
\label{CG-exchange}
C_{s_1,m_1,s_2,m_2}^{s,m}=(-1)^{s_1+s_2-s} C_{s_2,m_2,s_1,m_1}^{s,m}.
\ee

The multiplets with odd values of $S{-}2s$ are absent in the symmetric tensor product,
due to the  exchange symmetry of the Clebsch-Gordan coefficients
\eqref{CG-exchange}.
{}From the angular functions \eqref{fs'm} one can compose ``new'' integrals of the original Hamiltonian
via
\be
\label{I'}
I'_{S}=\sum_m f_{S,m}R_{S,m}
\qquad\textrm{with}\quad
S=2s,2s-2,\ldots,
\ee
each corresponding to a symmetric multiplet in the tensor product
of two spin-$s$ multiplets.
Note that the first integral from this set
just coincides with the square of the original integral,
$I'_{2s}=I_s^2$, as can easily be verified using \eqref{CG-1}.
Since $S$ is always integer, the related multiplet contains an
$m=0$ state, which is a constant of motion of the spherical Hamiltonian:
\be
\label{Fss}
\mathcal{F}^{S}_{s}(u)
=\sum_m C_{s,m,s,-m}^{S,0}\mathcal{J}^s_m(u).
\ee
Unwanted fractional powers of $\mathcal{I}$ cancel as before.
These two sets of integrals are, of course, equivalent.

A similar ``blending'' procedure can be applied to the mixing of two different integrals
$I_{s_1}$ and $I_{s_2}$ with integer value of $s_1{-}s_2$. The resulting integrals of $\mathcal{I}$ are
parameterized by the whole set of $2s_{\min}{+}1$ different angular momenta obeying the sum rule.

The construction straightforwardly generalizes also to
multilinear forms composed from the angular functions.
The expression \eqref{Jsm} expands to
\be
{\cal J}^{m_1\dots m_k}_{s_1\dots s_k}(u)
=\mathcal{I}(u)^{\frac12\sum_\ell s_\ell}\prod_{\ell=1}^k
{\widetilde f}_{s_\ell,m_\ell}(u)
\qquad
\textrm{with}
\quad
{\sum_{\ell=1}^km_\ell=0},
\ee
where the last relation implies that the total spin $\sum s_\ell$ must be an integer.
These observables can be combined into a single multiplet of integer spin $S$
by a $(k{-}1)$-fold application of the Clebsch-Gordan decomposition. The final set
of observables ${\widetilde f}_{S,m}$ forms an integral of the original Hamiltonian, while
its $m{=}0$ element corresponds to an integral of the spherical Hamiltonian.

So far, we have only considered products of the angular functions.
More generally however, one could also employ
fractions of them, with the same spin  projection of the numerator and
the denominator, such as ${\widetilde f}_{s_1,m}/{\widetilde f}_{s_2,m}$.
Of course, this entails introducing singularities, which might create problems
for the quantization due to inverse powers of moments.

It has to be mentioned that the variety of angular constants of motion constructed here
are not independent. It may even happen that
some of them vanish. Moreover, the compatibility of the integrals of
motion for $H$ does not at all yet imply the compatibility of the associated integrals
for $\mathcal{I}$, as can be seen from \eqref{Jsm}.

\newpage
\subsection*{Examples}

At the end of this section, we demonstrate our method by presenting
some simplest examples for the obtained constants of motion.

First we note that there exist two bilinear conserved quantities
\eqref{Jsm} and \eqref{Fss}, which have a rather simple form in
terms of the original angular coefficients.
The first one is the canonical ``singlet'', which is the same
both in the original and the rotated basis,
\be
\label{singlet}
\mathcal{F}^{0}_{s}(u)\sim
\sum_{m}(-1)^{s-m} {\widetilde f}_{s,m}{\widetilde f}_{s,-m}
 = \sum_{m}(-1)^{s-m} f_{s,m}f_{s,-m}.
\ee
The second one is given by the trivial superpositions of the states
\eqref{Jsm}, which is reduced by the orthogonality of the $d$-matrices
to
\be
\label{square}
\sum_m {\cal J}_s^m
\sim
\sum_{m}\mathcal{I}^{s-m}f_{s,m}^2.
\ee

For the integral $I_s$ of the Hamiltonian $H$ with conformal spin $s{=}\frac12$, the
general formula \eqref{Jsm} takes its simplest form, up to a normalization factor,
\begin{equation}
\mathcal{J}_{\frac12}^{\frac12}
\sim
\mathcal{I} f_{\frac12,-\frac12 }^2 +f_{\frac12,\frac12 }^2.
\end{equation}

Consider now the integral with conformal spin $s{=}1$ of the original Hamiltonian.
The related linear integral of $\mathcal{I}$ is (see \eqref{int-lin})
\be
{\cal J}_{1}\sim\mathcal{I} f_{1,1}+ f_{1,-1}.
\ee
In addition, there are two quadratic integrals given by \eqref{Jsm},
one of which (${\cal J}_{s=1}^{m=0}$) is the square of the above integral,
while the other one can be identified with either \eqref{singlet} or \eqref{square}.
The Hamiltonian itself can be considered as a particular case. For
$I_1=H$, the coefficient $f_{10}$ vanishes while the others become constants,
so the sole constant of $\mathcal{I}$ extracted from $H$ is $\mathcal{I}$ itself.

The first nontrivial case corresponds to the next conformal spin $s=\frac32$,
when there is no linear but two independent quadratic integrals.
The simplest choice then are the two functions  \eqref{singlet} and \eqref{square}.

\newpage
\subsection{Four-particle Calogero model}
\label{FourParticleCalogero}
The (rational) Calogero model \cite{calogero69,calogero71,moser},
which is an integrable $N$-particle one-dimensional system with pairwise
inverse-square interaction (and its various generalizations related
with different Lie algebras and Coxeter groups~\cite{algebra})
is a famous example of a conformal mechanical system
(for the review, see~\cite{polychronakos}).
Usually, Lax-pair and matrix-model approaches
are employed for the study of this system. These are common methods which
are applied to other integrable models not related to the conformal group.
At the same time, many properties of the rational Calogero model are
due to its conformal invariance, and they are shared with
other conformal mechanical models. For example,
the ``decoupling transformation" in the Calogero model~\cite{decoupling}
can be formulated purely in terms of conformal transformations~\cite{glp1,glp2,glp3}
(see also~\cite{lobach}). Note that the rational
$N$-particle Calogero model is a maximally superintegrable system,
i.e.~it possesses $N{-}1$ additional
functionally independent integrals apart from the Liouville
integrals being in involution~\cite{woj83}. Despite an impressive
list of references on this subject~\cite{supint2}, the
superintegrability of the Calogero model still seems to be mysterious.
Preliminary considerations have indicated a
direct connection between the additional constants of motion and the
``angular" part of the Calogero model~\cite{cuboct}.

Let us use the general method developed in the previous section
to construct the complete set of constants of motion
for the spherical mechanics of the four-particle Calogero model after
decoupling the center of mass (i.e.\ of the rational $A_3$ Calogero model).
This spherical mechanics also describes a particle on the two-dimensional
sphere, interacting by the Higgs-oscillator law with force centers located
in the vertices of a cuboctahedron. By this reason, the
system was termed ``cuboctahedric Higgs oscillator''
\cite{cuboct}.

We remind that the standard rational Calogero model, 

\be
\label{Calogero}
H=\frac{1}{2}\sum_{i=1}^N p_i^2 + \sum_{i<j}\frac{g^2}{(x_i-x_j)^2},
\ee
has $N$ Liouville constants of motion, given in terms of a Lax matrix by the expression \cite{polychronakos}
\be
\label{Liouville}
I_s=\text{Tr}\,L^{2s} \qquad\textrm{with}\quad s=\sfrac12,1,\dots,\sfrac{N}2,
\ee
where
\be
\label{Lax}
L_{jk}=\delta_{jk}p_k+(1{-}\delta_{jk})\frac{ig}{x_j-x_k}.
\ee
Hence,  $I_\frac12=\sum_i p_i$ and $I_1=\frac12 H$.
Furthermore, the quantities
\be
\label{woj}
I_s^{(1)}=\hat{\mathcal{I}} I_s \qquad\textrm{for}\quad \qquad s\ne 1
\ee
 coincide with Wojciechowski's additional integrals \cite{hkln}. Together with \eqref{Liouville}, they form a complete set of
functionally independent integrals making the system maximally superintegrable \cite{woj83}.

We choose $N{=}4$ and pass to new coordinates
\be
\begin{split}
y_0=\sfrac12(x_1{+}x_2{+}x_3{+}x_4),\qquad
&y_1=\sfrac12(x_1{+}x_2{-}x_3{-}x_4),
\\
y_2=\sfrac12(x_1{-}x_2{+}x_3{-}x_4),\qquad
&y_3=\sfrac12(x_1{-}x_2{-}x_3{+}x_4)
\end{split}
\ee
and associated momenta $\tilde p_i$ with $i=0,1,2,3$. This transformation
decouples the center-of-mass coordinate $y_0$ and momentum $\tilde p_0$ from
the others. After setting
\be
y_0=0, \quad \tilde p_0=0,
\ee
 the Hamiltonian takes the form of the rational $D_3{\sim}A_3$ Calogero model \cite{cuboct}
\be
\label{D3}
H=
\sfrac{1}{2}\sum_{i=1}^3\tilde p_i^2+\sum_{i,j=1}^3\left(\frac{g^2}{(y_i-y_j)^2}+\frac{g^2}{(y_i+y_j)^2}\right)
=  \sfrac{1}{2}p_r^2+\frac{{\mathcal{I}}(p_\theta , p_\varphi, \theta
,\varphi )}{2r^2}.
\ee
In the second equation,
we introduced spherical coordinates $(r,\theta,\varphi)$ on $\R^3(y_1,y_2,y_3)$
together with their conjugate momenta $(p_r,p_\theta,p_\varphi)$, so that

\be
\begin{split}
\mathcal{I}(p_\theta,p_\varphi, \theta,\varphi) &={p^2_\theta}+\frac{p^2_\varphi}{\sin^2\theta}+
\\
+&\frac{2g^2}{\sin^2\theta}\sum_{\pm} \left[\frac{1}{(\cos \varphi \pm\sin\varphi)^2}+\frac{1}{(\cot\theta\pm\sin\varphi )^2}+\frac{1}{(\cot\theta\pm\cos\varphi )^2} \right], \label{z4}
\end{split}
\ee

According to \eqref{Liouville} and \eqref{Lax},
the conformal Hamiltonian \eqref{D3} has two Liouville constants of motion
of conformal dimension three and four, given by
\begin{align}
\label{trL3}
\begin{split}
I_\frac32=\text{Tr}(L^3)= \sum_{i=1}^4 p_i^3 + \ldots
\quad
&= 3 \tilde p_1 \tilde p_2 \tilde p_3 + \ldots
\quad
=\sfrac{3}{2} p_r^3 \cos\theta\sin^2\theta\sin 2\varphi + \ldots\;,
\end{split}
\\
\label{trL4}
\begin{split}
I_2=\text{Tr}(L^4)=\sum_{i=1}^4 p_i^4 +  \ldots
\quad
&=\sfrac{1}{4}(\tilde p_1^4+\tilde p_2^4+\tilde p_3^4)+
\sfrac{3}{2}(\tilde p_1^2 \tilde p_1^2 + \tilde p_1^2 \tilde p_3^2 + \tilde p_2^2\tilde p_3^2 ) + \ldots =
\\
&=\sfrac{1}{4} p_r^4 \left(\sin^2 2\theta +  \sin^4\theta ~ \sin^2 2\varphi + 1 \right)+\ldots\;.
\end{split}
\end{align}
Here, we have written out only the terms of highest order in the momentum.
Comparing  \eqref{trL3} and \eqref{trL4} with \eqref{decomp},
we obtain the spherical functions $f_{s,-s}$ as the coefficients
of the monomials $p_r^{2s}$,
\be
f_{\frac{3}{2}}(\theta,\varphi)=\sfrac{3}{2}\cos\theta\sin^2\theta\sin 2\varphi,
\qquad
f_{2}(\theta,\varphi)=\sfrac{1}{4}\left(\text{sin}^2 2\theta +  \text{sin}^4\theta ~ \text{sin}^22\varphi\right).
\label{f34}
\ee
Here and in the following, we use for convenience the shorter notation
\be
f_s(\theta,\varphi):=f_{s,-s}(\theta,\varphi).
\ee
The Liouville integrals \eqref{trL3} and \eqref{trL4} are supplemented by the
two related Wojciechowski integrals $I_\frac32^{(1)}$ and $I_2^{(1)}$
\eqref{woj}, whose leading-term coefficients are
\be
g_{\frac{3}{2}}=\hat{\mathcal{I}}\,f_{\frac{3}{2}} \qquad\textrm{and}\qquad
g_{2}=\hat{\mathcal{I}}\,f_{2}.
\ee
Note that the $f_s$ depend on the angles only while the $g_s$ are linear
in the angular momenta.
Together with the Hamiltonian \eqref{D3}, we obtain a complete set
$\{H,I_\frac32,I_\frac32^{(1)},I_2,I_2^{(1)}\}$ of five independent integrals.

In order to derive the Poisson algebra of integrals, we compute first
the commutators between the related coefficients:
\be
\begin{aligned}
& \{f_{\frac{3}{2}},g_{\frac{3}{2}}\}= 18\,(f_{\frac{3}{2}}^2-f_2), \quad
&& \{f_2, g_2\} = 8\,(4f_2^2-\sfrac13{f_{\frac{3}{2}}^2}-f_2),
\\
& \{f_{\frac{3}{2}},f_2\}= 0,
&& \{g_{\frac{3}{2}}, g_2\} =
 4\,(2g_{\frac{3}{2}} f_2 - 3 f_{\frac{3}{2}} g_2),
\\
& \{f_{\frac{3}{2}},g_2\}= \{f_2,g_{\frac{3}{2}}\}
 = 8f_{\frac{3}{2}} (3  f_2 - 1).
&&
\end{aligned}
\ee

Since the map $I_s\to f_s$ is a Poisson algebra homomorphism \cite{hkln},
we immediately get the analogous relations for the constants of motion by inserting
powers of $2H$ in order to balance the conformal spins on both sides of the equations
(the coefficient for the Hamiltonian \eqref{D3} is a constant: $f_1=\sfrac12$).
Thus, the nontrivial Poisson brackets are
\be
\begin{aligned}
 & \{I_\frac32,I_\frac32^{(1)}\}= 18\,(I_\frac32^2-2 I_2 H),
  && \{I_2, I_2^{(1)}\} = 8\,(4I_2^2-\sfrac{2}{3} I_\frac32^2 H-4 I_2 H^2),
\\[25pt]
  &\{I_\frac32,I_2^{(1)}\}= \{I_2,I_\frac32^{(1)}\}= 8 I_\frac32(3 I_2- 4 H^2),
\quad
 && \{I_\frac32^{(1)}, I_2^{(1)}\} =4\,(2I_\frac32^{(1)} I_2 - 3 I_\frac32 I_2^{(1)}).
\end{aligned}
\ee
This is a particular realization of part of the quadratic algebra
related to the Hamiltonian \cite{kuznetzov} (see \cite{sasaki01}
for rational Calogero models based on arbitrary root systems).
It is expressed in terms of independent generators, therefore
higher orders appear on the right-hand sides.

We now derive a complete set of functionally independent constants of motion
for the spherical mechanics of the four-particle Calogero model.
The second expression in \eqref{det} immediately yields
the spherical constant of motion associated with
\eqref{trL4},
\be
{\cal J}_2 = -\sfrac{1}{\sqrt{6}}\bigl(\sfrac{1}{256} \hat{\mathcal{I}}^4 + \sfrac{5}{16} \mathcal{I} \hat{\mathcal{I}}^2
+ 4 \mathcal{I}^2\bigr) f_{2}.
\ee
Its explicit expression, which can be calculated using \eqref{z4} and \eqref{f34}, is highly complicated,
\be
\label{J2}
\begin{aligned}
{\cal J}_2=&\sfrac{1}{\sqrt{6}}\biggl[
\sfrac{1}{16} (3 \cos4\varphi-11)\,p_{\theta}^4-
\sfrac{3}{4}\cot\theta \sin 4\varphi\ p_{\theta}^3 p_{\varphi}-
\Bigl(\frac{11{+}9 \cos4\varphi}{8\sin^2\theta}{+}\sfrac{9}{4}\sin^2 2\varphi\Bigr)p_{\theta}^2 p_{\varphi}^2
\\
+&\sfrac{3}{4} \cot^3\theta \sin 4\varphi\ p_{\theta}p_{\varphi}^3+
\frac{3 \cos ^4\theta \cos 4 \varphi + 21 \sin ^4 \theta - 18 \sin ^2 \theta -11}{16\sin^4\theta}\,p_{\varphi}^4
\biggr]
\\[4pt]
+&g^2K_1(\theta,\varphi)\,p_{\theta}^2
+g^2K_2(\theta ,\varphi)\,p_{\theta }p_{\varphi}
+g^2K_3(\theta ,\varphi)\,p_{\varphi}^2+g^4K_4(\theta ,\varphi),
\end{aligned}
\ee
where the functions $K_1(\theta ,\varphi),K_2(\theta
,\varphi), K_3(\theta ,\varphi), K_4(\theta ,\varphi)$ are given by
{\scriptsize
\begin{align}
K_1(\theta,\varphi)& = \frac{1}{2^{14}\sqrt{6}\cos^2 2\varphi \
(\cos^2\theta -\sin^2\theta \cos^2\varphi)^2
(2\sin^2\theta \cos 2\varphi +3\cos 2\theta +1)}\ \times \\ \nonumber
\Bigl(&768\,(25 + 29 \cos 2 \theta) \sin^6\theta \cos 12 \varphi
+96\,(1370 + 2327 \cos 2 \theta +1542 \cos 4 \theta
+393 \cos 6 \theta) \sin^2\theta \cos 8 \varphi \\ \nonumber
&-(119258 + 175774 \cos 2 \theta +45096 \cos 4 \theta +57723 \cos 6 \theta
-10242 \cos 8 \theta +5607 \cos 10 \theta) \sin^{-2}\theta \cos 4 \varphi
\\ \nonumber
&+(1021064 + 365088 \cos 2\theta - 223008 \cos 4\theta
- 183840 \cos 6\theta - 61800 \cos 8\theta -655360\sin^{-2}\theta) \Bigr)\;,
\end{align}
\begin{align}
K_2(\theta,\varphi) &= \frac{3\cot \theta  \tan 2 \varphi }
{8\sqrt{6}\sin^2\theta\,(17\cos 4\theta+28\cos 2\theta-8\sin^4\theta\cos 4\varphi+19)^2}
\ \times \\ \nonumber
\Bigl(&351 \cos 10 \theta +1350 \cos 8 \theta +13779 \cos 6 \theta
+9992\cos 4\theta+35022\cos 2\theta-13824\sin^8\theta\cos^2\theta\cos 8\varphi
+5042 \\ \nonumber
&-64\,(81\cos 6\theta+702\cos 4\theta+1071\cos 2\theta+962)
\sin ^4\theta \cos 4 \varphi \Bigr)\:,
\end{align}
\begin{align}
K_3(\theta,\varphi) &= \frac{1 }{16\sqrt{6}\cos^2 2\varphi\,
(17\cos 4\theta+28\cos 2\theta-8\sin^4\theta\cos 4\varphi+19)^2}
\ \times \\ \nonumber
\Bigl(&162\,(13 \sin 2\varphi +\sin 6\varphi)^2\cos 8\theta
+24\,(3898{-}1569\cos 4\varphi{-}282\cos 8\varphi{+}\cos 12\varphi)\cos 6\theta
\\ \nonumber
&+36\,(6686{+}1931\cos 4\varphi{-}430\cos 8\varphi{+}5\cos 12\varphi)\cos 4\theta
+72\,(546{+}10587\cos 4\varphi{-}898\cos 8\varphi{+}5\cos 12\varphi)\cos 2\theta\\[6pt] \nonumber
&-(1087746{-}1625907\cos4\varphi{+}46158\cos 8\varphi{+}483\cos 12\varphi)
\\ \nonumber
&+262144\,(5{-}4 \cos 4\varphi)\,\sin^{-2}\theta
-32768\,(11{-}3 \cos 4\varphi)\,\sin^{-4}\theta \Bigr)\;,
\end{align}
\begin{align}
K_4(\theta,\varphi)&= \frac{-1}{64\sqrt{6} \left((60 \cos 2 \theta +33 \cos 4\theta +35) \cos2 \varphi -8 \sin ^4\theta
\cos 6 \varphi\right)^4}
\times \\\nonumber
&\Bigl[  64 (335698872 \cos 2 \theta
+204278376 \cos 4 \theta +100740648 \cos 6 \theta +30799596 \cos 8
\theta +3629304 \cos 10 \theta
\\\nonumber
&+515160 \cos 12 \theta -649944 \cos
14 \theta -194643 \cos 16 \theta +197597863) \cos 8 \varphi
\\\nonumber
&+384 \sin^4
\theta  \bigl(
(-16777208 \cos 2 \theta -15290507 \cos 4 \theta
-10272396 \cos 6 \theta -4824234 \cos 8 \theta -2019708 \cos 10
\theta \\\nonumber
&-312741 \cos 12 \theta -8174886) \cos 12 \varphi -768 \sin^8
\theta  (828 \cos 2 \theta +243 \cos 4 \theta +617) \cos 20 \varphi
\\\nonumber
&-32
\sin ^4 \theta  (290832 \cos 2 \theta +188916 \cos 4 \theta +81648
\cos 6 \theta +13851 \cos 8 \theta +166129) \cos 16 \varphi\bigr)
\\\nonumber
&+\sin^{-4} \theta \bigl(-(9941103400 \cos 2 \theta +11541549238
\cos 4 \theta +10411072176 \cos 6 \theta +8259070392 \cos 8 \theta\\\nonumber
&+4658511600 \cos 10 \theta +1965778311 \cos 12 \theta +569460204
\cos 14 \theta +67528026 \cos 16 \theta -29495988 \cos 18 \theta\\\nonumber
&-8028477 \cos 20 \theta +4163058670) \cos 4 \varphi +62158979032 \cos 2
\theta +46026533130 \cos 4 \theta +27521060688 \cos 6 \theta\\\nonumber
&+12943186248 \cos 8 \theta +4533912336 \cos 10 \theta +1033949913
\cos 12 \theta -11388780 \cos 14 \theta -94673178 \cos 16 \theta\\\nonumber
&-31001292 \cos 18 \theta -6738147 \cos 20 \theta +34904741074\bigr)\Bigr] \;.
\end{align}
}

The system of equations \eqref{recurrent} can be applied in order to
express the coefficients $f_{\frac32,m}$ in terms of the ``lowest'' one:
\be
f_{\frac32,-\frac12}=\sfrac{1}{2\sqrt{3}}\hat{\mathcal{I}} f_{\frac32},
\qquad
f_{\frac32,\frac12}=
\bigl(\sfrac{1}{8\sqrt{3}}\hat{\mathcal{I}}^2+\sfrac{\sqrt{3}}{2}\mathcal{I} \bigr) f_{\frac32},
\qquad
f_{\frac32,\frac32}=
\bigl(\sfrac{1}{48}\hat{\mathcal{I}}^2+\sfrac{7}{12}\mathcal{I} \bigr)\hat{\mathcal{I}} f_{\frac32}.
\ee
Then, using \eqref{z4} and \eqref{f34},
one obtains the spherical constants of motion (\ref{Jsm}) associated with \eqref{trL3},
namely ${\cal J}_{\frac{3}{2}}^{\frac{3}{2}}$ and ${\cal J}_{\frac{3}{2}}^{\frac{1}{2}}$.
Their explicit expressions are rather lengthy:
\begin{align}
\label{J12}
\begin{split}
{\cal J}_{\frac{3}{2}}^{\frac{1}{2}} = -&\sfrac{3}{32}\sin^2 2 \varphi \,p_\theta^6
- \sfrac{3}{16}\cot \theta  \sin 4 \varphi \,p_\theta^5 p_\varphi
\\
- &\sfrac{3}{128 \sin^2 \theta} \left( 6 \cos^2 \theta + ( 13 - 3 \cos 2 \theta ) \cos 4 \varphi \right) p_\theta^4 p_\varphi^2
\\
- &\sfrac{3}{128 \sin^4 \theta} \left(22 \sin^4 \theta
 - ( 43 - 53 \cos 2 \theta ) \cos 4 \varphi \cos^2 \theta + 6 \cos 2 \theta \right) \,p_\theta^2 p_\varphi^4
 \\
- &\sfrac{3}{32 \sin^5 \theta }( 7 - 9 \cos 2 \theta ) \cos^3 \theta \sin 4 \varphi \; p_\theta  p_\varphi^5
+ \sfrac{3}{2} \cot \theta \sin 4 \varphi\,p_\theta^3 p_\varphi^3
\\
- &\sfrac{3 \cos^2 \theta }{128 \sin^6 \theta} \left(  (5+ 11 \cos 4 \varphi )\sin^2 \theta
+ (2-9\cos2\theta\sin^2\theta)(1-\cos 4 \varphi ) \right)  p_\varphi^6
\\
 +&\ \textrm{terms of lower order in $p_\theta$ and $p_\varphi$},
\end{split}
\\
\label{J32}
\begin{split}
{\cal J}_\frac{3}{2}^\frac{3}{2}=-&\sfrac{9}{32}\sin^2 2 \varphi\, p_\theta^6
- \sfrac{9}{16}  \cot \theta \sin 4 \varphi\, p_\theta^5 p_\varphi
- \sfrac{9}{64}  \left( \sfrac{5 \cos 4 \varphi + 3}{\sin^2\theta }+10 \sin^2 2 \varphi \right)p_\theta^4 p_\varphi^2
\\
- &\sfrac{9}{64 \sin^4 \theta }
\left(5 \cos ^4\theta \cos 4 \varphi+ 10 \sin^2 \theta - 5 \sin^4 \theta  + 3 \right)
p_\theta^2 p_\varphi^4
+ \sfrac{9}{16}\cot^5\theta \sin 4 \varphi \, p_\theta  p_\varphi^5
\\
+ &\sfrac{9\cos^2 \theta}{64 \sin^6 \theta }
\left(\cos^4\theta \cos 4 \varphi- 6\sin^2 \theta - \sin^4 \theta  - 1 \right)
p_\varphi^6\
\\
+ &\ \textrm{terms of lower order in $p_\theta$ and $p_\varphi$}.
\end{split}
\end{align}
Clearly, ${\cal I}$, ${\cal J}_2$, ${\cal J}_{\frac{3}{2}}^{\frac{1}{2}}$ and
${\cal J}_{\frac{3}{2}}^{\frac{3}{2}}$ cannot be functionally
independent, since our spherical mechanics has a four-dimensional phase space.
Indeed, we uncover the following algebraic relation,
\be
{\cal J}_{\frac{3}{2}}^{\frac{3}{2}} = \sfrac{1}{3}{\cal J}_{\frac{3}{2}}^{\frac{1}{2}}
+2\sqrt{\sfrac{2}{3}}{\cal J}_2 \mathcal{I}+\sfrac{1}{3}\mathcal{I}^3+
4 g^2 \mathcal{I}^2.
\ee
This is the only relation among the four constants of motion,
since \eqref{J12} and \eqref{J32}
are not in involution with  \eqref{J2}. Even their free-particle
parts ($g{=}0$ projects to the terms of highest order in the momenta)
do not commute as is easy to verify.
Hence, we have found three functionally independent spherical constants
of motion for the $A_3$ Calogero model.
This confirms the superintegrability of that system.

\newpage
\setcounter{equation}{0}
\section{A PARTICLE NEAR THE HORIZON OF EXTREMAL BLACK HOLES}
\label{BlackHoles1}
\subsection{Conformal mechanics associated with the near horizon geometry}
\label{BlackHolesConformal}
The black hole solutions allowed in supersymmetric field
theories have an extremality property, that is, the inner and outer horizons
 of the black hole coalesce. In this case one can pass to the near-horizon limit, which brings us to new solutions of Einstein equations.
 In this limit (near-horizon extremal black hole) the solutions become conformal invariant.
The conformal invariance was one of the main reasons why the extremal black holes have
been payed so much attention to for the last fifteen years. Indeed, due to conformal
invariance black hole solutions are
a good research area for studying conformal field theories and
AdS/CFT correspondence (for the recent review see \cite{bkls}).
The simplest way to research this type of configurations is to study the motion of a (super)particle in this
background. The first paper that considered such a problem is
\cite{cdkktp}, where the motion of particle near horizon of extremal
Reissner-Nordstr\"om black hole has been considered. Later similar
problems in various extremal black hole backgrounds were studied by
several authors (see \cite{ikn,bk,galaj} and refs therein).

In general, some important features of a black hole are adequately captured in the model of a relativistic particle moving on the curved background \cite{gn}.
A classic example of such a kind is the discovery of a quadratic first integral for a massive particle in the Kerr space--time \cite{car} which preceded the construction of the second rank Killing tensor for the Kerr geometry \cite{wp}. In some instances the argument can be reversed.
In Ref. \cite{cdkktp} a massive charged particle moving near the horizon of the extremal Reissner-Nordstr\"om black hole was related to the
conventional conformal mechanics in one dimension \cite{Alfaro} by implementing a specific limit in which
the black hole mass $M$ is large, the difference between the particle mass and the absolute value of its charge
$(m-|e|)$ tends to zero with $M^2 (m-|e|)$ fixed. The angular variables effectively decouple in the
above mentioned limit and show
up only in an indirect way via the effective coupling constant characterizing the conformal mechanics.
 In that setting the absence of a normalizable ground state in the conformal mechanics
and the necessity to redefine its Hamiltonian \cite{Alfaro} were given a new black hole
interpretation \cite{cdkktp}.

It is
obvious, that particle moving on conformal-invariant background
 inherits the
property of (dynamical) conformal symmetry, that is, one can present
additional generators $K$ and $D$, which form, together with the
Hamiltonian $H$, the conformal algebra \ref{so120} (see page \pageref{so120}):
\be
\nonumber
\{H,D\}=H, \quad \{D,K \} =K, \quad \{H,K \}=2D.
\ee

Conformal mechanics associated with the near horizon geometry of an extremal black
hole is described by the triple  \cite{gn}
\be \label{ham}
\begin{split}
H=&r \left( \sqrt{{(r p_r)}^2 + L(p_a,\varphi^a)}
-q(p_a,\varphi^a) \right),
\\
K=&\frac{1}{r} \left( \sqrt{{(r p_r)}^2
+  L(p_a,\varphi^a)}+q(p_a,\varphi^a) \right)+ t^2 H + 2 t r p_r,
\\
D=&t H + r p_r,
\end{split}
\ee
which involves the Hamiltonian $H$, the generator of dilatations
$D$, and the generator of special conformal transformations $K$.

As proved in Chapter \ref{confmech}, one can state that a conformal mechanics can be presented
in a non-relativistic ``canonical" form \be \label{dec}
H=\frac{p^2_R}{2}+\frac{2 {\cal  I }(u)}{R^2},\qquad
\Omega=dp_R\wedge dR+\frac 12\omega_{\alpha
\beta}(u)~du^\alpha\wedge du^\beta. \ee where $R$ and $p_R$ are the new
effective radial coordinate and the
 momentum,  and ${\cal I}=HK-D^2$ is the Casimir element of $so(2,1)$ algebra. We will show how the new radial coordinate and
 momentum $R$ and $p_R$ are related to the old ones shortly.

 There is no general canonical transformation known which transforms arbitrary conformal  mechanics to the form
(\ref{dec}). For the particular case of the near-horizon motion of
the particle in the extremal Reissner-Nordstr\"om background such transformation has been suggested
 in \cite{bgik}, while recently it was extended to the case of general four-dimensional
near horizon extremal black hole in \cite{gn}.
 In the latter the authors, taking  into account the integrability of
 this system, suggested the generic canonical transformation,
 assuming  that the angular system ($\frac 12 \omega_{\alpha\beta}du^\alpha\wedge du^\beta, {\cal I}(u)$) is
 formulated in action-angle variables.
  They exemplified  their scheme,  constructing the action-angle variables for a neutral particle
  near the horizon of extremal
  Reissner-Nordstr\"om black hole, as well as discussed the case of the charged particle near the horizon
  of extremal  Dilaton-Axion (Cl\'ement-Gal'tsov)  black hole\cite{cg},
  without actually constructing the action-angle variables for the second
  system. Then, the action-angle
formulation for the angular part of a near-horizon  particle
dynamics in the extremal Kerr black hole background has also been
presented \cite{bny}.

The functions $L(p_a,\varphi^a)$ and $q(p_a,\varphi^a)$ entering Eq. (\ref{ham}) depend
on the details of a particular black hole under consideration: see
\cite{bgik,galaj} for the near horizon extremal Reissner-Nordstr\"om
black hole, \cite{cg1} for the rotating extremal dilaton--axion black
hole,  \cite{g2} for the extremal Kerr  solution, and \cite{go} for
the  extremal Kerr-Newman and Kerr-Newman-AdS-dS black holes.

Let us denote $D_0={D|}_{t=0}$, $K_0={K|}_{t=0}$, where  $t$ is the temporal coordinate. Note that $H$, $D_0$ and $K_0$ obey the structure relations of $so(2,1)$  as well. The latter fact allows one to separate the radial canonical pair from the rest by introducing the new radial coordinate \cite{hkln}
\be\label{Rtr}
R=\sqrt{2 K_0}, \qquad p_R=\frac{2 D_0}{\sqrt{2 K_0}} \quad \Rightarrow \quad \{p_R, R\}=1
\ee
such that
\be\label{Hrtr}
H=\frac 12 p_R^2+\frac{2\mathcal{I}}{R^2},
\ee
where $\mathcal{I}$ is the Casimir element of $so(2,1)$
\be\label{i}
\mathcal{I}=H K-D^2=H K_0-D_0^2 = L(p_a,\varphi^a)-q(p_a,\varphi^a)^2.
\ee

In general, $\mathcal{I}$ is at most quadratic in momenta canonically conjugate to the remaining angular variables.

However, with respect to the Poisson bracket the new
radial variables $(R, p_R)$ do not commute with $p_a, \varphi^a$.
In order to split them, we perform a canonical transformation
$(r,p_r, \varphi^a, p_{a})\to
(R,p_R, {\widetilde\varphi^a}, {\widetilde p_{a}})$, which is defined by (\ref{Rtr})
and by the
following transformation of the remaining variables (for related earlier studies see \cite{bgik,lp,galaj,gn})
\be
\begin{split}
&{\widetilde\varphi^a}=\varphi^a+\frac{\partial U}{\partial p_a},\qquad
{\widetilde p_a}=p_a-\frac{\partial U}{\partial \varphi^a},
\\
&U(rp_r,p_a,\varphi^a)\equiv
\frac12\int_{x=rp_r}dx\log\left(\sqrt{x^2/4 +L(p_a,\varphi^a)}
+f( p_a)\right).
\end{split}
\label{tphi}
\ee
As a result, $(R,p_R)$ and $({\widetilde\varphi}^a, {\widetilde p}_a)$ constitute canonical pairs.

Thus,  by applying a proper canonical transformation one can bring the  model of a massive relativistic particle moving near the horizon of an extremal black hole to the conventional conformal mechanics form. Important information about the system, which was originally defined in $d$ dimensions, is thus imprinted in the $(d-2)$--dimensional spherical mechanics.

In this way the spherical sectors of the conformal mechanics on the Reissner-Nordstr\"om, Dilaton-Axion and  Kerr backgrounds \cite{bny,sag},
as well as on the Myers--Perry background with equal rotation parameters \cite{jpcs}  were analyzed.
Spherical mechanics describing these systems looks as follows \cite{gns14}:
\\

{\bf Reissner-Nordstr\"om BH}. The spherical mechanics associated with the near horizon Reissner-Nordstr\"om black hole is governed by the Hamiltonian
\be
{\cal
I}=p^2_\theta+\frac{(p_\varphi+ep\cos\theta)^2}{\sin^2\theta}+(mM)^2-(eq)^2,\quad
\omega = dp_\theta\wedge d\theta+dp_\varphi\wedge d\varphi,
\ee
where $m$ and $e$ are the mass and the electric
charge of a particle, while $M$ , $q$ and $g$ are the mass,  the electric
and magnetic charges of the black hole, respectively.
This is precisely the spherical Landau problem (a particle on a two-dimensional sphere in the presence of a constant magnetic field generated by the Dirac monopole) shifted by the constant
${\cal I}_0 =(mM)^2-(eq)^2$. A link between the two systems was discussed in \cite{bgik,galaj}.\\

{\bf Clement-Gal'tsov BH}. This solution of the Einstein--Maxwell--dilaton--axion theory can be viewed as interpolating between the near horizon extremal Reissner-Nordstr\"om black hole and the near horizon extremal Kerr black hole  \cite{cg1}. The corresponding spherical mechanics reads
\be
{\cal I}=p^2_\theta +\frac{(p_{\varphi}\cos\theta -e)^2}{\sin^2\theta}+m^2.
\ee
Here $m$ and $e$ are the mass and the electric charge of a particle. This system  coincides  with the planar rotator \cite{gn,sag}.\\

{\bf Kerr BH}.
Spherical mechanics associated with the near horizon Kerr geometry is defined by the integrable but not exactly solvable system \cite{bny}
 \be {\cal
I}=p_\theta^2+\left[{\left(\frac{1+\cos^2 {\theta}}{2 \sin\theta}
\right)}^2-1\right] p^2_\varphi +\left(\frac{1+ \cos^2 {\theta}}{2}
\right) {(m r_0)}^2,\label{calint} \ee
where $m$ is the mass of a particle and $r_0$ is the horizon radius.\\

{\bf Kerr-Newman-AdS-dS BH.}
The Kerr-Newman-AdS-dS black hole is a solution of the Einstein-Maxwell
equations with a non-vanishing cosmological constant \cite{carter}.
Its near horizon limit has been constructed in \cite{strominger}, while the conformal mechanics on this background was built in  \cite{go}.
The Hamiltonian of the corresponding spherical mechanics reads
\be\label{rela}
{\cal I}=
\frac{p^2_\theta}{\alpha(\theta)}+\left(\frac{\Gamma(\theta)}{\gamma(\theta)}-k^2\right)\left[p_\varphi +e\lambda(\theta)\right]^2  +U(\theta).
\ee
It describes a particle probe on a two-dimensional curved space with the metric
\be
ds^2=\alpha(\theta){d\theta^2}+\frac{d\varphi^2}{\Gamma(\theta)/\gamma(\theta)-k^2},
\ee
which moves the in potential and magnetic fields defined by the expressions
\be
 U(\theta)=m^2\Gamma(\theta) -\frac{e^2k^2f^2(\theta)}{\Gamma(\theta)/\gamma(\theta)-k^2},\qquad
 \lambda(\theta)d\varphi=\frac{\Gamma(\theta) f(\theta)}{\Gamma(\theta)-k^2\gamma(\theta)}d\varphi .
\ee
Here we denoted
\bea
&&
\Gamma(\theta)=\frac{r_0^2}{1+\nu_{+}^2}\left(1+\nu_{+}^2\cos^2\theta\right),  \quad
\alpha(\theta)=\left(\frac{r_+^2}{ r_0^2}\right)\frac{1+\nu^{2}_+}{1-\nu_{0}^2\cos^2\theta},\nonumber \\
&&\gamma(\theta)=\left[\frac{r_+(1+\nu^{2}_{+})}{1-\nu^{2}_0}\right]
\frac{(1-\nu_{0}^2\cos^2\theta)\sin^2\theta}{1+\nu_{+}^2\cos^2\theta} ,\\
&&f(\theta)=
\frac{1+\nu^2_+}{\nu_+(1-\nu^2_0)}\frac{\frac{q_e}{2}(1-\nu_{+}^2\cos^2\theta)+q_m\cos\theta}{1+\nu_{+}^2\cos^2\theta}
\nonumber
\eea
and used the following  notation for the constant parameters
\be
\nu_+\equiv\frac{a}{r_+},\quad \nu_0\equiv\frac{a}{l},\quad k\equiv 2\left(\frac{r_0}{r_+}\right)^2\frac{1+\nu_{0}^2}{\nu_+ (1+\nu_{+}^2)^2},
\quad
r_0^2=r^2_+\frac{(1+\nu^2_+ )(1-r_+^2/l^2)}{1+6r_+^2/l^2-3r_+^4/l^4-q^2/l^2}.
\ee
Above  $m$  and $e$ are the mass and the electric charge of a particle, $r_+$ is the horizon radius and $l^2$ is linked to the cosmological constant via $\Lambda=-3/l^2$
The  parameters $M$, $a$, $q_e$ and $q_m$ are related to the mass, angular momentum, electric and magnetic charges of
the black hole, respectively (for explicit relations see e.g. \cite{strominger})
\begin{equation}
a^2 = \frac{r_+^2(1+3r_+^2/ l^2)-q^2}{1-r_+^2/ l^2},\qquad
M = \frac{r_+[(1+r_+^2/ l^2)^2-q^2/ l^2]}{1-r_+^2/ l^2},
\end{equation}
This system reduces to the near-horizon Kerr particle when $q_e=q_m=0$ and
$l^2\rightarrow\infty$.\\

{\bf $5d$ Myers--Perry  BH.}
In the case of the five-dimensional near horizon Myers--Perry black hole
one reveals a three-dimensional integrable system governed by the Hamiltonian
$$
{\cal I}=\frac 14 p_\theta^2+
\frac{\rho_0^4}{4 {(a+b)}^2} \left[ \frac{p_{\tilde\phi}^2}{a^2\sin^2{\theta}}+\frac{p_{\psi}^2}{b^2\cos^2{\theta}}
-\frac{1}{\rho_0^2} \left(\frac{b}{a} p_{\phi}+\frac{a}{b} p_{\psi}\right)^2 \right]-
$$
\be
\frac14\left(\sqrt{\frac{b}{a}}{p_{\phi}+\sqrt{\frac{a}{b}}p_{\psi}}\right)^2+m^2 \rho_0^2, \qquad \rho_0^2=ab+a^2 \cos^2{\theta}+b^2 \sin^2{\theta}.
\label{H1}\ee
This system is integrable but not exactly solvable for arbitrary values of  rotational parameters $a$, $b$.
For the special case that the rotation parameters are equal to each other $a=b$ it becomes exactly-solvable and maximally superintegrable
\be\label{H4}
{\cal I}=\frac 14\left[ p_\theta^2+\frac{p_\phi^2}{\sin^2{\theta}}+\frac{p_\psi^2}{\cos^2{\theta}}-
\frac 32 (p_\phi+p_\psi)^{2}+8(mr_0)^2\right].
\ee
Fixing the momenta $p_\phi, p_\psi$ we arrive the one dimensional system on the circle given by the
modified P\"oshle-Teller potential.\\

{\bf $5d$ Myers-Perry-AdS-dS BH.}
A generalization of the five--dimensional rotating black hole solution by Myers and Perry to include a cosmological constant was constructed in \cite{hht}.
Its near-horizon limit  was built in  \cite{lmp}. The corresponding spherical mechanics reads
\bea\label{H3}
&&
\mathcal{I}=\frac 12 \Delta_\theta p_\theta^2+\frac 12 \left( \frac{\rho_0^4}{\Delta_\theta \sin^2{\theta}}-\frac{(1+r_0^2/l^2) b^2 \rho_0^2}{\Delta_\theta} -\frac{4 a^2 {(r_0^2+b^2)}^2}{4r_0^2 } \right) p_\phi^2+
\nonumber\\[2pt]
&&
+\frac 12 \left( \frac{\rho_0^4}{\Delta_\theta \cos^2{\theta}}-\frac{(1+r_0^2/l^2) a^2 \rho_0^2}{\Delta_\theta} -\frac{4 b^2 {(r_0^2+a^2)}^2}{4r_0^2 } \right) p_\psi^2-
\nonumber\\[2pt]
&&
-\left(\frac{ (1+r_0^2/l^2) a b \rho_0^2}{\Delta_\theta}+\frac{4a b(r_0^2+a^2)(r_0^2+b^2)}{4r_0^2 } \right) p_\phi p_\psi+g^2 \cos^2{\theta}.
\eea
Here
$g^2$ is a coupling constant which vanishes for $a=b$, $m$ is the particle mass and we denoted
\bea
&&
\Delta = \frac{1}{r^2}(r^2 + a^2)(r^2 + b^2)
(1 + r^2/l^{2}) - 2M, \qquad
\Delta_\theta=1 - (a^2 \cos^2\theta)/l^2 - (b^2 \sin^2\theta)/l^2,
\nonumber\\[2pt]
&&
\rho^2_0 = ab + a^2\cos^2\theta + b^2\sin^2\theta, \qquad
\Xi_a=1-a^2/l^{2}, \qquad
\Xi_b=1 - b^2/l^{2}.
\eea
The parameters $M$, $a$, and $b$ are linked to the mass and the angular momenta (for explicit relations see e.g. \cite{lmp}).
$l^2$ is taken to be positive for AdS and negative for dS and is related to the cosmological constant via $\Lambda=-6/l^2$.\\

{\bf Higher-dimensional rotating BH \cite{jhep13,jpcs,gal}.}

For the extremal black hole with equal rotation parameters the Hamiltonian of a spherical mechanics was derived in \cite{gal}.

In the case of $d=2n+1$ dimensions one finds
\bea\label{Hodd}
&&
\mathcal{I}=\sum_{i,j=1}^{n-1}(\delta_{ij}-\mu_i \mu_j) p_{\mu_i} p_{\mu_j}+
\sum_{i=1}^{n} \frac{p_{\phi_i}^2}{\mu_i^2} -\frac{(n+1)}{n} {\left(\sum_{i=1}^{n} p_{\phi_i} \right)}^2,
\eea
where $(\mu_i,p_{\mu_i})$, $i=1,\dots,n-1$ and $(\phi_j,p_{\phi_j})$, $j=1,\dots,n$ form canonical pairs obeying the conventional Poisson brackets $\{ \mu_i,p_{\mu_j}\}=\delta_{ij}$, $\{\phi_i,p_{\phi_j} \}=\delta_{ij}$ and $\mu_n^2$ entering the second sum in (\ref{Hodd}) is found from the unit sphere equation  $\sum_{i=1}^{n} \mu_i^2=1$.

For $d=2n$ the Hamiltonian, which governs the corresponding spherical mechanics, reads
\be
\label{Heven}
\begin{split}
\mathcal{I}=&\sum_{i,j=1}^{n-1}((2n-3) \rho_0^2 \delta_{ij}-\mu_i \mu_j) p_{\mu_i} p_{\mu_j}+
\\
+&\sum_{i,j=1}^{n-1}\left(\frac{(2n-3)\rho_0^2}{\mu_i^2} \delta_{ij}-\frac{{(2n-3)}^2 \rho_0^2}{2(n-1)}-\frac{2}{n-1}
\right) p_{\phi_i} p_{\phi_j}+m^2 \rho_0^2,
\\
\rho_0^2=&\frac{2(n-1)}{2n-3}-\sum_{i=1}^{n-1} \mu_i^2,
\end{split}
\ee
where $(\mu_i,p_{\mu_i})$ and $(\phi_j,p_{\phi_j})$, $j=1,\dots,n-1$ form canonical pairs  and $m^2$ is a coupling constant. Note that, as compared to the previous case, the number of the azimuthal coordinates
is decreased by one.

Because the azimuthal angular variables $\phi_i$ are cyclic, it is natural to consider a reduction in which they are discarded. This is achieved by
setting in (\ref{Hodd}) and (\ref{Heven}) the momenta canonically conjugate to $\phi_i$ to be coupling constants
\be\label{red}
p_{\phi_i} \quad \rightarrow \quad g_i.
\ee
Note that, after such a reduction, both (\ref{Hodd}) and (\ref{Heven}) yield dynamical systems, which contain $(n-1)$ configuration space degrees of freedom.

Considering the reduction over the cyclic variables and investigating the integrability, we established in \cite{jhep13} that the spherical mechanics corresponding to the $(2n+1)$-dimensional black hole is a maximally superintegrable and exactly solvable system, i.e. it is completely similar to the five-dimensional black hole with the coinciding rotational parameters.
In contrast with this case, the spherical mechanics corresponding to the $2n$-dimensional black hole lacks only one constant of the motion to become maximally superintegrability and is not exactly solvable. The solution of its equations of motion is given by elliptic integrals and is similar to that derived for the Kerr background in \cite{bny}.

\newpage
\subsection{Near horizon metrics}
\label{nearHorizonGeneral}
\subsubsection*{Kerr black hole }
The Kerr solution \cite{kerr} is the stationary axially symmetric solution of the vacuum Einstein equations,
which describes the rotating black hole with mass $M$  and angular momentum $J$.
It was discovered in 1963 as a solution of the vacuum Einstein equations describing the rotational black holes.
Its uniqueness, proven  by Carter \cite{car71}, as well as the separability of variables of  the particle moving
 in the Kerr background \cite{car} gave to the Kerr solution a special role in General Relativity.

A very particular case of the Kerr solution, when  the Cauchy's
and event horizons coincide is called extremal Kerr  solution \cite{bny}. In
this special case the angular momentum $J$ of the Kerr black hole
is related with the mass of the Kerr black hole $M$ by the
expression $J=\gamma M^2/c$ (in the following formulae we put the gravitational constant $\gamma=1$ and the speed of light $c=1$). In 1999 Bardeen and
Horowitz derived the near-horizon limit of the extremal Kerr
solution and found that the isometry group of the limiting metric
is $SO(1,2)\times U(1)$ \cite{bh}. They conjectured that the
extremal Kerr throat solution might admit a dual conformal theory
description in the spirit of AdS/CFT duality. The extensive study
of AdS/Kerr duality was initiated almost a decade later in
\cite{str} which continues today (see e.g. \cite{cms} and refs
therein). It is clear to the moment, that the near-horizon extremal
Kerr solution and its generalizations
  play a distinguished role in supergravity. Particularly,    their
 thermodynamical properties and connection to the string  theory allow one
 to expect that the quantum gravity should be closely related with these objects.

 The Kerr solution of the vacuum Einstein equations is defined by the metric \cite{bny}
\be
\label{kerr}
\begin{split}
{d s}^2&=-\left(\frac{\Delta-a^2 \sin^2 \theta }{\rho^2} \right) dt^2 +\frac{\rho^2} \Delta  d r^2+{\rho^2} d \theta^2 +
\\
&+\left(\frac{{(r^2+a^2)}^2-\Delta a^2 \sin^2 \theta }{\rho^2} \right) \sin^2 \theta  d \varphi^2
-\frac{2 a  (r^2+a^2-\Delta) \sin^2 \theta }{\rho^2} dt d \varphi,
\end{split}
\ee
where
\be
 \Delta =r^2+a^2-2 Mr, \qquad {\rho^2}=r^2+a^2 \cos^2  \theta ,\qquad a= \frac{J}{M}.
\ee

The extremal solution of the Kerr metric corresponds to the choice  $M^2=J$, so that the event horizon is at  $r=M$.
The limiting near-horizon metric  is given by the expression \cite{bh}
\be\label{kn}
{d s}^2=\left(\frac{1+\cos^2  \theta }{2} \right) \left[-\frac{r^2}{r_0^2} dt^2+\frac{r_0^2}{r^2} dr^2+r_0^2 d \theta^2 \right]+
\frac{2 r_0^2 \sin^2 \theta }{1+\cos^2  \theta  } {\left[d \varphi+\frac{r}{r_0^2} dt \right]}^2,\qquad r_0\equiv \sqrt{2}M.
\ee
The Kerr metric  admits the second rank Killing tensor \cite{wp}, which allows to
integrate the geodesic equation for a massive particle in Kerr space-time by quadratures \cite{car}.
The  limiting
Killing tensor  becomes  reducible, in the sense that
it can be constructed from the Killing vectors corresponding to the $SO(2,1) \times U(1)$ isometry group.

\newpage
\subsection{Four-dimensional black holes}
\label{4dimens}
Here we will construct  the action-angle
variables for the angular parts of the following two four-dimensional exactly
solvable systems:
\begin{itemize}
\item
Charged particle moving near the horizon of extremal Reissner-Nordstr\"om  black hole with magnetic charge
\item
Particle moving near the horizon of extremal Dilaton-Axion (Cl\'ement-Gal'tsov) black hole
\end{itemize}

The study of this problems not only fills the gap in the paper \cite{gn}, but also presents its own interest.

Let us reinforce the above
mentioned observations on the near-horizon dynamics of particle in the
background of extremal Kerr black hole \cite{bny}: the use of
action-angle variables allowed to find there a critical point
$|p_\varphi|=2 mM  $ (with $m$ being the mass of the probe particle,
$ M$ being mass of extremal Kerr black hole), where the trajectories become
closed. We will show that there are similar singular points in the
dynamics of charged particle moving near the horizons of extremal
Reissner-Nordstr\"om and Cl\'ement-Gal'tsov black holes. They are defined
by the relation $p_\varphi=\pm s$, where  $s=ep$ for the Reissner-Nordstr\"om case (with $e$  being the electric charge of
probe particle, and $p$ being  magnetic charge of extremal Reissner-Nordstr\"om black hole), and $s=e$ for the case of extremal
Cl\'ement-Gal'tsov black hole (with $e$ being the electric charge of probe particle). \\[18pt]

Let us shortly repeat the steps required for the construction of action angle variables of the spherical mechanics \ref{i} related to the four-dimensional extremal black hole systems. In this particular case we have a two dimensional system $(\omega_0, {\cal I})$,
\be
{\cal I}=L(\theta,p_\theta, p_\varphi )-q(p_\varphi)^2,\qquad \omega_0=dp_\theta\wedge d\theta+ dp_\varphi\wedge d\varphi,
\ee
To construct the action-angle variables, we should introduce the generating function
\be
S({\cal I}, p_\varphi, \theta, \varphi )
= p_\varphi \varphi
+\int_{\begin{array}{c}
{\cal I}={\rm const}\\
 p_\varphi={\rm const}
 \end{array}} p_\theta({\cal I}, p_\varphi, \theta) d\theta = p_\varphi \varphi + S_0({\cal I}, p_\varphi, \theta).
\label{s0}\ee
First, we define, by its use, the action variables
\be
I_{1}({\cal I}, p_\varphi)=\frac{1}{2\pi}\oint p_\theta({\cal I}, p_\varphi , \theta) d\theta,\quad I_2=p_\varphi,,
\ee

Then, interting the first expression we get ${\cal I}$ as a function of the action variables: ${\cal I}={\cal I}(I_1,I_2)$. Using this expression, we find the corresponding angle variables

\be
\Phi_{1,2}=\frac{\partial{S({\cal I}(I_1,I_2), I_2, \theta, \varphi )}}{\partial I_{1,2}}
\ee

\newpage
\subsection*{Reissner-Nordstr\"om black hole}
Here we construct the action-angle variables for the angular part of the conformal mechanics describing the motion of charged particle near horizon of extremal Reissner-Nordstr\"om black hole (which defines the electrically and magnetically charged static black hole configuration).

The extremal Reissner-Nordstr\"om black hole solution of Einstein--Maxwell equations reads
\cite{galaj}:
\be
{d s}^2=-{\left(1-\frac{M}{r}\right)}^2 dt^2+{\left(1-\frac{M}{r}\right)}^{-2} dr^2+r^2 d\Omega^2\ , \quad
A=-\frac{q}{r} dt+p \cos\theta d\varphi\ .
\ee
Here $M$, $q$, $p$ are the mass, the electric and magnetic charges, respectively, and
$d\Omega^2=d\theta^2+\sin^2\theta d\varphi^2$ is the standard metric on a sphere. For the extremal solution one has $M=\sqrt{q^2+p^2}$.
Throughout the paper we use units for which $G=1$.

The near horizon limit is most easily accessible in isotropic coordinates ($r~\rightarrow ~r-M$)
which cover the region outside the horizon only
\be\label{RN1}
{d s}^2=-{\left(1+\frac{M}{r}\right)}^{-2} dt^2+{\left(1+\frac{M}{r}\right)}^2 \left( dr^2+r^2 d\Omega^2\right)\ .
\ee
When $r\rightarrow 0$ the metric takes the form
\be\label{RN-nearhorizon}
{d s}^2=-{\left(\frac{r}{M}\right)}^2 dt^2+{\left(\frac{M}{r}\right)}^2 dr^2+M^2 d\Omega^2\ ,
\ee
while implementing the limit in the two--form field strength, one finds the background vector field
\be\label{Pot}
\quad
A=\frac{q}{M^2} r dt+p \cos\theta d\varphi\ .
\ee
The last two lines give the Bertotti-Robinson solution of Einstein--Maxwell equations.

Notice that
in the literature on the subject it is customary to use other coordinates where the horizon is at $r=\infty$.
In particular, the use of these coordinates facilitates the analysis in \cite{cdkktp}. In this paper we refrain from
using such a coordinate system.

From (\ref{RN-nearhorizon}) it follows that in the near horizon limit the space--time geometry is the product
of a two-dimensional sphere of radius $M$ and a two-dimensional pseudo Riemannian space--time with the metric
\be\label{ads2}
{d s}^2=-{\left(\frac{r}{M}\right)}^2 dt^2+{\left(\frac{M}{r}\right)}^2 dr^2\ .
\ee
The latter proves to be the metric of $AdS_2$. To summarize, the background geometry is that of the $AdS_2 \times S^2$ space--time with 2--form flux.

Having fixed the background fields, we then consider the action of a relativistic particle
on such a background
\be\label{start}
S=-\int d t \left(m\sqrt{ {(r/M)}^2-{(M/r)}^2 {\dot r}^2 -M^2 ({\dot\theta}^2
+\sin^2 \theta {\dot\varphi}^2)   }+eqr/M^2 +e p \cos \theta \dot\varphi~\right)\ .
\ee
Here $m$ and $e$ are the mass and the electric charge of a particle, respectively.

The particle dynamics is most easily analyzed within the Hamiltonian formalism. Introducing the momenta $(p_r,p_\theta,p_\varphi)$ canonically
conjugate to the configuration space variables $(r,\theta,\varphi)$, one finds the Hamiltonian
\be\label{h}
H=(r/M)\left(\sqrt{m^2+{(r/M)}^2 p_r^2 +{(1/M)}^2(p_\theta^2+\sin^{-2}\theta {(p_\varphi+e p \cos\theta)}^2)} +e q/M\right)\ ,
\ee
which generates  time translations.
In agreement with the isometries of the background metric one also finds the conserved quantities
\be
\begin{split}
\label{kd}
K=&M^3/r \left(\sqrt{m^2+{(r/M)}^2 p_r^2 +{(1/M)}^2(p_\theta^2+\sin^{-2}\theta {(p_\varphi+e p \cos\theta)}^2)} -e q/M\right)+
\\
&+t^2 H+2tr p_r\ , \qquad D=tH+r p_r\ ,
\end{split}
\ee
which generate special conformal transformations and dilatations, respectively. Together with the Hamiltonian they form $so(2,1)$ algebra.

From these expressions we immediately get the angular part of our system
\be
{\cal
I}=p^2_\theta+\frac{(p_\varphi+s\cos\theta)^2}{\sin^2\theta}+(mM)^2-(eq)^2,\quad
\omega = dp_\theta\wedge d\theta+dp_\varphi\wedge d\varphi.
\label{br}\ee
It is precisely the spherical Landau problem (Hamiltonian system, describing the motion of the particle on the sphere
in the presence of constant magnetic field generated by Dirac monopole), shifted on the constant
${\cal I}_0 =(mM)^2-(eq)^2$.
Here and through this section we will use the notation
\be
s=ep\;,
\ee
which is precisely the  Dirac's ``monopole number".

For the construction of action-angle variables of the obtained system,
let us introduce the generating function (\ref{s0}), where the second term looks as follows
\be  S_0({\cal I}, p_\varphi , \theta)= \int_{{\cal I}={\rm
const}} d\theta\sqrt{{\cal I}-
(mM)^2+(eq)^2-\frac{(p_\varphi + s\cos\theta)^{2}}{\sin^2\theta}}=\nonumber\ee
 \be
=\left\{
\begin{array}{cc}
2|s|\arcsin\frac{|s|}{\sqrt{{\widetilde {\cal I}}}} \cot\frac\theta2-2\sqrt{{\widetilde {\cal I}}+s^2}\arctan \frac{\sqrt{{\widetilde {\cal I}}+s^2}\cot\frac\theta2}{\sqrt{{\widetilde {\cal I}}-s^2\cot^2\frac\theta2}}
&{\rm for}\; p_\varphi=s\\
-2|s|\arcsin\frac{|s|}{\sqrt{{\widetilde {\cal I}}}} \tan\frac\theta2+2\sqrt{{\widetilde {\cal I}}+s^2}\arctan \frac{\sqrt{{\widetilde {\cal I}}+s^2}\tan\frac\theta2}{\sqrt{{\widetilde {\cal I}}-s^2\tan^2\frac\theta2}}
&{\rm for}\; p_\varphi=-s\\

\sqrt{{\widetilde {\cal I}}+s^2} \left[
2\arctan t-\sum_{\pm}
\sqrt{((1\pm b)^2- a^2)}\arctan{\frac{(1\pm b)t\pm a}{\sqrt{(1\pm b)^2-a^2}}}\right],&{\rm for}\; |p_\varphi|\neq |s|
\end{array} \right. .
\ee
Here we introduce the notation
\be
\begin{split}
{\widetilde {\cal I}}&={\cal I}-(m M)^2+(e q)^2,
\\
a^2&=\frac{{\widetilde{\cal I}}^2 -(p^2_\varphi-s^2){\widetilde{\cal I}}}{({\widetilde{\cal I}}+s^2)^2},
 \quad b=-\frac{s p_\varphi}{{\widetilde{\cal I}}+s^2},
\\
t&=\frac{a-\sqrt{a^2-(\cos\theta -b)^2}}{\cos\theta -b}.
\end{split}
\ee
{\sl Hence, the equation $p_\varphi=\pm s$ defines critical points, where the system  changes its behaviour.}

For the non-critical values $p_\varphi \neq \pm s$ we get, by the use of standard methods \cite{arnold}, the following expressions for the action-angle variables,
\be
\label{actions1}
\begin{split}
I_1&=\sqrt{{\widetilde{\cal I}}+s^2} -\frac {| s+p_\varphi| +| s-p_\varphi |}{2},
\\
\Phi_1&=-\arcsin
\frac{({\widetilde{\cal I}}+s^2)\cos\theta+s p_\varphi}{{\widetilde{\cal I}}^2 -(p^2_\varphi-s^2){\widetilde{\cal I}}}
\\
I_2&=p_\varphi,
\\
\Phi_2&=\varphi+\gamma_1\Phi_1
+\gamma_2 \arctan\left(\frac{a-(1-b)t}{\sqrt{(1-b)^2-a^2}}\right)
+\gamma_3 \arctan\left(\frac{a+(1+b)t}{\sqrt{(1+b)^2-a^2}}\right)
\end{split}
\ee

where

\be
(\gamma_1,\gamma_2,\gamma_3)=\left\{
\begin{array}{lrrr}
sgn(I_2)(1,&-1,&1)& {\rm for }\;|I_2|>|s|\\
sgn(s)(0,&-1,&-1)& {\rm for }\;|I_2|<|s|
\end{array}
\right.
\ee
Respectively, the Hamiltonian reads
\be
\begin{split}
{\cal I}&=\left(I_1+\frac {| s+I_2| +| s-I_2 |}{2}\right)^2
+(mM)^2-(eq)^2-s^2=
\\
=&\left\{\begin{array}{cc}
(I_1+|I_2|)^2+(mM)^2-(eq)^2-s^2 &{\rm for }\; |I_2|>|s|\\
(I_1+|s|)^2+(mM)^2-(eq)^2-s^2 &{\rm for }\; |I_2|<|s|\\
\end{array}\right.
\end{split}
\ee

The effective frequencies $\Omega_{1,2}=\partial {\cal I}/\partial I_{1,2}$ looks as follows
\be
\begin{split}
{\Omega_1}&=\left\{\begin{array}{cc}
2(I_1+|I_2|) &{\rm for }\; |I_2|>|s|\\
2(I_1+|s|) &{\rm for }\; |I_2|<|s|
\end{array}\right. ,
\\
{\Omega_2}&=\left\{\begin{array}{cc}
2 (I_1+|I_2|) {\rm sgn} I_2&{\rm for }\; |I_2|>|s|\\
0 &{\rm for }\; |I_2|<|s|\\
\end{array}\right. .
\end{split}
\ee
It is seen, that  in subcritical  regime, $|I_2|< s $, the frequency $\Omega_2$ becomes zero, while
frequency $\Omega_1$  depends on $I_1$ only. In overcritical regime, when $|I_2|>s $, the frequencies $\Omega_1$ and $\Omega_2$
coincides modulo to sign:the frequency $\Omega_2$ is positive
for positive values of $I_2$ (which is precisely  angular momentum $p_\varphi $), and vice versa.
This is essentially different from the periodic motion in the spherical part of the ``Kerr particle" observed in \cite{bny}, where the critical point separated two phases, both of which corresponded to the two-dimensional motion, but with opposite sign of $\Omega_2$.

So, in both regimes the trajectories are closed, and the motion is effectively one-dimensional one.
It reflects the existence of the additional constant of motion in the system  (\ref{br}),
reflecting the  $so(3)$ invariance of the spherical Landau problem.
  In other words, it is superintegrable one.
 In action-angle variables the additional constant of motion reads
$I_{add}=\sin(\Phi_1-\Phi_2)$ (cf.\cite{gonera,hlnsy}).
%Let us notice, that to perform semiclassical quantization, we should impose  the conditions
%\be
%I_1=n_\theta+ \frac12,\quad n_\theta=0,1,2,\ldots\qquad I_2= n_\varphi, \quad n_\varphi =0,\pm 1,\pm2,\ldots
%\ee
%\subsubsection*{Special  case $|p_\varphi|=|s|$}
\\[15pt]

Now, let us write down the expressions for action-angle variables at the critical point $p_\varphi=\pm s$,
\be
I_1=2({\sqrt{\widetilde{\cal I}+s^2}}-|s|)
\;,\quad
\Phi_1=\left\{
\begin{array}{cc}
-2\arctan \frac{\sqrt{\widetilde{\cal I}+s^2}\cot\frac\theta2}{\sqrt{{\widetilde{\cal I}}-s^2\cot^2\frac\theta2}}
&{\rm for}\; p_\varphi=s\\
 2\arctan \frac{\sqrt{\widetilde{\cal I}+s^2}\tan\frac\theta2}{\sqrt{{\widetilde{\cal I}}-s^2\tan^2\frac\theta2}}
&{\rm for}\; p_\varphi=-s
\end{array} \right.\ee
Respectively, the Hamiltonian reads
\be
{\cal I}=(\frac{I_1}{2}+|s|)^2+(m M)^2-(e q)^2-s^2
\ee
Notice, that the obtained action variable is not the corresponding limit of (\ref{actions1}).

\newpage
\subsection*{Cl\'ement-Gal'tsov black hole}
Now, let us  consider the motion of a particle near the horizon of extremal rotating
Cl\'ement-Gal'tsov(dilaton--axion) black hole \cite{cg1}.
The conformal generators of this particle system (with mass $m$ and ``effective monopole number" $s$,which was refereed in \cite{cg,cg1} as  ``effective electric charge" $e$) read
\cite{cg}
\be
\begin{split}
H=&r \left(\sqrt{m^2+{(rp_r)}^2 +p_\theta^2+\sin^{-2}\theta
{[p_\varphi-s\cos\theta]}^2   } -p_\varphi \right),
\\
K=&\frac{1}{r} \left(\sqrt{m^2+{(rp_r)}^2 +p_\theta^2+\sin^{-2}\theta
{[p_\varphi-s \cos\theta]}^2   } +p_\varphi \right),
\\
D=&r p_r.
\end{split}
\ee
The Casimir of  conformal algebra is given by the expression
\be
{\cal I}=p^2_\theta +\frac{(p_{\varphi}\cos\theta -s)^2}{\sin^2\theta}+m^2.
\label{calIscal}\ee
The second term in  generating function for the action-angle variables (\ref{s0})
 can be explicitly integrated in elementary functions
(its explicit expression could be found, e.g. in Appendix in \cite{lny})
 \be   S_0=\int_{{\cal I}={\rm
const}} d\theta\sqrt{{ \cal I}-
m^2-\frac{(p_{\varphi}\cos\theta -s)^2}{\sin^2\theta}}=
\nonumber\ee
\be
=\left\{
\begin{array}{cc}
-2|s|\arcsin\frac{|s| \tan\frac\theta2}{\sqrt{{\cal I}-m^2}}+2\sqrt{{\cal I}-m^2+s^2}\arctan \frac{\sqrt{{\cal I}-m^2+s^2}
\tan\frac\theta2}{\sqrt{{\cal I}-m^2-s^2\tan^2\frac\theta2}}
&{\rm for}\; p_\varphi=s\\
2|s|\arcsin\frac{|s| \cot\frac\theta2}{\sqrt{{\cal I}-m^2}}-2\sqrt{{\cal I}-m^2+s^2}\arctan \frac{\sqrt{{\cal I}-m^2+s^2}
\cot\frac\theta2}{\sqrt{{\cal I}-m^2-s^2\cot^2\frac\theta2}}
&{\rm for}\; p_\varphi=-s\\
\sqrt{{\cal I}-m^2+p^2_\varphi} \left[
2\arctan t-\sum_{\pm}
\sqrt{((1\pm b)^2- a^2)}\arctan{\frac{(1\pm b)t \pm a}{\sqrt{(1\pm b)^2-a^2}}}\right],&{\rm for}\; |p_\varphi|\neq |s|
\end{array} \right.\ee
where we introduced the notation
\be
\begin{split}
&a^2\equiv\frac{({\cal I}-m^2)^2 +({\cal I}-m^2)(p^2_\varphi-s^2)}{({\cal I}-m^2+p^2_\varphi )^2},
\\
&b\equiv\frac{s p_\varphi}{{\cal I}-m^2+p^2_\varphi},
\\
&t=\frac{a-\sqrt{a^2-(\cos\theta -b)^2}}{\cos\theta -b}.
\end{split}
\ee
{\sl Hence, the equation $p_\varphi=\pm s $ defines critical points, where the system  changes its behaviour.}

For the non-critical values $p_\varphi \neq \pm s$ we get, by the use of standard methods \cite{arnold}, the
following expressions for the action-angle  variables,
\be
\begin{split}
&I_1=\sqrt{{\cal I}-m^2+p^2_\varphi} -\frac {| s+p_\varphi| +| s-p_\varphi |}{2},
\\
&\Phi_1=-\arcsin
\frac{({\cal I}-m^2+p^2_\varphi)\cos\theta-s p_\varphi}{\sqrt{({\cal I}-m^2)^2 +({\cal I}-m^2)(p^2_\varphi-s^2)}}
\label{actions2}
\end{split}
\ee
\be
\begin{split}
&I_2=p_\varphi,
\\
&\Phi_2=\varphi+\gamma_1\Phi_1
+\gamma_2 \arctan\left(\frac{a-(1-b)t}{\sqrt{(1-b)^2-a^2}}\right)
+\gamma_3 \arctan\left(\frac{a+(1+b)t}{\sqrt{(1+b)^2-a^2}}\right)
\end{split}
\ee
where
\be
(\gamma_1,\gamma_2,\gamma_3)=\left\{
\begin{array}{lrrr}
sgn(I_2)(1,&-1,&1)& {\rm for }\;|I_2|>|s|\\
sgn(s)(0,&1,&1)& {\rm for }\;|I_2|<|s|
\end{array}
\right.
\ee
Respectively, the Hamiltonian reads
\be
{\cal I}=\left( I_1+\frac{|s+I_2| +|I_2 -s|}{2} \right)^2-I^2_2+m^2 =\left\{\begin{array}{cc}
(I_1+|I_2|)^2-I^2_2+m^2 &{\rm for }\; |I_2|>|s|\\
(I_1+|s|)^2-I^2_2+m^2 &{\rm for }\; |I_2|<|s|\\
\end{array}\right. .
\ee

In the critical points $p_\varphi=\pm s$, the action-angle  variables read
\be
I_1=2(\sqrt{{\cal I}-m^2+s^2}-|s|),\qquad\Phi_1=\left\{
\begin{array}{cc}
 2\arctan \frac{\sqrt{{\cal I}-m^2+s^2}\tan\frac\theta2}{\sqrt{{\cal I}-m^2-s^2\tan^2\frac\theta2}}
&{\rm for}\; p_\varphi=s\\
-2\arctan \frac{\sqrt{{\cal I}-m^2+s^2}\cot\frac\theta2}{\sqrt{{\cal I}-m^2-s^2\cot^2\frac\theta2}}
&{\rm for}\; p_\varphi=-s
\end{array} \right.
\ee
Inverting the first expression, we shall get the expression for Hamiltonian
\be
{\cal I}=(\frac{I_1}{2}+|s|)^2+m^2-s^2
\ee
Let us notice, that the action variable at the critical point is different from  the corresponding limit of (\ref{actions2}).

To clarify the meaning of critical point let us calculate the effective frequencies  of the system,
 $\Omega_{1,2}=\partial {\cal I}/\partial I_{1,2}$,
\be
{\Omega_1}=\left\{\begin{array}{cc}
2(I_1+|I_2|) &{\rm for }\; |I_2|>|s|\\
2(I_1+|s|) &{\rm for }\; |I_2|<|s|\\
\end{array}\right. ,
\qquad
{\Omega_2}=\left\{\begin{array}{cc}
2I_1sgnI_2&{\rm for }\; |I_2|>|s|\\
-2 I_2 &{\rm for }\; |I_2|<|s|\\
\end{array}\right. .
\ee
It is seen from this expressions, that in contrast
with previous case, the system does not possess the hidden symmetries.
The motion is nondegenerated in noncritical regimes.
So, in contrast with Reissner-Nordstr\"om case, the system is essentially
two-dimensional one, and its trajectories are unclosed.
The frequencies $\Omega_1$ are the same in the both cases, while  $\Omega_2$ are essentially different.
Moreover,  frequency $\Omega_2$ behaves in essentially different ways in subcritical and overcritical regimes.
 In the first case it is proportional to $I_1$, and in the second case to $I_2$.

\newpage
\setcounter{equation}{0}
\section{HIGHER-DIMENSIONAL GENERALIZATIONS}
\label{higherdim}
\subsection{Near horizon metrics}
\subsubsection*{Myers--Perry black hole in  $d=2n+1$ }

A vacuum solution of the Einstein equations describing the Myers--Perry black hole in $d=2n+1$ dimensions for the special case that all $n$ rotation parameters are equal reads \cite{mp}
\bea\label{ed2}
&&
ds^2=\frac{\Delta}{U} {\left(dt -a \sum_{i=1}^{n} \mu_i^2 d \phi_i\right)}^2-\frac{U}{\Delta} dr^2
-\frac{1}{r^2} \sum_{i=1}^{n} \mu_i^2 {\left(a dt-(r^2+a^2) d\phi_i\right)}^2
\nonumber\\[2pt]
&& \qquad-(r^2+a^2) \sum_{i=1}^{n} d \mu_i^2+\frac{a^2 (r^2+a^2)}{r^2} \sum_{i<j}^{n} \mu_i^2 \mu_j^2 {\left(d\phi_i-d\phi_j \right)}^2,
\\[2pt]
&&
\Delta=\frac{{(r^2+a^2)}^n}{r^2}-2M, \quad U={(r^2+a^2)}^{n-1}, \quad \mu_n^2=1-\sum_{i=1}^{n-1} \mu_i^2,
\nonumber
\eea
where $M$ stands for the mass and $a$ is the rotation parameter. In what follows we focus on the extremal solution, for which
\be
M=\frac{n^n {r_0}^{2n-2}}{2}, \qquad a^2=(n-1) r_0^2.
\ee
These conditions follow from the requirement that $\Delta(r)$ has a double zero at the horizon radius $r=r_0$.

The isometry group of (\ref{ed2}) is $U(1)\times U(n)$. The first factor corresponds to time translations, while the second factor describes the enhanced symmetry $U(1)^n \to U(n)$, which occurs if all
rotation parameters of the black hole are set equal.
In order to make $U(n)$ explicit, one parametrizes $n$ spatial two--planes, in which the black hole may rotate, by the coordinates (see, e.g., Ref. \cite{vsp})
\be\label{planes}
x_i=r \mu_i \cos{\phi_i}, \qquad y_i=r \mu_i \sin{\phi_i},
\ee
where $i=1,\dots,n$, and constructs the vector fields
\be
\label{vf}
\begin{split}
&\xi_{ij}=x_i \frac{\partial}{\partial x_j}-x_j \frac{\partial}{\partial x_i}+y_i \frac{\partial}{\partial y_j}-y_j \frac{\partial}{\partial y_i},
\\
&\rho_{ij}=x_i \frac{\partial}{\partial y_j}-y_j \frac{\partial}{\partial x_i}+x_j \frac{\partial}{\partial y_i}-y_i \frac{\partial}{\partial x_j}.
\end{split}
\ee
These are antisymmetric and symmetric in their indices, respectively, and obey the structure relations of $u(n)$\footnote{The conventional structure relations of $u(n)$ are derived form (\ref{str}) by considering another basis $E_{ab}=\frac 12 (\xi_{ab}+i \rho_{ab})$, the Casimir elements of $u(n)$ being $C_1=E_{i_1 i_1}$, $C_2=E_{i_1 i_2} E_{i_2 i_1}$, \dots, $C_n=E_{i_1 i_2} E_{i_2 i_3} \dots E_{i_n i_1}$.}

\be
\label{str}
\begin{split}
&[\xi_{ij},\xi_{rs}]=\delta_{jr} \xi_{is} +\delta_{is} \xi_{jr} -\delta_{ir} \xi_{js}-\delta_{js} \xi_{ir},
\\
&[\rho_{ij},\rho_{rs}]=-\delta_{jr} \xi_{is} -\delta_{ir} \xi_{js} -\delta_{is} \xi_{jr}-\delta_{js} \xi_{ir},
\\
&[\xi_{ij},\rho_{rs}]=\delta_{jr} \rho_{is} +\delta_{js} \rho_{ir} -\delta_{ir} \rho_{js}-\delta_{is} \rho_{jr}.
\end{split}
\ee
It is straightforward to verify that (\ref{vf}) are the Killing vectors of the original black hole metric.
Another way to reveal the $U(n)$--symmetry  is to introduce the complex coordinates
\be
z_j=r \mu_j e^{i \phi_j}
\ee
and rewrite the metric in terms of them. In the complex notation the unitary symmetry is manifest.

In order to construct the near horizon metric, one redefines the coordinates \cite{gal}
\be
r \quad \rightarrow \quad r_0 + \epsilon r_0 r, \qquad t \quad \rightarrow \quad \frac{n r_0 t}{2(n-1)\epsilon}, \qquad \phi_i \quad \rightarrow \quad \phi_i+\frac{r_0 t}{2 a \epsilon}
\ee
and then sends $\epsilon$ to zero. This yields
\be
\label{Nhm1}
\begin{split}
&ds^2=r^2 dt^2-\frac{dr^2}{r^2}-2n(n-1)\sum_{i=1}^{n} d\mu_i^2-2 \sum_{i=1}^{n} \mu_i^2 {(r dt+n \sqrt{n-1} d \phi_i)}^2+
\\
&\qquad \quad +2n{(n-1)}^2 \sum_{i<j}^{n} \mu_i^2 \mu_j^2 {(d \phi_i-d\phi_j)}^2, \qquad \mu_n^2=1-\sum_{i=1}^{n-1} \mu_i^2.
\end{split}
\ee
It is straightforward to verify that (\ref{Nhm1}) is a vacuum solution of the Einstein equations.
The near horizon metric has a larger symmetry. In addition to $U(1)\times U(n)$ transformations considered above, the isometry group of (\ref{Nhm1}) includes the
dilatation
\be\label{tp1}
t'=t+\lambda t, \qquad r'=r-\lambda r,
\ee
and the special conformal transformation
\be\label{sp}
t'=t+(t^2+\frac{1}{r^2}) \sigma, \qquad r'=r-2 tr\sigma, \qquad
{\phi'}_i=\phi_i-\frac{2}{r n \sqrt{n-1} } \sigma,
\ee
which all together form $SO(2,1)\times U(n)$, the first factor being the conformal group in one dimension.

\newpage
\subsubsection*{Myers--Perry black hole in $d=2n$}

A vacuum solution of the Einstein equations describing the Myers--Perry black hole in $d=2n$ dimensions for the special case that all $n-1$ rotation parameters are equal, reads \cite{mp}
\be
\label{ed1}
\begin{split}
ds^2&=\frac{\Delta}{U} {\left(dt -a \sum_{i=1}^{n-1} \mu_i^2 d \phi_i\right)}^2-\frac{U}{\Delta} dr^2
-\frac{{(r^2+a^2)}^{n-2}}{r U} \sum_{i=1}^{n-1} \mu_i^2 \left(a dt-(r^2+a^2) d \phi_i \right)^2
\\
&-(r^2+a^2) \sum_{i=1}^{n-1} d\mu_i^2-r^2 d \mu_n^2 +\frac{a^2 {(r^2+a^2)}^{n-1}}{r U} \sum_{i<j}^{n-1} \mu_i^2 \mu_j^2 {\left(d\phi_i-d\phi_j \right)}^2,
\\
\Delta&=\frac{1}{r} {(r^2+a^2)}^{n-1}-2M, \quad U=\frac{1}{r} {(r^2+a^2)}^{n-2} (r^2+a^2 \mu_n^2), \quad \mu_n^2=1-\sum_{i=1}^{n-1} \mu_i^2,
\end{split}
\ee
where $M$ is the mass and $a$ is the rotation parameter. As compared to the previous case, the number of the azimuthal coordinates is decreased by one.
For the extremal solution $\Delta$ has a double zero at the horizon radius $r=r_0$. In particular, from $\Delta(r_0)=0$ and $\Delta'(r_0)=0$ one finds
\be
M=\frac{r_0^{2n-3} {[2(n-1)]}^{n-1}}{2}, \qquad a^2=(2n-3) r_0^2.
\ee

The isometry group of (\ref{ed1}) includes time translations and the enhanced rotational symmetry $U(1)^{n-1} \to U(n-1)$, which is a consequence of setting
all the rotation parameters equal. The unitary symmetry is manifest in the complex coordinates
\be
z_j=\mu_j e^{i \phi_j}=x_j+i y_j.
\ee
The corresponding Killing vector fields are realized as in Eq. (\ref{vf}) with $x_i$ and $y_i$ taken from the previous line.

In order to implement the near horizon limit, one redefines the coordinates
\be
r \quad \rightarrow \quad r_0 + \epsilon r_0 r, \qquad t \quad \rightarrow \quad \frac{2(n-1) r_0 t}{(2n-3)\epsilon}, \qquad \phi_i \quad \rightarrow \quad \phi_i+ \frac{r_0 t}{a \epsilon},
\ee
and then
sends $\epsilon$ to zero, which yields \cite{gal}
\be
\label{Nhm}
\begin{split}
ds^2 &= \rho_0^2 \left(r^2 dt^2-\frac{dr^2}{r^2} \right)-2(n-1)\sum_{i=1}^{n-1} d\mu_i^2-d\mu_n^2+
\\
&+\frac{2 (n-1)}{\rho_0^2} \sum_{i<j}^{n-1} \mu_i^2 \mu_j^2 {(d \phi_i-d\phi_j)}^2
-\frac{4}{{(2n-3)}^2 \rho_0^2} \sum_{i=1}^{n-1} \mu_i^2 {(r dt+(n-1) \sqrt{2n-3}d \phi_i)}^2,
\\
\rho_0^2 &=\frac{1+(2n-3) \mu_n^2}{2n-3}, \qquad \mu_n^2=1-\sum_{i=1}^{n-1} \mu_i^2.
\end{split}
\ee
It is straightforward to verify that (\ref{Nhm}) is a vacuum solution of the Einstein equations.
Like in $d=2n+1$, the near horizon metric exhibits additional conformal symmetry, which
is realized as in Eqs. (\ref{tp1}) and (\ref{sp}) with the obvious alteration of the special conformal transformation
\be
{\phi'}_i=\phi_i-\frac{2}{r (n-1) \sqrt{2n-3} } \sigma
\ee
acting on the azimuthal angular variables. Thus, for $d=2n$ the near horizon symmetry is $SO(2,1)\times U(n-1)$.

\newpage
\subsection{Extremal Myers-Perry black hole in $d=2n+1$}

For the spherical mechanics (\ref{Hodd}) associated with the extremal rotating black hole in $d=2n+1$ dimensions the reduction (\ref{red}) yields\footnote{We denote the reduced Hamiltonian by the same letter $\mathcal{I}$. This does not cause confusion, because, from now on, we abandon the parent formulations (\ref{Hodd}) and (\ref{Heven}).}
\bea\label{Hodd1}
&&
\mathcal{I}=\sum_{i,j=1}^{n-1}(\delta_{ij}-\mu_i \mu_j) p_{\mu_i} p_{\mu_j}+
\sum_{i=1}^{n}\frac{g_i^2}{\mu_i^2}, \qquad \mu_n^2=1-\sum_{i=1}^{n-1} \mu_i^2.
\eea
Since the first term in (\ref{Hodd1}) involves the inverse metric on an $(n-1)$--dimensional sphere,
the model can be interpreted as a particle moving on $\mathcal{S}^{n-1}$ in the external field.

The analysis of integrability of (\ref{Hodd1}) is facilitated in spherical coordinates. Introducing one angle at a time
\be
\mu_n=\cos\theta_{n-1},\quad \mu_i=x_i \sin\theta_{n-1}, \qquad \sum_{i=1}^{n-1} x^2_i=1
\ee
and computing the metric induced on the sphere $\sum_{a=1}^n d \mu_a^2$ and its inverse, one can bring (\ref{Hodd1}) to the form
\be
\label{Hodd2}
\begin{split}
&\mathcal{I}=p_{\theta_{n-1}}^2+\frac{g_n^2}{\cos^2{\theta_{n-1}}}+\frac{1}{\sin^2{\theta_{n-1}}} \left(
\sum_{i,j=1}^{n-2}(\delta_{ij}-x_i x_j) p_i p_j+
\sum_{i=1}^{n-1}\frac{g_i^2}{x_i^2}\right),
\\
&x_{n-1}^2=1-\sum_{i=1}^{n-2} x^2_i,
\end{split}
\ee
where $p_i$ are momenta canonically conjugate to $x_i$, $i=1,\dots,n-2$. Thus, the canonical pair $(\theta_{n-1},p_{\theta_{n-1}})$ is separated, while the expression in braces gives the first integral of the Hamiltonian (\ref{Hodd2}). Because its structure is analogous to (\ref{Hodd1}), one can proceed along the same lines
\be
x_{n-1}=\cos\theta_{n-2},\quad x_a=y_a \sin\theta_{n-2}, \qquad \sum_{a=1}^{n-2} y^2_a=1
\ee
until one achieves a complete separation of the variables. The resulting Hamiltonian is a kind of matryoshka doll
\bea\label{Hodd3}
&&
\mathcal{I}=F_{n-1},
\eea
where $F_{n-1}$ is derived from the recurrence relation
\be\label{F}
F_i=p_{\theta_i}^2+\frac{g_{i+1}^2}{\cos^2{\theta_{i}}}+\frac{F_{i-1}}{\sin^2{\theta_{i}}},
\ee
with $i=1,\dots,n-1$ and $F_0=g_1^2$. The functionally independent integrals of motion in involution $F_i$ ensure the integrability of (\ref{Hodd1}). To avoid confusion, let us stress that, given $n$, the Hamiltonian (\ref{Hodd3}) describes a system with $(n-1)$ configuration space degrees of freedom.
Note that in a different context this model has been discussed in \cite{vsp}. Worth mentioning also is that, if a system with the Hamiltonian $F_{i-1}$ has some integrals of motion, these automatically are the integrals of motion of a larger system governed by the Hamiltonian $F_i$. For $n=2$ Eq. (\ref{Hodd2}) reproduces the celebrated P\"oschl--Teller model \cite{pt}.

Although the integrability of (\ref{Hodd1}) is obvious in spherical coordinates, the fact that the model is maximally superintegrable is less evident. In order to prove it, we resort to the parent formulation (\ref{Hodd}) and analyze how the reduction (\ref{red})
affects the symmetries (\ref{vf}) \footnote{A realization of $U(n)$ in (\ref{Hodd}) is derived from Eq. (\ref{vf}) by the standard substitution
$\frac{\partial}{\partial \mu_i} \rightarrow p_{\mu_i}$, $\frac{\partial}{\partial \phi_i} \rightarrow p_{\phi_i}$, which links the Killing vectors to the first integrals of the Hamiltonian mechanics. The Hamiltonian (\ref{Hodd}) proves to be a combination of the first two Casimir elements
$\xi_{ij}^2+\rho_{ij}^2$ and ${(\rho_{ii})}^2$.}. First of all, we notice that $\rho_{ii}$ (no summation over repeated indices) generates rotation in the $i$--th plane. Within the canonical framework it is represented by $\rho_{ii}=2 p_{\phi_i}$. Then the very nature of the reduction mechanism (\ref{red}) suggests that those generators in (\ref{vf}), which Poisson commute with $\rho_{ii}$, will be symmetries of the reduced Hamiltonian (\ref{Hodd1}). Because (\ref{Hodd}) was constructed from the Casimir elements of $u(n)$, it is straightforward to verify that the combinations (no summation over repeated indices)
\be\label{I}
I_{ij}=\xi_{ij}^2+\rho_{ij}^2
\ee
with $i<j$  generate the desired symmetries.

Before we proceed to treat the general case, it proves instructive to illustrate the construction by the examples of $n=3$ and $n=4$, which correspond to  seven--dimensional and nine--dimensional black hole configurations. For $n=3$ the Hamiltonian reads\footnote{Here and in what follows the subscript attached to the Hamiltonian refers to the number of configuration space degrees of freedom in the model.}
\bea\label{Hodd4}
&&
{\mathcal{I}}_2=p_{\theta_{2}}^2+\frac{g_3^2}{\cos^2{\theta_{2}}}+\frac{1}{\sin^2{\theta_{2}}} \left(p_{\theta_1}^2+\frac{g_1^2}{\sin^2{\theta_1}}+\frac{g_2^2}{\cos^2{\theta_1}}
\right).
\eea
In order to construct the integrals of motion, one makes use of (\ref{planes}) and (\ref{vf})
\bea
&&
\xi_{12}=-p_{\theta_1} \cos{\phi_{12}} +\left(p_{\phi_1} \cot{\theta_1} +p_{\phi_2} \tan{\theta_1}\right) \sin{\phi_{12}},
\nonumber\\[4pt]
&&
\xi_{13}=-\left( p_{\theta_1} \cos{\theta_1} \cot{\theta_2} +p_{\theta_2}  \sin{\theta_1}\right) \cos{\phi_{13}} +
\left(p_{\phi_1} \frac{\cot{\theta_2}}{\sin{\theta_1}}+p_{\phi_3} \sin{\theta_1} \tan{\theta_2}\right) \sin{\phi_{13}},
\nonumber\\[4pt]
&&
\xi_{23}=\left( p_{\theta_1}  \sin{\theta_1} \cot{\theta_2} -p_{\theta_2} \cos{\theta_1}\right) \cos{\phi_{23}}+
\left(p_{\phi_2} \frac{\cot{\theta_2}}{\cos{\theta_1}}+p_{\phi_3} \cos{\theta_1} \tan{\theta_2}\right) \sin{\phi_{23}},
\nonumber\\[4pt]
&&
\rho_{12}=p_{\theta_1} \sin{\phi_{12}} +\left(p_{\phi_1} \cot{\theta_1} +p_{\phi_2} \tan{\theta_1}\right) \cos{\phi_{12}},
\nonumber\\[4pt]
&&
\rho_{13}=\left( p_{\theta_1} \cos{\theta_1} \cot{\theta_2} +p_{\theta_2}  \sin{\theta_1}\right) \sin{\phi_{13}} +
\left(p_{\phi_1} \frac{\cot{\theta_2}}{\sin{\theta_1}}+p_{\phi_3} \sin{\theta_1} \tan{\theta_2}\right) \cos{\phi_{13}},
\nonumber\\[4pt]
&&
\rho_{23}=-\left( p_{\theta_1}  \sin{\theta_1} \cot{\theta_2} -p_{\theta_2} \cos{\theta_1}\right) \sin{\phi_{23}}+
\left(p_{\phi_2} \frac{\cot{\theta_2}}{\cos{\theta_1}}+p_{\phi_3} \cos{\theta_1} \tan{\theta_2}\right) \cos{\phi_{23}},
\nonumber\\[4pt]
&&
\rho_{11}=2 p_{\phi_1}, \qquad \rho_{22}=2 p_{\phi_2}, \qquad \rho_{33}=2 p_{\phi_3},
\eea
where we abbreviated $\phi_{ij}=\phi_i-\phi_j$,  which after implementing the reduction (\ref{red}) yield
\bea\label{I1}
&&
{\tilde I}_{12}=p_{\theta_1}^2+\frac{g_1^2}{\sin^2{\theta_1}}+\frac{g_2^2}{\cos^2{\theta_1}},
\nonumber\\[4pt]
&&
{\tilde I}_{13}={(p_{\theta_1} \cos{\theta_1} \cot{\theta_2}+p_{\theta_2} \sin{\theta_1})}^2+{\left(g_1 \frac{\cot{\theta_2}}{\sin{\theta_1}}+g_3 \sin{\theta_1} \tan{\theta_2} \right)}^2,
\nonumber\\[4pt]
&&
{\tilde I}_{23}={(p_{\theta_1} \sin{\theta_1} \cot{\theta_2}-p_{\theta_2} \cos{\theta_1})}^2+{\left(g_2 \frac{\cot{\theta_2}}{\cos{\theta_1}}+g_3 \cos{\theta_1} \tan{\theta_2} \right)}^2.
\eea
It is straightforward to verify that the vectors $\partial_A {\tilde I}_{ij}$,  where $A=(\theta_1,\theta_2,p_{\theta_1},p_{\theta_2})$ are linearly independent and, hence, the first integrals are functionally independent. Because the Hamiltonian is constructed from ${\tilde I}_{ij}$ \footnote{Recall that the parent Hamiltonian (\ref{Hodd}) was constructed from the Casimir elements of $u(n)$. Up to a constant, the sum $\sum_{i<j=1}^n {\tilde I}_{ij}$ is what is left after the reduction.}
\bea
&&
{\mathcal{I}}_2={\tilde I}_{12}+{\tilde I}_{13}+{\tilde I}_{23}+g_3 (g_3-2 g_1 - 2 g_2),
\eea
one has three functionally independent integrals of motion for a system with two degrees of freedom and, hence, the model is maximally superintegrable. Note that the algebra formed by ${\tilde I}_{ij}$ is nonlinear.
It is convenient to treat the Hamiltonian  ${\mathcal{I}}_2$ (with the additive constant $g_3 (g_3-2 g_1 - 2 g_2)$ being discarded) and ${\tilde I}_{12}$ as the first integrals in involution, while ${\tilde I}_{23}$ is the additional first integral, which renders the model maximally superintegrable.

The case $n=4$ is treated likewise. From Eqs. (\ref{Hodd3}) and (\ref{F}) one derives the Hamiltonian
\bea
&&
{\mathcal{I}}_3=p_{\theta_3}^2+\frac{g_4^2}{\cos^2{\theta_3}}+\frac{1}{\sin^2{\theta_3}} \left[ p_{\theta_2}^2+\frac{g_3^2}{\cos^2{\theta_2}}+\frac{1}{\sin^2{\theta_2}} \left(p_{\theta_1}^2+\frac{g_2^2}{\cos^2{\theta_1}}+\frac{g_1^2}{\sin^2{\theta_1}} \right)\right],
\nonumber\\[2pt]
&&
\eea
while the first integrals prove to be exhausted by those in (\ref{I1}) and three more functions
\bea\label{I3}
&&
{\tilde I}_{14}={\left(p_{\theta_1} \frac{\cos{\theta_1} \cot{\theta_3}}{\sin{\theta_2}}+p_{\theta_2} \sin{\theta_1} \cos{\theta_2} \cot{\theta_3}+p_{\theta_3} \sin{\theta_1} \sin{\theta_2}\right)}^2+\left(g_1 \frac{\cot{\theta_3}}{\sin{\theta_1} \sin{\theta_2}}+
\right.
\nonumber\\[2pt]
&& \qquad \quad
{\left.
g_4 \sin{\theta_1} \sin{\theta_2}  \tan{\theta_3} \frac{}{} \right)}^2,
\nonumber\\[2pt]
&&
{\tilde I}_{24}={\left(p_{\theta_1} \frac{\sin{\theta_1} \cot{\theta_3}}{\sin{\theta_2}}-p_{\theta_2} \cos{\theta_1} \cos{\theta_2} \cot{\theta_3}-p_{\theta_3} \cos{\theta_1} \sin{\theta_2}\right)}^2+\left(g_2 \frac{\cot{\theta_3}}{\cos{\theta_1} \sin{\theta_2}}+
\right.
\nonumber\\[2pt]
&& \qquad \quad
{\left.
g_4 \cos{\theta_1} \sin{\theta_2}  \tan{\theta_3} \frac{}{} \right)}^2,
\nonumber\\[2pt]
&&
{\tilde I}_{34}={\left(p_{\theta_2} \sin{\theta_2} \cot{\theta_3} -p_{\theta_3} \cos{\theta_2} \frac{}{}\right)}^2+{\left(g_3 \frac{\cot{\theta_3}}{\cos{\theta_2}}+g_4 \cos{\theta_2} \tan{\theta_3} \right)}^2.
\eea
As in the preceding case, the Hamiltonian is a combination of ${\tilde I}_{ij}$
\be
{\mathcal{I}}_3=\sum_{i<j}^4{\tilde I}_{ij}+{(g_3-g_4)}^2-2 g_1 (g_3+g_4)-2 g_2 (g_3+g_4).
\ee
Because for a system with $n$ configuration space degrees of freedom the maximal number of functionally independent integrals of motion is $2n-1$, the set (\ref{I1}) and (\ref{I3}) is overcomplete and only five functions prove to be independent.

That for generic $n$ the model is maximally superintegrable can now be proved by induction. For $n=2$ the systems involves only one configuration space degree of freedom and the Hamiltonian is the only integral of motion. For $n=3$ we choose ${\mathcal{I}}_2$, ${\tilde I}_{12}$ and ${\tilde I}_{23}$ to be the functionally independent first integrals.
When passing from $n=3$ to $n=4$, the integrals of motion of the former model are automatically the integrals of motion of the latter. To complete the set, we choose ${\mathcal{I}}_3$ and ${\tilde I}_{34}$. Obviously, this process can be continued to any order. Given a superintegrable system with the Hamiltonian ${\mathcal{I}}_{n-1}$, $n-1$ configuration space degrees of freedom and $2(n-1)-1$ functionally independent integrals of motion, one introduces one more configuration space degree of freedom and two new integrals of motion ${\mathcal{I}}_{n}$ and
\be
{\tilde I}_{n-1, n}={\left(p_{\theta_{n-2}} \sin{\theta_{n-2}} \cot{\theta_{n-1}} -p_{\theta_{n-1}} \cos{\theta_{n-2}} \frac{}{}\right)}^2+{\left(g_{n-1} \frac{\cot{\theta_{n-1}}}{\cos{\theta_{n-2}}}+g_n \cos{\theta_{n-2}} \tan{\theta_{n-1}} \right)}^2,
\ee
which all together describe a system with $n$ configuration space degrees of freedom and $2n-1$ functionally independent integrals of motion.

Let us construct the action--angle variables for the system. Following the standard procedure \cite{arnold}, one introduces the generating function
\be\label{Sodd}
S^{odd}(F_i,|g_i|,\theta_i)=\sum_{i=1}^{n-1}\int p_{\theta_i}(F_1,\dots,F_{n-1}, \theta_i) \mathrm{d}\theta_i,
\ee
where $p_{\theta_i}(F_1,\dots,F_{n-1}, \theta_i)$  are to be expressed from  (\ref{F}).
For the action variables one has
\be\label{Iodd}
I_i=\frac{1}{2 \pi} \oint \mathrm{d}\theta_i \left[ \sqrt{F_i-\frac{F_{i-1}}{\sin^2\theta_i}-\frac{g_{i+1}^2}{\cos^2\theta_i}}\right] =\frac12(\sqrt{F_i}-\sqrt{F_{i-1}}-|g_{i+1}|),
\ee
which can be inverted to yield
\be
F_i=\left(2 \sum_{k=1}^i I_k + \sum_{k=1}^{i+1}|g_{k}|\right)^2.
\label{Fodd}\ee
The angle variables are defined by
\be
\Phi^{odd}_i=\frac{\partial S^{odd}}{\partial I_i}=\sum_{k=i}^{n-1} \arcsin X_k+2 \sum_{k=i+1}^{n-1} \arctan Y_k,
\label{Phiodd}\ee
where we abbriviated
\be
\begin{array}{ccl}
&&X_k  =  \frac{\left(F_k+F_{k-1}-g_{k+1}^2\right)-2 F_k \sin^2 \theta_k}{\sqrt{\left(-F_k+F_{k-1}-
g_{k+1}^2\right)^2-4 F_k g_{k+1}^2}}\\
&&Y_k
 =2\frac{
{\left(F_k+F_{k-1}-{g}_{k+1}^2\right)}
\sqrt{{F_k \sin^2 \theta_k \cos^2 \theta_k - F_{k-1} \cos^2 \theta_k
 -{g}_{k+1}^2\sin^2 \theta_k}}-{ \sin^2 \theta_k}\sqrt{{F_k\left(F_k+F_{k-1}-{g}_{k+1}^2\right)^2- F^2_kF_{k-1}}}}
 {\sqrt{F_{k-1}}\left(F_k+F_{k-1}-{g}_{k+1}^2)-2 F_k \sin^2 \theta_k\right)}

%%\\Z_k & = & -\frac{2F_k \cos^2 \theta_k }{\left(4 F_k-F_{k-1}+4 \hat{g}_{k+1}^2\right)-8F_k \cos^2 \theta_k }\sqrt{\frac{\left(4 F_k-F_{k-1}+4 \hat{g}_{k+1}^2\right)^2}{ 4F_k \hat{g}_{k+1}^2}-16} +\\
%%& & +\frac{\left(4 F_k-F_{k-1}+4 \hat{g}_{k+1}^2\right)}{\left(4 F_k-F_{k-1}+4 \hat{g}_{k+1}^2\right)-8F_k \cos^2 \theta_k }\sqrt{\frac{4F_k \sin^2 \theta_k \cos^2 \theta_k - F_{k-1} \cos^2 \theta_k -4\hat{g}_{k+1}^2\sin^2 \theta_k }{4 \hat{g}_{k+1}^2}}
\end{array}
\label{XY}\ee
Being rewritten in the action--angle variables, the Hamiltonian reads
\be
{\mathcal{I}}= \left(2 \sum_{k=1}^{n-1} I_k + \sum_{k=1}^n|{g}_{k}|\right)^2,
\ee
which coincides with the Hamiltonian of a free particle on an $(n-1)$--dimensional sphere up to the shift of the action variables \cite{hlnsy,hlnsy2}. Thus, the only difference with that case is the shift in the range of
$\sum_k I_k$ from $[0,\infty )$ to $[\sum_{k=1}^n|\hat{g}_{k}|, \infty ) $. Thus, the system possesses $SO(n+1)$ symmetry and is, obviously, maximally superintegrable.

Let us discuss how hidden constants of the motion can be revealed within the action--angle formulation. Evolution of the angle variables is governed by the equation (see, e.g., Refs. \cite{hlnsy,gonera})
\be
\frac{d \Phi^{odd}_i}{d t}=2\left(2 \sum_{k=1}^{n-1} I_k + \sum_{k=1}^n|{g}_{k}|\right).
\ee
The expressions $\cos(\Phi^{odd}_i-\Phi^{odd}_j+ {\rm const})$ define constants of the motion for any $i,j=1,\dots,n-1$ and only
$n-2$ of these are  functionally independent
\be
G_i=\cos\left(\Phi^{odd}_i-\Phi^{odd}_{i+1}\right)
%=\cos\left(\arcsin X_i + 2 \arctan Y_{i+1}\right)
=\frac{\sqrt{1-X_i^2}(1-Y_{i+1}^2)-2 X_i Y_{i+1}}{1+Y_{i+1}^2},
\label{hiddenAA}\ee
where $i=1,\dots,n-2$. Because the $(n-1)$--dimensional system has $(2n-3)$ functionally independent constants of the motion, it is maximally superintegrable.
The fact that the Hamiltonian is expressed via the action variables in terms of elementary functions implies also that the system is exactly solvable.

\newpage
\subsection{Extremal Myers-Perry black hole in $d=2n+2$}

For the spherical mechanics (\ref{Heven}) associated with the extremal rotating black hole in $d=2n$ dimensions the reduction (\ref{red}) yields
\bea\label{Final}
&&
\mathcal{I}=\sum_{i,j=1}^{n-1}((2n-3) \rho_0^2 \delta_{ij}-\mu_i \mu_j) p_{\mu_i} p_{\mu_j}+\sum_{i=1}^{n-1} \frac{(2n-3) \rho_0^2 g_i^2 }{\mu_i^2}+\nu \sum_{i=1}^{n-1} \mu_i^2,
\eea
where $\nu$ and $g_i$ are coupling constants and $\rho_0^2$ is given in (\ref{Heven}).

Like above, the proof of superintegrability of (\ref{Final}) is facilitated by introducing the spherical coordinates
\be
\mu_i=x_i \sin{\theta_{n-1}}, \qquad \sum_{i=1}^{n-1} x_i^2=1 \quad \Rightarrow \quad \sum_{i=1}^{n-1} \mu_i^2=\sin^2{\theta_{n-1}}.
\ee
In order to transform the kinetic term in (\ref{Final}), one inverts the metric then computes the line element in spherical coordinates and then inverts it again.
This yields
\be\label{HH}
\mathcal{I}=2(n-1) p^2_{\theta_{n-1}}+\nu \sin^2{\theta_{n-1}}+\left(\frac{2(n-1)}{\sin^2{\theta_{n-1}}}-2n+3 \right)
\left(
\sum_{i,j=1}^{n-2}(\delta_{ij}-x_i x_j) p_i p_j+
\sum_{i=1}^{n-1}\frac{g_i^2}{x_i^2}\right),
\ee
where $p_i$ are momenta canonically conjugate to $x_i$, $i=1,\dots,n-2$. Beautifully enough, the rightmost factor in (\ref{HH}) is the Hamiltonian of a particle on ${\mathcal{S}}^{n-2}$, which was studied in detail in the preceding Section. This sector provides $2(n-2)-1$ functionally independent integrals of motion, which correlates with the $U(n-1)$ symmetry of the parent formulation (\ref{Heven}). Because (\ref{HH}) involves one more canonical pair $(\theta_{n-1},p_{\theta_{n-1}})$ and only one extra integral of motion (the Hamiltonian (\ref{HH}) itself), the full theory lacks for only one integral of motion to be maximally superintegrable.

Let us construct action--angle variables for the system. In order to simplify the bulky formulae below, from now on we change the notation
$n \to n+1$, which corresponds to a black hole in $d=2(n+1)$ dimensions. To avoid confusion, the corresponding Hamiltonian will be denoted by ${\cal I}_{0}$
\be
{\cal I}_0=2 n p^2_{\theta_{n}}+\nu\sin^2\theta_{n} + \left(\frac{2 n}{\sin^2 \theta_{n}}-2n+1\right)F_{n-1},
\ee
with
$F_{n-1}$ given in (\ref{F}).
One starts with the generating function
\be
S^{even}=\sum_{i=1}^{n}\int p_{\theta_i}({\cal I}_0, F_1,\dots,F_{n-1}, \theta_i) \mathrm{d}\theta_i=
\int p_{\theta_n}({\cal I}_0, F_{n-1}, \theta_n)d\theta_n +S^{odd},
\label{Seven}\ee
where $S^{odd}$ has the structure similar to (\ref{Sodd}),
and the expression for $p_{\theta_n}$  is derived from the Hamiltonian ${\cal I}_{0}$.
The action variables $I_1,\ldots I_{n-1}$ coincide with those in the odd--dimensional case, while
for $I_{n}$ one gets
\be
I_{n}=\sqrt{\frac{-a^- \nu} {8n}}a^+ {\cal F}_1\left(\frac12,1,-\frac12,2,a^+,\frac{a^+}{a^-}\right),
\label{55}\ee
where ${\cal F}_1$ is Appell's first hypergeometric function (see e.g. \cite{dwight}) and
\be
a^\pm=1-\frac{{\cal I}_0}{2 \nu}-\frac{2n-1}{2 \nu}F_{n-1} \pm \sqrt{\left(1-\frac{{\cal I}_0}{2 \nu}-\frac{2n-1}{2 \nu}F_{n-1}\right)^2+\frac{{\cal I}_0}{\nu}-\frac{F_{n-1}}{\nu}-1}.
\label{56}\ee
Inverting this expressions, we would get the
Hamiltonian written in terms of the action variables. Unfortunately, this cannot be done in elementary functions.
While the system under consideration is integrable, it fails to be exactly solvable.

The angle variable conjugated to $I_n$  reads
\be
\Phi^{even}_{n}=\frac{\partial {\cal I}_0}{\partial I_{n}}\frac1{\sqrt{8 \nu n a^+}} {\cal F}\left(\arcsin\sqrt{\frac{a^+}{a^+-\cos^2 \theta_{n}}}, 1-\frac{a^-}{a^+}\right),
\ee
while other $(n-1)$ angle variables are defined by the expressions
\be
\Phi^{even}_i=\Phi^{odd}_i- A \Pi\left(1-\frac1{a^+},\arcsin\sqrt{\frac{a^+}{a^+-\cos^2 \theta_{n}}}, 1-\frac{a^-}{a^+}\right)+ B  {\cal F}\left(\arcsin\sqrt{\frac{a^+}{a^+-\cos^2 \theta_{n}}}, 1-\frac{a^-}{a^+}\right),
\ee
where $\Phi^{odd}_i$ were defined in the preceding section, ${\cal F}(\phi | m)$ is the elliptic integral of the first kind,
$\Pi(n; \phi | m)$ is the elliptic integral of the third kind, and we abbriviated
\be A=\sqrt{\frac{8 n F_{n-1}}{\nu}}\frac{1}{\sqrt{a^+}(a^+-1)},\quad
B=A+\frac{\sqrt{2 F_{n-1}}}{\sqrt{n \nu a^+}}\left(\frac{\partial {\cal I}_0}{\partial F_{n-1}}+2 n-1\right).
\ee
It follows  from (\ref{55}) and (\ref{56}), that the  ratio of the effective frequencies ${\omega_1}=\partial{\cal I}/\partial{I_{n}}$ and ${\omega_2}=\partial{\cal I}/\partial{F_{n-1}}$ is not a rational number. Furthermore, it is a function of the action variables.
Hence, although $(\omega_2\Phi_n - \omega_1\Phi_i)$ commute with the Hamiltonian ${\cal I}_0$, they are not periodic. As a result, using these functions one cannot define additional globally defined constants of the motion (for a related discussions see \cite{gonera,hlnsy}).
All hidden symmetries of the model are thus contained in (\ref{hiddenAA}). Because the
$n$-dimensional system has $n+(n-2)=2n-2$ constants of the motion, it lacks for only one first integral to be maximally superintegrable system.

 \newpage
\setcounter{equation}{0}
\section*{SUMMARY}
\label{summary}
\addcontentsline{toc}{section}{SUMMARY}

Here we recollect the main results of this thesis.
\begin{itemize}
\item We have suggested using the action-angle variables for the study of a (quasi)particle in quantum ring. We have presented the action-angle variables for three two-dimensional singular oscillator systems, which play the role of the confinement potential for the quantum ring. The first one is the usual (Euclidean) singular oscillator in constant magnetic field, the other two are spherical generalizations of the first one - singular Higgs oscillator and singular $CP^1$ oscillator in constant magnetic field.

\item We have suggested a procedure of constructing new integrable systems form the known ones, by adding a “radial” part to the “angular” Hamiltonian. Using this method we have constructed a class of integrable generalizations of oscillator and Coulomb systems on $N$-dimensional Euclidian space $R^N$, sphere $S^N$ and hyperboloid $H^N$. We have computed the explicit expressions for action-angle variables for systems with oscillator and Coulomb potentials on Euclidean space, on the sphere and on the hyperboloid.

As an example, we have constructed the spherical ($S^N$) and pseudospherical ($H^N$) generalization of the two-dimensional superintegrable models introduced by Tremblay, Turbiner and Winternitz and by Post and Winternitz. We have demonstrated the superintegrability of these systems and have written down their hidden constants of motion.

\item We have developed the methods of study of conformal mechanics, based on separation of the radial and angular parts of the Hamiltonian of the system.

We have studied the angular part as a new Hamiltonian system with finite motion and suggested a method to construct the constants of motion of the new system from the constants of motion of the initial conformal system.

We have  illustrated the effectiveness of this method on the example of the rational $A_3$ Calogero model and its spherical mechanics (which defines the cuboctahedric Higgs oscillator). For the latter we have constructed a complete set of functionally independent constants of motion, proving its superintegrability.

\item We have closely explored the conformal mechanics associated with near-horizon motion of massive relativistic particle in the field of extremal black holes in arbitrary dimensions by separating the radial and angular parts of conformal mechanics and studying the angular part using the action-angle variables. We have proved, that by applying a proper canonical transformation one can bring the above mentioned model to the conventional conformal mechanics form. We have writted down the explicit expressions of the above mentioned canonical transformation for a large class of extremal black holes.

\item We have studied in details the near-horizon motion of massive relativistic particle in the field of extremal Reissner-Nordstr\"om  black hole with magnetic momentum. We have shown that the angular part of this system is equivalent to the spherical Landau problem, and has a hidden constant of motion. We have found a ``critical point" that divides the different phases of effective periodic  motion.

We have analysed the near-horizon motion of massive relativistic particle in the field of extremal Cl\'ement-Gal'tsov (Dilaton-Axion) black hole. In contrast with Reissner-Nordstr\"om case, the angular part of this system does not possess hidden constant of motion. We have found a critical  point that divide the phases (both effectively two-dimensional ones) of rotations  in opposite directions.

\item We have inspected the integrability of spherical mechanics models associated with the near horizon extremal Myers-Perry black hole in arbitrary dimension for the special case that all rotation parameters are equal.

As in the previous cases, we have extracted the spherical part of the initial Hamiltonian and studied it as a new system. We have performed a step-by-step transformation to generalized spherical coordinates, constructing a new constant of motion on every step.

We have proved the superintegrability of the new system and demonstrated that the spherical mechanics associated with the black hole in odd dimensions is maximally superintegrable, while its even-dimentsional counterpart lacks for only one constant of the motion to be maximally superintegrable.

\end{itemize}

\newpage
\section*{ACKNOWLEDGMENTS}

It is a pleasure to thank the many people who have knowingly or otherwise helped me in the preparation of this dissertation. I must first mention the name of my Academic supervisor, Professor Armen Nersessian. I owe great deal to him, who has helped me from shaping the idea of this research to getting it to final completion. 

I would like to express many thanks to supervisor of my Bachelor thesis  and permanent co-author of many papers Professor Tigran Hakobyan for his professional kin eye and readiness to help me.

I am grateful to  Professor Merab Gogberashvili from Tbilisi State University, for  co-supervising my thesis within the framework of the Regional Training Network in Theoretical Physics, funded by Volkswagenstiftung under contract No 86260.

I express my gratitude  to my co-authors,  Professor Stefano Bellucci from INFN-Laboratori Nazionali di Frascati,  Professor Anton Galajinsky from Tomsk Polytechnic University, Professor Olaf Lechtenfeld from Leibniz Universitat Hannover, and to my co-author and friend Dr. Vahagn Yeghikyan. Most of the results presented in this thesis, have been  obtained in collaboration with them. Special thanks to Professor Galajinsky, who constructed  near-horizon geometries of higher-dimensional extreme Perry-Myers black holes, and  found, in terms of initial coordinates,  the explicit expression of hidden constants of motion  of the particle moving in these backgrounds.

I am so much grateful to the Authority of the Faculty of Physics of Yerevan State University for providing a stimulating environment for the research.

My special thanks to the people who were always wonderfully helpful in discussing the topic with me during the seminars and assisting in all possible ways. My greatest thanks go to Professors Edvard  Chubaryan, Roland  Avagyan, Vadim Ohanyan, David Karakhanyan, Hayk Sarkisyan, Gohar  Harutyunyan, and all the staff of Academician Gourgen Sahakyan's Chair of Theoretical Physics. Discussions with all these people helped me observe the topic from different angles and shape my ideas.

Last but not least I thank the reviewers of my thesis, Professor Ruben Manvelyan, Professor Levon Mardoyan and Professor Sergey Krivonos from the Joint Institute for Nuclear Research (Dubna), who carefully read my thesis and made many useful comments.

%%%%% CLEAR DOUBLE PAGE!
\newpage{\pagestyle{empty}\cleardoublepage}


\newpage
\begin{thebibliography}{100}
\addcontentsline{toc}{section}{BIBLIOGRAPHY}

	\bibitem{arnold} V. I. Arnold. {\sl Mathematical methods in classical mechanics}, Nauka Publ., Moscow, (1973).

	\bibitem{goldstein} H. Goldstein. {\sl Classical Mechanics}, Addison-Wesley Press, Cambridge, (1950).

	\bibitem{lny} O. Lechtenfeld, A. Nersessian and V. Yeghikyan. {\sl Action-angle variables for dihedral systems on the circle}, Physics Letters    {\bf A 374} (2010) 4647. %[arXiv:1005.0464 [hep-th]].

	\bibitem{nano} V. Ya. Prinz, V.A. Seleznev, A.K. Gutakovsky, A.V. Chehovskiy, V.V. Preobrazhenskii, M.A. Putyato, T.A. Gavrilova. {\sl Free-standing and overgrown InGaAs=GaAs nanotubes, nanohelices and their arrays}, Physica E {\bf 6} (2000) 828;
	\\
	V Ya Prinz, D Grutzmacher, A Beyer, C David, B Ketterer, E Deckardt. {\sl A new technique for fabricating three-dimensional micro- and nanostructures of various shapes}, Nanotechnology {\bf 12} (2001), 399;
	\\
	O. G. Schmidt, K. Eberl, {\sl Nanotechnology: Thin solid films roll up into nanotubes} Nature {\bf 410}, 168(2001)

	\bibitem{lens} D. Leonard, M. Krishnamurthy, C. M. Reaves, S. P. Denbaars, P. M. Petroff. {\sl Direct formation of quantum-sized dots from uniform coherent islands of InGaAs on GaAs surfaces}, Applied Physics Letters {\bf 63}, 3203 (1993);
	\\
	D. Leonard, K. Pond, and P. M. Petroff. {\sl Critical layer thickness for self-assembled InAs islands on GaAs}, Physical Review {\bf B 50}, 11687 (1994)

	\bibitem{higgs} P. W. Higgs. {\sl Dynamical symmetries in a spherical geometry. I}, Journal of Physics {\bf A12}, 309 (1979).
	%%CITATION = JPAGB,A12,309;%%
	\\
	H. I. Leemon. {\sl Dynamical Symmetries In A Spherical Geometry. 2}, Journal of Physics  {\bf A12} (1979) 489.
	%%CITATION = JPAGB,A12,489;%%

	\bibitem{Bellucci} S. Bellucci and A. Nersessian. {\sl (Super)Oscillator on CP(N) and Constant Magnetic Field},
	Physical Review    D {\bf 67}, 065013 (2003)
	\\
	Erratum-ibid. Physical Review D {\bf 71}, 089901 (2005)
	% [arXiv:hep-th/0211070].
	%%CITATION = PHRVA,D67,065013;%%
	\\
	  S. Bellucci, A. Nersessian, A. Yeranyan,  {\sl Quantum mechanics model on Kaehler conifold}, Physical Review  {\bf D70}, 045006 (2004);
	  %%CITATION = PHRVA,D70,045006;%%
	\\
	S. Bellucci, A. Nersessian, A. Yeranyan, {\sl Quantum oscillator on CP(n) in a constant magnetic field}, Physical Review  {\bf D70}, 085013 (2004);
	  %%CITATION = PHRVA,D70,085013;%%
	\\
	S. Bellucci, A. Nersessian, A. Yeranyan. {\sl Hamiltonian reduction and supersymmetric mechanics with Dirac monopole}, Physical Review D {\bf 74}, 065022 (2006);
	\\
	S. Bellucci, A. Nersessian. {\sl Supersymmetric Kahler oscillator in a constant magnetic field}, Contributed
	to International Seminar on Supersymmetries and Quantum Symmetries SQS 03, Dubna, Russia, 24-29 Jul
	2003 %arXiv:hep-th/0401232.

	\bibitem{qr} A. Lorke, R. J. Luyken, A. V. Govorov, J. P. Kotthaus, J. M. Garcia, P. M. Petroff. {\sl Spectroscopy of Nanoscopic Semiconductor Rings}, Physical Review  Letters {\bf 84}, 2223 (2000).
	\\
	A. Fuhrer, S. Luscher, T. Ihn, T. Heinzel, K. Ensslin, W. Wegscheider, M. Bichler. {\sl Energy spectra of quantum rings}, Nature {\bf 413}, 822 (2001).

	\bibitem{qdch} T. Chakraborty and P. Pietlainen. {\sl Electron-electron interaction and the persistent current in a quantum ring}, Physical Review {\bf B50}, 8460 (1994).

	\bibitem{qdsw} W.-C. Tan and J.C.Inkson, {\sl Electron states in a two-dimensional ring - an exactly soluble model}, Semiconductor Science and Technology, {\bf 11}, 1635 (1996).

	\bibitem{comp} J. Simonin, C. R. Proetto, Z. Barticevic, G. Fuster. {\sl Single-particle electronic spectra of quantum rings: A comparative study}, Physical Review {\bf B70}, 205305 (2004).

	\bibitem{hsarkisyan} A. K. Atayan, E. M. Kazaryan, A. V. Meliksetyan, and H. A. Sarkisyan. {\sl Interband Magnetoabsorption in Cylindrical Quantum Layer with Smorodinsky-Winternitz Confinement Potential}, Journal of Computational and Theoretical Nanoscience {\bf 7}, (2010) 1165.

	\bibitem{TTW} F. Tremblay, A.V. Turbiner and P. Winternitz. {\sl An infinite family of solvable and integrable quantum systems on a plane},
	Journal of Physics  {\bf A 42} (2009) 242001 %[arXiv:0904.0738 [math-ph]];
	\\
	%%CITATION = ARXIV:0904.0738;%%
	F. Tremblay, A.V. Turbiner and P. Winternitz. {\sl Periodic orbits for an infinite family of classical superintegrable systems}, Journal of Physics {\bf A 43} (2010) 015202 %[arXiv:0910.0299 [math-ph]].

	\bibitem{TTWothers} C. Quesne. {\sl Superintegrability of the Tremblay-Turbiner-Winternitz quantum Hamiltonians on a plane for odd k},  Journal of Physics    {\bf A 43} (2010) 082001 %[arXiv:0911.4404 [math-ph]];
	\\
	C. Quesne. {\sl Chiral super-Tremblay-Turbiner-Winternitz Hamiltonians and their dynamical superalgebra},  Journal of Physics    {\bf A 43} (2010) 495203 %[arXiv:1008.4269 [math-ph]];
	\\
	E.G. Kalnins, J.M. Kress and  W. Miller Jr. {\sl Superintegrability and higher order integrals for quantum systems}, Journal of Physics {\bf A 43} (2010) 265205;
	\\
	A.J. Maciejewski,  M. Przybylska and H. Yoshida. {\sl Necessary conditions for classical super-integrability of a certain family of potentials in constant curvature spaces},  Journal of Physics    {\bf A 43} (2010) 382001. %[arXiv:1004.3854 [math-ph]].

	\bibitem{PW} S. Post and P. Winternitz. {\sl An infinite family of superintegrable deformations of the Coulomb potential},
	Journal of Physics    {\bf A 43} (2010) 222001 %[arXiv:1003.5230 [math-ph]].

	\bibitem{Alfaro} V. de Alfaro, S. Fubini and G. Furlan. {\sl Conformal Invariance In Quantum Mechanics}, Nuovo Cim. A 34 (1976) 569.

	\bibitem{bkls} I. Bredberg, C. Keeler, V. Lysov, A. Strominger. {\sl Cargese lectures on the Kerr/CFT correspondence}, Nuclear Physics B - Proceedings Supplements {\bf 216} (2011) 194.
	%[arXiv:1103.2355].

	\bibitem{cdkktp} P. Claus, M. Derix, R. Kallosh, J. Kumar, P.K. Townsend, A. Van Proeyen. {\sl Black Holes and Superconformal Mechanics},  Physical Review  Letters {\bf 81} (1998) 4553.
	%[hep-th/9804177]

	\bibitem{ikn} E. Ivanov, S. Krivonos, J. Niederle. {\sl Conformal and Superconformal Mechanics Revisited},  Nuclear Physics {\bf B 677} (2004) 485. %[hep-th/0210196].

	\bibitem{bk} S. Bellucci and S. Krivonos. {\sl N=2 supersymmetric particle near extreme Kerr throat}, Journal of High Energy Physics {\bf 1110} (2011) 014. %arXiv:1106.4453.

	\bibitem{galaj} A. Galajinsky. {\sl Particle dynamics on AdS(2) x S**2 background with two-form flux},
	  Physical Review   D {\bf 78} (2008) 044014
	  %[arXiv:0806.1629 [hep-th]].

	\bibitem{bnsy} S. Bellucci, A. Nersessian, A. Saghatelian, V. Yeghikyan. {\sl Quantum ring models and action-angle variables}, Journal of Computational and Theoretical Nanoscience {\bf 8}, 769 (2011)
	%[arXiv:1008.3865]

	\bibitem{monop} L. G. Mardoyan, L. S. Petrosyan and H. A. Sarkisyan. {\sl The charge - dyon bound system in the spherical quantum well},   Physical Review  {\bf A68}, 014103 (2003)
	% [arXiv:quant-ph/0304129].
	  %%CITATION = QUANT-PH/0304129;%%
	\\
	L. Mardoyan, A. Nersessian, H. Sarkisyan and V. Yeghikyan. {\sl Dipole transitions and Stark effect in the charge-dyon system},   Journal of Physics    {\bf A 40}, 5973 (2007)
		% [arXiv:cond-mat/0609768];
		%%CITATION = JPAGB,A40,5973;%%
	\\
	S. Bellucci, V. Ohanyan. {\sl Two-center quantum MICZ-Kepler system and the Zeeman effect in the charge-dyon system}, Physics Letters  {\bf A372}, 5765 (2008).

	\bibitem{si} C. Castelnovo, R. Moessner, S. L. Sondhi. {\sl Magnetic monopoles in spin ice}, Nature, {\bf 451}, 42 (2008).

	\bibitem{anosc} A. Nersessian, V. Yeghikyan. {\sl Anisotropic inharmonic Higgs oscillator and related (MICZ-)Kepler-like systems},  Journal of Physics {\bf A 41} 155203 (2008).

	\bibitem{fiht} G. M. Fichtenholz. {\sl Foundations of Differential and Integral Calculus}, 6th edition, 1966,   Nauka Publ,  Moscow(in Russian)

	\bibitem{aramyan} K. S. Aramyan. {\sl Integrable model of a two-dimensional singular spherical oscillator in a constant magnetic field}
	  Theoretical and Mathematical Physics {\bf 156}, 1075(2008)
		%%CITATION = TMFZA,156,131;%%

	\bibitem{mn} L. Mardoyan and  A. Nersessian. {\sl Oscillator potential for the four-dimensional Hall effect}, Physical Review   {\bf B72}, 233303 (2005);
	\\
	S. Bellucci, L. Mardoyan, A. Nersessian. {\sl Hyperboloid, instanton, oscillator}, Physics Letters {\bf B636}, 137 (2006).

	\bibitem{hkln} T. Hakobyan, S. Krivonos, O. Lechtenfeld and A. Nersessian. {\sl Hidden symmetries of integrable conformal mechanical systems},  Physics Letters {\bf A 374} (2010) 801. %arXiv:0908.3290

	\bibitem{hlnsy} T. Hakobyan, O. Lechtenfeld, A. Nersessian, A. Saghatelian, V. Yeghikyan. {\sl Integrable generalizations of oscillator and Coulomb systems via action-angle variables},  Physics Letters   {\bf A 376} (2012) 679
	%[arXiv:1108.5189][hep-th].

	\bibitem{hlnsy2} T.Hakobyan, O.Lechtenfeld, A.Nersessian, A.Saghatelian V.Yeghikyan. {\sl Action-angle variables and novel superintegrable systems}, Physics of Particles and Nuclei, 43 (2012) 577-582

	\bibitem{sch} E. Schr\"odinger. {\sl Method of Determining Quantum Mechanical Eigenvalues and Eigenfunctions}, Proc.  Roy.  Irish Soc.  {\bf 46} (1941) 9;
	\\
	E. Schr\"odinger. {\sl Further Studies on Solving Eigenvalue Problems by Factorization}, Proc.  Roy.  Irish Soc.  {\bf 46} (1941) 183.

	\bibitem{flugge} S. Fl\"ugge. {\sl Practical quantum mechanics 1}, Springer, 1971

	\bibitem{gonera} C. Gonera. {\sl Note on superintegrability of TTW model}, Physics Letters A {\bf 376} (2012) 2341.
	%arXiv:1010.2915 [math-ph].
	%%CITATION = ARXIV:1010.2915;%%

	\bibitem{hlns} T. Hakobyan, O. Lechtenfeld, A. Nersessian, A. Saghatelian. {\sl Invariants of the spherical sector in conformal mechanics},   Journal of Physics  {\bf A 44} (2011) 055205.
	%[arXiv:1008.2912].

	\bibitem{calogero69} F. Calogero. {\sl Solution of a Three-Body Problem in One Dimension},  Journal of Mathematical Physics {\bf 10} (1969) 2191.

	\bibitem{calogero71} F. Calogero. {\sl Solution  of  the  One-Dimensional  N-Body Problems with Quadratic and/or Inversely Quadratic Pair  Potentials},  Journal of Mathematical Physics {\bf 12} (1971) 419.

	\bibitem{moser} J. Moser. {\sl Three integrable Hamiltonian systems connected with isospectral deformations}, Advances in Mathematics {\bf 16} (1975) 197.

	\bibitem{algebra} J. Wolfes. {\sl On the three-body linear problem with three-body interaction},  Journal of Mathematical Physics {\bf 15} (1974) 1420;
	\\
	F. Calogero and C. Marchioro. {\sl Exact solution of a one-dimensional three-body scattering problem with two-body and/or three-body inverse-square potentials},  Journal of Mathematical Physics {\bf 15} (1974) 1425;

	\bibitem{polychronakos} M. Olshanetsky and A. Perelomov. {\sl Classical integrable finite-dimensional systems related to Lie algebras},  Physics Reports {\bf 71} (1981) 313;
	\\
	M. Olshanetsky and A. Perelomov. {\sl Quantum integrable systems related to lie algebras},  Physics Reports {\bf 94} (1983) 313;
	\\
	A.P. Polychronakos. {\sl The physics and mathematics of Calogero particles},  Journal of Physics  {\bf A 39} (2006) 12793.

	\bibitem{cuboct} T. Hakobyan,  A. Nersessian, V. Yeghikyan. {\sl Cuboctahedric Higgs oscillator from the Calogero model},  Journal of Physics  {\bf A 42} (2009) 205206.
	%[arXiv:0808.0430].

	\bibitem{sagh-calogero} A. Saghatelian. {\sl Constants of Motion of the Four-Particle Calogero Model}, Physics of Atomic Nuclei, 75, 10 (2012) 1288-1293

	\bibitem{woj83} S. Wojciechowski. {\sl Superintegrability of the Calogero-Moser system}, Physics Letters   {\bf A 95} (1983) 279.

	\bibitem{feher} L. Feher, I. Tsutsui, T. Fulop. {\sl Inequivalent quantizations of the three-particle Calogero model constructed by separation of variables}, Nuclear Physics   {\bf B 715} (2005) 713.

	\bibitem{bks} S. Bellucci, S. Krivonos and A. Sutulin. {\sl N=4 supersymmetric 3-particles Calogero model}, Nuclear Physics {\bf B 805} (2008) 24.

	\bibitem{decoupling} N. Gurappa and P.K. Panigrahi. {\sl Equivalence of the Calogero-Sutherland model to free harmonic oscillators}, Physical Review {\bf B 59} (1999) R2490;
	\\
	P. Ghosh. {\sl Super-Calogero-Moser-Sutherland systems and free super-oscillators: a mapping}, Nuclear Physics   {\bf B 595} (2001) 519;
	\\
	T. Brzezi\'nski, C. Gonera, and P. Ma\'slanka. {\sl On the equivalence of the rational Calogero-Moser system to free particles}, Physics Letters   {\bf A 254} (1999) 185.

	\bibitem{glp1} A. Galajinsky, O. Lechtenfeld and K. Polovnikov. {\sl Calogero models and nonlocal conformal transformations}, Physics Letters   {\bf B 643} (2006) 221;

	\bibitem{glp2} A. Galajinsky, O. Lechtenfeld and K. Polovnikov. {\sl N=4 superconformal Calogero models}, Journal of High Energy Physics {\bf 0711} (2007) 008;

	\bibitem{glp3} A. Galajinsky, O. Lechtenfeld and K. Polovnikov. {\sl N=4 mechanics, WDVV equations and roots}, Journal of High Energy Physics {\bf 0903} (2009) 113.

	\bibitem{hln} T. Hakobyan, O. Lechtenfeld and A. Nersessian. {\sl The spherical sector of the Calogero model as a reduced matrix model},  Nuclear Physics   {\bf B 858} (2012) 250. %arXiv:1110.5352.

	\bibitem{braden} H.W. Braden. {\sl  Rigidity, functional equations and the Calogero-Moser model}, Journal of Physics  {\bf A 34} (2001) 2197.

	\bibitem{wigner} E.P. Wigner.
	{\sl Gruppentheorie und ihre Anwendungen auf die Quantenmechanik der Atomspektren}, Vieweg Verlag, Braunschweig, 1931.

	\bibitem{angular1} L.C. Biedenharn and J.D. Louck.
	{\sl Angular Momentum in Quantum Physics}, Addison-Wesley, Reading, 1981.

	\bibitem{angular2} D.A. Varshalovich, A.N. Moskalev, and V.K. Khersonskii.
	{\sl Quantum Theory of Angular Momentum}, Moscow, 1988.

	\bibitem{lobach} T. Hakobyan and A. Nersessian. {\sl Lobachevsky geometry of (super)conformal mechanics},  Physics Letters   {\bf A 373} (2009) 1001.

	\bibitem{supint2} M.F. Ranada. {\sl Superintegrability of the Calogero-Moser system: Constants of motion, master symmetries, and time-dependent symmetries},  Journal of Mathematical Physics {\bf 40} (1999)  236;
	\\
	S. Benenti, C. Chanu, and G. Rastelli. {\sl The super-separability of the three-body inverse-square Calogero system},  Journal of Mathematical Physics {\bf 41} (2000) 4654;
	\\
	J.T. Harwood, R.G. McLenaghan, and R.G. Smirnov. {\sl Invariant classification of orthogonally separable Hamiltonian systems in Euclidean space},  Communications in Mathematical Physics {\bf 259} (2005) 679;
	\\
	R.G. Smirnov, P. Winternitz. {\sl A class of superintegrable systems of Calogero type},  Journal of Mathematical Physics {\bf 47} (2006) 093505;
	\\
	R. Sasaki and K. Takasaki. {\sl Explicit solutions of the classical Calogero and Sutherland systems for any root system},  Journal of Mathematical Physics \textbf{47} (2006) 012701.

	\bibitem{kuznetzov} V. Kuznetsov. {\sl Hidden symmetry of the quantum Calogero-Moser system}, Physics Letters {\bf A 218} (1996) 212.

	\bibitem{sasaki01} R. Caseiro, J.-P. Fransoise and R. Sasaki. {\sl  Quadratic Algebra associated with Rational Calogero-Moser Models}, Journal of Mathematical Physics {\bf 42} (2001) 5329.

	\bibitem{gn} A. Galajinsky and A. Nersessian. {\sl Conformal mechanics inspired by extremal black holes in d=4}, Journal of High Energy Physics {\bf 1111} (2011) 135. %arXiv:1108.3394.

	\bibitem{car} B. Carter. {\sl Global Structure of the Kerr Family of Gravitational Fields}, Physical Review  {\bf 174} (1968) 1559.

	\bibitem{wp} M. Walker, R. Penrose. {\sl On quadratic first integrals of the geodesic equations for type {22} spacetimes}, Communications in Mathematical Physics {\bf 18} (1970) 265.

	\bibitem{bgik} S. Bellucci, A. Galajinsky, E. Ivanov and S. Krivonos. {\sl AdS(2) / CFT(1), canonical transformations and superconformal mechanics}, Physics Letters   {\bf B 555} (2003) 99.
	  %[hep-th/0212204].

	\bibitem{cg} G. Cl\'ement, D. Gal'tsov. {\sl Conformal mechanics on rotating Bertotti-Robinson spacetime},  Nuclear Physics   {\bf B 619} (2001) 741
	%hep-th/0105237.

	\bibitem{bny} S. Bellucci, A. Nersessian and V. Yeghikyan. {\sl Action-angle variables for the particle near extreme Kerr throat},
	Modern Physics Letters {\bf A 27} (2012) 1250191. %arXiv:1112.4713.

	\bibitem{cg1} G. Cl\'ement, D. Gal'tsov. {\sl Bertotti-Robinson type solutions to Dilaton-Axion Gravity}, Physical Review  D {\bf 63} (2001) 124011
	%gr-qc/0102025.

	\bibitem{g2} A. Galajinsky. {\sl Particle dynamics near extreme Kerr throat and supersymmetry}, Journal of High Energy Physics {\bf 1011} (2010) 126
	%arXiv:1009.2341.

	\bibitem{go} A. Galajinsky, K. Orekhov. {\sl N=2 superparticle near horizon of extreme Kerr-Newman-AdS-dS black hole}, Nuclear Physics   {\bf B 850} (2011) 339
	%arXiv:1103.1047.

	\bibitem{lp} C. Leiva and M. Plyushchay. {\sl Conformal Symmetry of Relativistic and Nonrelativistic Systems and AdS/CFT Correspondence}, Annals of Physics {\bf 307} (2003) 372
	%hep-th/0301244.

	\bibitem{sag} A. Saghatelian. {\sl Near-horizon dynamics of particle in  extreme Reissner-Nordstr\"om and Cl\'ement-Gal'tsov  black hole backgrounds: action-angle variables}, Classical and Quantum Gravity {\bf 29} (2012) 245018. %arXiv:1205.6270.

	\bibitem{jpcs} Galajinsky A., Nersessian A., Saghatelian A. {\sl Action-angle variables for spherical mechanics related to near horizon extremal Myers-Perry black hole}, Journal of Physics: Conference Series. V. 474 (2013) P. 012019.
	
	\bibitem{gns14} Galajinsky A., Nersessian A., Saghatelian A. {\sl  Spherical Mechanics for a Particle Near the Horizon of Extremal Black Hole}, Physics of Particles and Nuclei Letters, Vol. 11 (2014), No. 7.

	\bibitem{carter} B. Carter. {\sl Hamilton-Jacobi and Schrodinger separable solutions of Einstein's equations}, Communications in Mathematical Physics 1968. V. 10  P. 280.

	\bibitem{strominger} Hartman T., Murata K., Nishioka T., Strominger A. {\sl CFT Duals for Extreme Black Holes }, Journal of High Energy Physics. 2009. V. 0904  P. 019.

	\bibitem{hht} Hawking S.W., Hunter C.J., Taylor-Robinson M.M. {\sl Rotation and the AdS CFT correspondence}, Physical Review  D. 1999. V. 59  P. 064005

	\bibitem{lmp} Lu H., Mei J., Pope C.N.  {\sl Kerr/CFT Correspondence in Diverse Dimensions}, Journal of High Energy Physics. 2009. V. 0904  P. 054.

	\bibitem{jhep13} Galajinsky A., Nersessian A., Saghatelian A. {\sl Superintegrable models related to near horizon extremal Myers-Perry black hole in arbitrary dimension}, Journal of High Energy Physics. 2013. V. 1306, P. 002

	\bibitem{gal} Galajinsky A. {\sl Near horizon black holes in diverse dimensions and integrable models}, Physical Review D.2013. V.87 P.024023

	\bibitem{kerr} R. P. Kerr.  {\sl Gravitational field of a spinning mass as an example of algebraically special metrics},
	Physical Review  Letters   {\bf 11} (1963) 237.

	\bibitem{car71} B. Carter. {\sl Axisymmetric black hole has only two degrees of freedom}, Physical Review   Letters   {\bf 26}, 331 (1971).

	\bibitem{bh} J.M. Bardeen and G.T. Horowitz. {\sl The Extreme Kerr Throat Geometry: A Vacuum Analog of $AdS_2~x~S^2$},  Physical Review  D {\bf 60} (1999) 104030. %hep-th/9905099.

	\bibitem{str} M. Guica, T. Hartman, W. Song, A. Strominger. {\sl The Kerr/CFT Correspondence},  Physical Review   D\  {\bf 80}, 124008 (2009).
	%[arXiv:0809.4266].

	\bibitem{cms} A. Castro, A. Maloney, A. Strominger. {\sl Hidden Conformal Symmetry of the Kerr Black Hole},  Physical Review D  {\bf 82},  024008 (2010)
	%[arXiv:1004.0996].

	\bibitem{mp} R.C. Myers and  M.J. Perry. {\sl Black holes in higher dimensional space-times},  Annals of Physics {\bf 172} (1986) 304.

	\bibitem{vsp} M. Vasudevan, K.A. Stevens and D.N. Page. {\sl Separability of the Hamilton-Jacobi and Klein-Gordon Equations in Kerr-de Sitter Metrics},  Classical and Quantum Gravity {\bf 22} (2005) 339. %gr-qc/0405125.

	\bibitem{pt} G. P\"oschl, E. Teller. {\sl Bemerkungen zur Quantenmechanik des anharmonischen Oszillators},  Zeitschrift fur Physik {\bf 83} (1933) 143.

	\bibitem{dwight} H.B. Dwight. {\sl Tables of integrals and other mathematical data}, 4th edition, The Macmillan Company, N.Y., 1961.


\end{thebibliography}
\end{document}